\documentclass[a4paper,fleqn,usenatbib,useAMS]{mnras}

\usepackage[T1]{fontenc}
\usepackage{ae,aecompl}
\usepackage{times}
\usepackage{amsmath}
\usepackage{amssymb}
\usepackage{float}
\usepackage{lmodern}
\usepackage{pdflscape}
\usepackage{booktabs}
\usepackage{subcaption}
\usepackage{epsfig}
\usepackage[titletoc,title]{appendix}
\usepackage[section]{placeins}

\def\apj{ApJ}
\def\mnras{MNRAS}
\def\nat{Nat}

\def\araa{ARA\&A}                
\def\aap{A\&A}                   
\def\apjs{ApJS}                  
\def\pasp{PASP}                  
\def\apjl{ApJ}                   

\def\ssr{SSRv}
\begin{document}

\newcommand{\DefineRemark}[2]{%
  \expandafter\newcommand\csname rmk-#1\endcsname{#2}%
}
\newcommand{\Remark}[1]{\csname rmk-#1\endcsname}

\DefineRemark{low_beta}{$1.75^{+0.15}_{-0.19}$}
\DefineRemark{mid_beta}{$1.46^{+0.22}_{-0.22}$}
\DefineRemark{high_beta}{$1.53^{+0.12}_{-0.13}$}
\DefineRemark{low_opening}{$9.3^{+19.6}_{- 3.9}$}
\DefineRemark{mid_opening}{$22.5^{+12.5}_{-10.7}$}
\DefineRemark{high_opening}{$12.5^{+12.3}_{- 7.4}$}
\DefineRemark{low_inc}{$9.6^{+22.2}_{- 3.7}$}
\DefineRemark{mid_inc}{$22.5^{+10.6}_{- 9.8}$}
\DefineRemark{high_inc}{$16.5^{+14.2}_{- 7.8}$}
\DefineRemark{low_kappa}{$0.10^{+0.26}_{-0.37}$}
\DefineRemark{mid_kappa}{$-0.18^{+0.23}_{-0.19}$}
\DefineRemark{high_kappa}{$0.14^{+0.26}_{-0.34}$}
\DefineRemark{low_gamma}{$3.06^{+1.26}_{-1.44}$}
\DefineRemark{mid_gamma}{ $3.16^{+1.25}_{-1.38}$}
\DefineRemark{high_gamma}{$2.32^{+1.54}_{-0.90}$}
\DefineRemark{low_xi}{$0.19^{+0.51}_{-0.14}$}
\DefineRemark{mid_xi}{$0.15^{+0.17}_{-0.09}$}
\DefineRemark{high_xi}{$0.17^{+0.34}_{-0.13}$}
\DefineRemark{low_Mbh}{$7.20^{+0.56}_{-0.76}$}
\DefineRemark{mid_Mbh}{$6.72^{+0.48}_{-0.42}$}
\DefineRemark{high_Mbh}{$6.89^{+0.51}_{-0.57}$}
\DefineRemark{low_fellip}{$0.40^{+0.37}_{-0.23}$}
\DefineRemark{mid_fellip}{$0.17^{+0.16}_{-0.11}$}
\DefineRemark{high_fellip}{$0.35^{+0.20}_{-0.19}$}
\DefineRemark{low_fflow}{$0.31^{+0.25}_{-0.20}$}
\DefineRemark{mid_fflow}{$0.26^{+0.17}_{-0.17}$}
\DefineRemark{high_fflow}{$0.28^{+0.18}_{-0.20}$}
\DefineRemark{low_thetae}{$27.3^{+43.2}_{-18.9}$}
\DefineRemark{mid_thetae}{$12.3^{+12.1}_{- 8.3}$}
\DefineRemark{high_thetae}{$17.9^{+27.8}_{-12.4}$}
\DefineRemark{low_taumean}{$3.01^{+1.47}_{-1.56}$}
\DefineRemark{mid_taumean}{$3.21^{+1.54}_{-1.72}$}
\DefineRemark{high_taumean}{$3.09^{+1.65}_{-1.26}$}
\DefineRemark{low_taumedian}{$1.21^{+0.60}_{-0.64}$}
\DefineRemark{mid_taumedian}{$1.60^{+0.96}_{-0.85}$}
\DefineRemark{high_taumedian}{$1.48^{+0.86}_{-0.64}$}

\title[Modelling the AGN broad line region using single-epoch spectra I]{Modelling the AGN broad line region using single-epoch spectra I. The test case of Arp 151}
\author[Raimundo et al.]{S. I. Raimundo$^{1}$\thanks{E-mail: sandra.raimundo$@$nbi.ku.dk}, A. Pancoast$^{2}$, M. Vestergaard$^{1, 3}$, M. R. Goad$^{4}$, A. J. Barth$^{5}$
\\
$^{1}$ DARK, Niels Bohr Institute, University of Copenhagen, Lyngbyvej 2, 2100 Copenhagen, Denmark\\
$^{2}$ Harvard-Smithsonian Center for Astrophysics, 60 Garden Street, Cambridge, MA 02138, USA\\
$^{3}$ Steward Observatory, University of Arizona, 933 N. Cherry Avenue, Tucson, AZ 85721, USA\\
$^{4}$ Department of Physics and Astronomy, University of Leicester, University Road, Leicester LE1 7RH, UK\\
$^{5}$ Department of Physics and Astronomy, 4129 Frederick Reines Hall, University of California, Irvine, CA, 92697-4575, USA}

\maketitle

\begin{abstract}
We show that individual (single-epoch) spectra of AGN can constrain some of the geometry and dynamics of the AGN broad line region. 
Studies of the cosmic influence of supermassive black holes are limited by the current large uncertainties in the determination of black hole masses. 
One dominant limitation is the unknown geometry, dynamics and line-of-sight inclination of the broad line region, used to probe the central black hole mass. Recent progress has been made to constrain the spatial and kinematic structure of the broad line region using dynamical modelling of AGN monitoring data and an underlying physical model for the broad line region.
In this work we test the ability of a modified version of this dynamical modelling code to constrain the broad line region structure using single-epoch spectra. 
We test our modelling code on single-epoch spectra of nearby Arp 151 by comparing our results with those obtained with monitoring data of this same object. 
We find that a significant fraction of the broad line region parameters can indeed be adequately constrained, with uncertainties that are comparable to, or at most a factor of $\sim$ a few higher than those obtained from modelling of monitoring data. 
Considering the wealth of available single-epoch spectroscopic observations, this method is promising for establishing the overall AGN population trends in the geometry and dynamics of the broad line region.
This method can be applied to spectra of AGN at low and high redshift making it valuable for studies of cosmological black hole and AGN evolution.
\end{abstract} 


\section{Introduction}
Our current theoretical model for Active Galactic Nuclei (AGN) associates the energy emitted by these objects with mass accretion onto a supermassive black hole \citep{lynden-bell69}.
The black hole mass is an essential parameter to understand the growth of black holes and their impact on the surrounding medium. A black hole mass measurement provides a snapshot of the mass growth stage of the black hole, allowing us to probe and constrain the evolution of black holes and their host galaxies (e.g. \citealt{shankar04}, \citealt{vestergaard04}, \citealt{raimundo09}). The maximum energy output of the AGN and its physical scale are also set by the mass of the black hole (see \citealt{fabian12} for a review), providing a reference scale for the possible impact of AGN feedback on the host galaxy. 

Black hole masses can rarely be measured directly. We often rely on indirect measurements, such as black hole and AGN scaling relationships, to determine the black hole mass (e.g. \citealt{vestergaard02}, \citealt{vestergaard&peterson06}, \citealt{kormendy&ho13}, \citealt{peterson14}). 
However, there are some objects for which the black hole mass can be determined directly and that are used to define and calibrate the scaling relations. In the local Universe the black hole mass can be determined directly by the measurement of stellar or gas dynamics affected by the black hole's gravitational potential (e.g. \citealt{kormendy&ho13}). These methods require nearby targets for which the achievable spatial resolution approaches the small scales where the gravitational potential of the black hole dominates. 
For black holes that are actively accreting, i.e. AGN, the black hole mass can be determined via reverberation mapping studies (\citealt{blandford&mckee82}, \citealt{peterson93}, \citealt{peterson14}). In the case of reverberation mapping the limitations are not associated with the resolvable spatial scales. They are instead associated with the required time cadence and long duration of the observations. Until the quality of monitoring data are sufficient to provide high-fidelity velocity-delay maps, there is also a need to assume a black hole mass scaling relation to account for the unknown BLR structure and kinematics.

In addition to the observational limitations outlined above, reverberation mapping studies require some prior knowledge or physical assumptions on the gas distribution in the vicinity of the black hole.
Reverberation mapping relies on the monitoring of the broad emission lines observed in some AGN.
These broad emission lines have velocities v$_{\rm FWHM}$ $ > 1500 - 2000$ km\,s$^{-1}$ and are emitted by rapidly orbiting gas in the vicinity of the black hole, in the so called broad line region (BLR) (e.g. \citealt{peterson00}). This gas is ionised by continuum photons from the accretion disc, originally produced during the process of gas accretion onto the supermassive black hole. Due to the very small distance between the BLR and the black hole (typically a few light-days to a few light-months for the most luminous AGN), this gas is a probe of the black hole dominated gravitational potential, and hence of the black hole mass. 
Converting the observed properties of the broad emission lines into a black hole mass estimate depends on the assumed gas distribution within the BLR and its inclination with respect to our line of sight (e.g. \citealt{peterson14}). 
However, the properties of the gas within the BLR are not well constrained (e.g. \citealt{collin-souffrin06}) and it is not clear what is the intrinsic geometric structure of the BLR of each AGN. 

There is some indication that the broad emission line profile contains information on the distribution of gas velocities in the BLR, where it was produced. This has been suggested based on observed correlations between line properties and AGN inclination (inferred from radio jets) (e.g. \citealt{wills&browne86}, \citealt{vestergaard00}), and from the modelling of single-epoch broad line profiles (e.g. \citealt{capriotti80}, \citealt{kwan&carroll82}, \citealt{eracleous94}, \citealt{rosenblatt94}, \citealt{kollatschny&zetzl13}, \citealt{storchi-bergmann17}). However, a single spectrum will only show a snapshot of the integrated emission and gas velocity distribution across the entire BLR.

Most progress in this area has been achieved through multi-epoch spectroscopy as used in reverberation mapping studies, to monitor the time-dependent changes in the line profile. 
Due to the time delayed response of the BLR with respect to (accretion disc) continuum variations, reverberation mapping campaigns can follow a variation (e.g. a strong increase) in the accretion disc emission as it propagates in time and energy through the BLR.
The effect of the ionising photons as they reach gas in the BLR can be traced in the spectra: each portion of the BLR responds according to its distance to the ionising continuum, its structure and its physical conditions as well as its dynamics \citep{peterson97}. This will be imprinted on the spectral profiles of the broad lines as the gas in the BLR responds to the continuum variations. The dynamics in particular will be probed via the velocity broadening and velocity changes to the line profiles. Measuring the delayed response of different portions of the BLR provides a method to infer the structure of the BLR, by using time resolved information as opposed to spatially resolved information.

Reverberation mapping studies have indeed been successful at determining the characteristic size of the BLR, i.e. the responsivity weighted distance (e.g. \citealt{peterson93}).
More recently, the first velocity delay maps that describe how the gas velocities change with time delay during the monitoring campaigns, were obtained from reverberation mapping techniques. 
By comparing the measured time lag as a function of velocity across the broad line with simple models for the kinematic state of the BLR 
it has been possible to identify that the BLR in nearby AGN span a range of dynamics, including virialized orbital motion, inflows and outflows (\citealt{bentz09}, \citealt{bentz10}, \citealt{denney10}, \citealt{grier13}, \citealt{pei17}, \citealt{xiao18}).

Additional progress has been achieved by direct modelling of the monitoring data using a geometrical and dynamical model such as by \cite{brewer11}, \citealt{pancoast11}, \citealt{li13}, \cite{pancoast14a} and \cite{pancoast18}. This model is able to derive quantitative constraints on the BLR geometry and dynamics parameters by assuming an underlying physical model for the BLR. Until now, 17 distinct low-redshift broad line AGN have been modelled using this technique (\citealt{brewer11}, \citealt{pancoast12}, \citealt{pancoast14b}, \citealt{grier17}, \citealt{williams18}). Although the physical parameters determined vary between AGN, the overall results from these studies seem to indicate that the BLR is a close to face-on thick disc-like structure with dynamics ranging from gas in bound elliptical orbits to gas in inflowing/outflowing trajectories, depending on the AGN. Until now these models have used high quality monitoring data (typically from reverberation mapping campaigns) to constrain the BLR geometry and dynamics parameters, which limits the number of AGN that can be modelled. 

Due to the necessary large observing time resources required, reverberation mapping campaigns have only been carried out in a small sub-sample of all known broad line AGN. There are $\sim$ 60 low-redshift AGN that have been the target of reverberation mapping campaigns to measure their black hole masses (e.g. \citealt{peterson04}, \citealt{bentz13}, \citealt{du14}, \citealt{derosa15}, see \citealt{bentz&katz15} for an updated list). Recent efforts to extend these studies to higher redshift (e.g. \citealt{shen16}, \citealt{grier17b}, \citealt{du18}) and higher luminosities (e.g. \citealt{lira18}, \citealt{du18}) have increased the sample size to more than 100. However, the feasibility of these studies decreases for high redshift and more luminous AGN that require observing times that can span years or decades (e.g. \citealt{kaspi07}, \citealt{lira18}). This makes it difficult to extend the BLR dynamical modelling analysis to a large sample of AGN.

Considering the significant resources required to obtain monitoring data, we have initiated a study to investigate to what extent single spectra (i.e. spectra measured at a single time instant as opposed to multiple epochs) can constrain the structural and dynamical parameters that define the BLR. 
The goal of this paper is to apply the direct modelling approach described by \cite{pancoast14b} and \cite{pancoast18} to single-epoch spectra.
We apply the model to Arp 151, an AGN that has previously been modelled and studied using multiple sets of monitoring data. By using Arp 151 as a test case and comparing our findings with previous results, we can evaluate how much information can be obtained from a single spectrum alone (i.e. in the absence of timing information).
If successful, this can be a powerful method to obtain the properties of the BLR of a much larger population of AGN than those targeted by reverberation mapping campaigns, as there is a very large database of AGN with single-epoch spectra available in the literature.
In this paper, we develop a modified version of the model presented by \cite{pancoast14a}, originally developed for AGN monitoring data. Our goal is to test if single-epoch spectra can constrain the structure of the BLR assuming an underlying physical model and an estimate for the size of the BLR. In Section \ref{sec:model} we give a brief description of past implementations of the model and introduce the key modifications that we carry out. We also describe the data used to constrain and test the model. In Section \ref{sec:results} we apply our modified model to single-epoch spectra of the well-studied AGN Arp 151. We compare and validate our results against previous modelling of the reverberation mapping dataset on Arp 151. In Section \ref{sec:discussion} we identify those model parameters that can be usefully constrained and to what degree of uncertainty, using our new model implementation. We also outline and discuss how our new approach can be extended and applied to AGN BLR population studies.
\section{Model and data description}
The broad emission lines observed in AGN spectra are produced when the gas distribution in the BLR reprocesses the incident ionising radiation.
The profile of the line produced will be a function of wavelength, and is affected by the spatial distribution, velocity and physical conditions of the gas in the BLR at the time of emission. The BLR properties can be parameterised using a model with a set of free parameters to account for the possible variety of BLR gas distributions and kinematics. In this section, we describe the model adopted to parameterise the BLR geometry and dynamics, our approach to test the model and the observational data used to constrain it.
\label{sec:model}
\subsection{Model description}
We use the broad line region phenomenological modelling code described by \cite{pancoast14a} and \cite{pancoast18}. 
The model defines the source of ionising photons and the structure and velocity of the gas in the broad line region via a set of free parameters. The AGN accretion disc is set as the ionising continuum source and modelled as a point source emitting isotropically. The broad line region gas emission is modelled using a large number of point particles with parameterised spatial and velocity distributions. The global particle distribution is flexible and can be, among others, a sphere, disc, shell or torus. These particles instantaneously reprocess the incident ionising continuum to produce line emission. The position and velocity of the particles will affect the time delay between changes in the continuum and later changes in the line emission, and the wavelength of the emitted line flux. The model first generates a continuum light curve based on Gaussian processes so that the ionising flux is know at arbitrary time instants. For a particular set of geometry and dynamics parameters and light curve parameters the model generates a representation of the broad line region. Each particle in the broad line region will reprocess the continuum emission according to a characteristic time delay. Therefore, an emission line will be composed of reprocessed radiation that corresponds to different instants of the ionising continuum light curve. Expected emission line profiles are then generated for that specific BLR representation. The emission line profiles and the continuum light curves generated by the model are then compared with observational data to constrain the model parameters.
\begin{figure}
\centering
\includegraphics[width=0.45\textwidth]{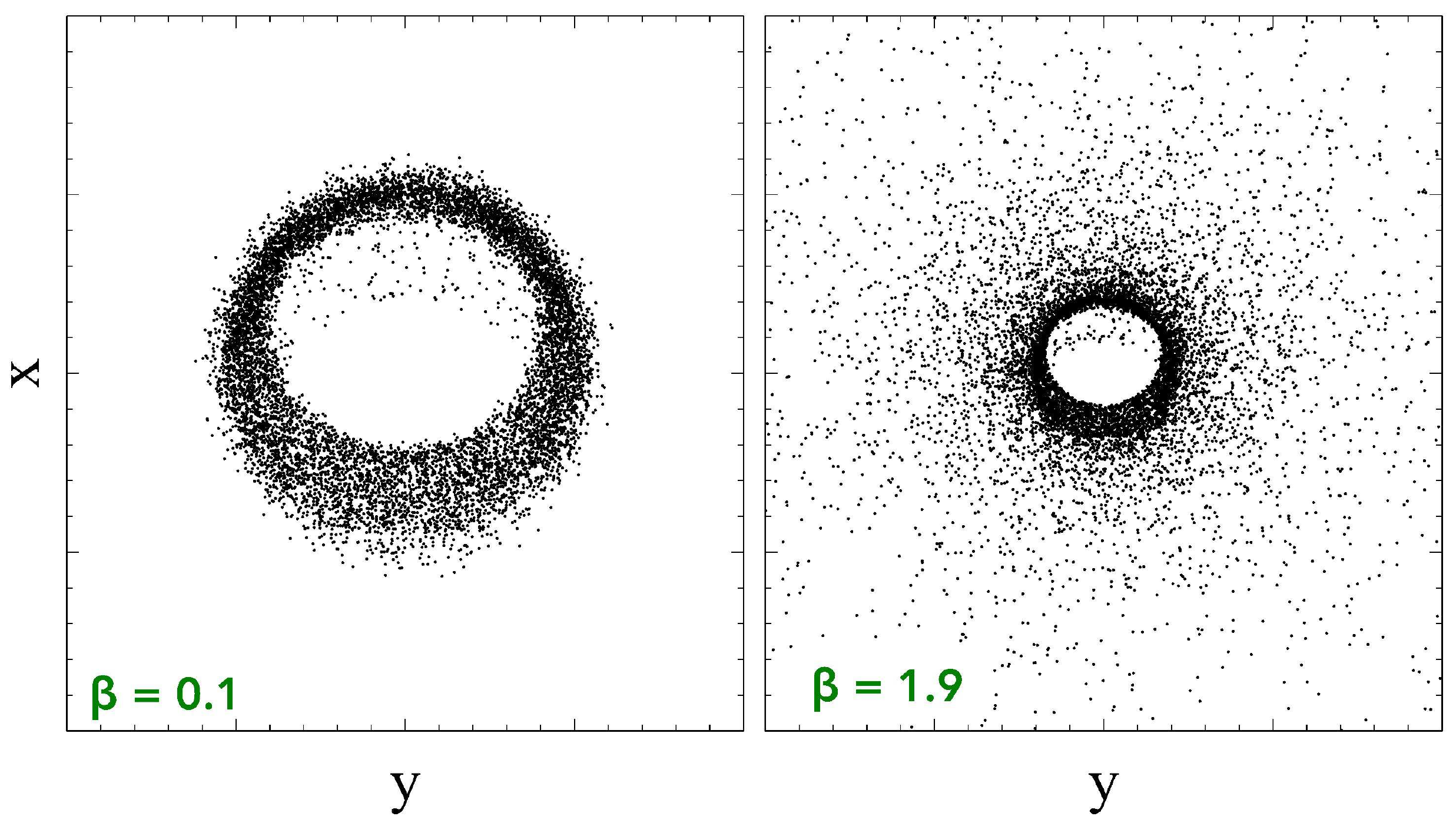}
\caption{Example of a BLR geometry for two different values of $\beta$. The black dots represent BLR clouds seen in the xy plane. The $\beta$ parameter describes how the clouds are distributed radially, hence $\beta < 1$ means that the distribution along the radial direction from the centre is Gaussian-like in shape. Left panel: $\beta$ = 0.1; Right panel: $\beta$ = 1.9. Both panels assume an inclination angle $\theta_{i}$ = 20$^{\circ}$ and an opening angle $\theta_{o}$ = 30$^{\circ}$.}
\label{beta_plots}
\end{figure}
The parameters are constrained via Bayesian inference using DNest3, a Diffusive Nested Sampling algorithm (\citealt{brewer09}, \citealt{brewer_dnest}). DNest3 is a Markov Chain Monte Carlo method based on Nested Sampling, an algorithm for Bayesian computation \citep{skilling06}. We use this method to explore the parameter space. It has the particular advantage of being able to efficiently explore complex multi-dimensional parameter spaces including handling strong correlations between the parameters.

The output of the model will be a multi-dimensional posterior probability distribution from which we extract the posterior probability distribution for each of the BLR parameters. In past implementations, the observational data consisted of a time series of broad emission line profiles and a time series of AGN continuum flux measurements such as those obtained from AGN monitoring campaigns. We describe our modified approach to dealing with single-epoch spectra in Section~\ref{sec:modifications}.
\subsubsection{Model parameters}
\label{sec:model_par}
The model uses a set of free parameters with associated prior probability distributions to characterise the geometry and dynamics of the broad line region. Below we give a summary of the parameters and show their assumed prior probability distributions in Table~\ref{parameter_table}. A more in depth parameter description is presented by \cite{pancoast14a}.

The particle radial distribution is modelled using a Gamma function, with the parameter $\beta$ controlling the shape of the distribution: $\beta < 1$ corresponds to a narrow Gaussian-like distribution, $\beta = 1$ to an exponential profile and $\beta > 1$ to a profile steeper than an exponential. The Gamma distribution is offset from the origin set by a minimum radius for the BLR. An example of the effect of $\beta$ in the BLR cloud distribution is shown in Fig~\ref{beta_plots}. The angular thickness of the BLR, as seen from the black hole, is defined by the opening angle $\theta_{o}$. This angle corresponds to the half angular thickness of the BLR, $\theta_{o} = 90^{\circ}$ for a sphere and approaching $\theta_{o} = 0^{\circ}$ for a thin disc. The BLR is oriented with an inclination angle $\theta_{i}$ with respect to the observer's line of sight: $\theta_{i} = 0^{\circ}$ for a face-on disc and $\theta_{i} = 90^{\circ}$ for an edge-on disc. 

There are three parameters that can break the symmetry in the BLR geometry: $\kappa$, $\gamma$ and $\xi$. 

\begin{table}
\centering
\begin{tabular}{c | c }
Parameter name & Prior \\
\hline
\textbf{$\beta$} & Uniform[0, 2] \\
\textbf{$\theta_{i}$}  & Uniform(cos($\theta_{i}$)(0, $\pi$/2))\\
\textbf{$\theta_{o}$}  & Uniform[0, $\pi/2$]\\
\textbf{$M_{BH}$} & LogUniform(2.78$\times$10$^{4}$, 1.67$\times$10$^{9}$M$_{\odot}$)\\
\textbf{$\kappa$} &  Uniform[-0.5, 0.5] \\
\textbf{$\gamma$} &  Uniform[1, 5] \\
\textbf{$\xi$} &  Uniform[0, 1] \\
\textbf{$f_{\rm ellip}$} & Uniform[0, 1]\\
\textbf{$f_{\rm flow}$} &  Uniform[0, 1]\\
\textbf{$\theta_{e}$} & Uniform[0,$\pi$/2]\\[0.2cm]
\multicolumn{2}{ l }{New feature in the modified model:}\\[0.1cm]
\textbf{$\tau_{\rm mean}$} & Gaussian($\mu_{\tau}$, $\sigma_{\tau}$) [days] \\
\hline\\
\end{tabular}
\caption{List of free parameters in the modelling and their corresponding prior probability.}
\label{parameter_table}
\end{table}

The parameter $\kappa$ defines if the particles emit isotropically or whether there is preferential emission towards the ionising source or away from it. For $\kappa = -0.5$ there is preferential emission towards the ionising source, for $\kappa = 0.5$ there is preferential emission away from the ionising source and for $\kappa = 0$ the emission is isotropic. Physically it represents an asymmetry in the BLR emission that could be caused by gas within the BLR blocking some of the emission or by an optical depth effect. 
The parameter $\gamma$ is used to control the particle displacement angle between the disc plane and the opening angle of the BLR and can have values between 1 and 5. For $\gamma = 1$ the particles are distributed uniformly and for $\gamma > 1$ the particles will be more displaced towards the opening angle of the BLR, i.e. more concentrated in the outer face of the BLR disc. Physically, $\gamma = 5$ may correspond to a situation in which the AGN continuum flux is obscured along the disk mid-plane. The parameter $\xi$ indicates the transparency of the mid-plane of the BLR. A value of $\xi = 1$ represents a transparent mid-plane and  $\xi = 0$ an opaque mid-plane. 

The black hole mass, $M_{\rm BH}$, sets the gravitational potential of the point particles and no other forces (such as radiation pressure, for example) are considered. 
The dynamics of the BLR is described by three main parameters that control the type of particle orbit: $f_{\rm ellip}$, $f_{\rm flow}$ and $\theta_{e}$. Here, $f_{\rm ellip}$ constrains the fraction of point particles with near-circular elliptical orbits. The expression $1 - f_{\rm ellip}$ corresponds to the fraction of point particles with orbits centred at around the radial inflowing or outflowing escape velocity. 
The binary parameter $f_{\rm flow}$ determines whether a point particle is on an inflowing orbit ($f_{\rm flow} < 0.5$) centred at the inflowing escape velocity or on an outflowing orbit ($f_{\rm flow} \geq 0.5$) centred on the outflowing escape velocity. Finally, the inflowing and outflowing velocity distributions can be rotated by $\theta_{e}$ towards the circular orbit velocity, so that an increasingly larger fraction of particles are in bound orbits as $\theta_{e} \rightarrow$ 90$^{\circ}$ - see \cite{pancoast14a} for further details.  

As in \citealt{pancoast14a} we also do not subtract the narrow emission line component before modelling the data, to avoid introducing uncertainties. The narrow emission line is modelled with a narrow Gaussian function, with the total narrow line flux and systematic central wavelength as free parameters, as described in Section 3.3 of \cite{pancoast18}.
 
Two additional parameters (C$_{\rm add}$ and C$_{\rm mult}$) are used to normalise the continuum. This makes the modelling independent of the absolute flux level of the continuum and of the emission line. The relevant timescales are characterised by the mean ($\tau_{\rm mean}$) and median ($\tau_{\rm median}$) time lag between the continuum and the line emission. In the original implementation of the model, these two parameters do not have an associated prior probability distribution but are calculated a posteriori from the final BLR geometry and dynamics distribution. We address the time lags differently in this work.
\subsubsection{Model modifications to analyse single-epoch spectra}
\label{sec:modifications}
The model first described by \cite{pancoast11} and \cite{pancoast14a} and in its most recent version by \cite{pancoast18} was originally built to use multi-epoch monitoring data to constrain the model parameters. That is, the data input for the model is a time series of broad emission line profiles and a time series of AGN continuum flux measurements. Our goal here is to instead use standard single-epoch spectra as the observational constraint: a spectrum with information on the continuum and the broad line emission measured at a single instant. 
We modify the model described by \cite{pancoast18} to use single-epoch spectra as data input, instead of monitoring data. The main modification we need to make is to add information of the absolute physical scale of the BLR, which we do via a prior probability distribution on the mean time delay between continuum and line emission.
Below we describe the modifications implemented on the model.

\begin{table}
\centering
\begin{tabular}{c | c | c | c}
 Parameter & No prior & Prior with  & Prior with \\
 & & $\sigma_{\tau} = 0.5\mu_{\tau}$ & $\sigma_{\tau} = 3\mu_{\tau}$ \\[0.05cm]
  $[1]$ & $[2]$ & $[3]$ & $[4]$\\
 \hline
$\tau_{\rm mean}$  (days)  &   $3.43^{+0.29}_{-0.32}$ &   $3.41^{+0.28}_{-0.27}$ &   $3.45^{+0.30}_{-0.27}$ \\ 
$\tau_{\rm median}$  (days)  &   $1.86^{+0.23}_{-0.24}$ &   $1.83^{+0.24}_{-0.22}$ &   $1.83^{+0.27}_{-0.25}$ \\ 
$\beta$ &   $1.35^{+0.13}_{-0.13}$ &   $1.37^{+0.13}_{-0.14}$ &   $1.38^{+0.14}_{-0.16}$ \\ 
$\theta_o$ (degrees) &   $26.4^{+ 3.9}_{- 5.9}$ &   $27.0^{+ 5.3}_{- 5.3}$ &   $26.9^{+ 5.9}_{- 5.6}$ \\ 
$\theta_i$ (degrees) &   $25.8^{+ 4.2}_{- 5.7}$ &   $26.4^{+ 5.0}_{- 5.2}$ &   $26.3^{+ 5.5}_{- 5.3}$ \\ 
$\kappa$ &   $-0.29^{+0.11}_{-0.10}$ &   $-0.32^{+0.09}_{-0.08}$ &   $-0.31^{+0.09}_{-0.09}$ \\ 
$\gamma$ &   $3.97^{+0.75}_{-1.10}$ &   $4.15^{+0.63}_{-1.07}$ &   $4.11^{+0.64}_{-1.01}$ \\ 
$\xi$ &   $0.09^{+0.12}_{-0.06}$ &   $0.10^{+0.08}_{-0.06}$ &   $0.09^{+0.08}_{-0.06}$ \\ 
$\log_{10}(M_{\rm BH}/M_\odot)$ &   $6.58^{+0.20}_{-0.12}$ &   $6.56^{+0.17}_{-0.13}$ &   $6.58^{+0.16}_{-0.15}$ \\ 
$f_{\rm ellip}$ &   $0.13^{+0.11}_{-0.09}$ &   $0.09^{+0.12}_{-0.07}$ &   $0.11^{+0.13}_{-0.08}$ \\ 
$f_{\rm flow}$ &   $0.27^{+0.15}_{-0.18}$ &   $0.25^{+0.17}_{-0.17}$ &   $0.25^{+0.18}_{-0.17}$ \\ 
$\theta_e$ (degrees) &   $12.7^{+11.3}_{- 8.6}$ &   $12.0^{+10.3}_{- 7.7}$ &   $12.9^{+11.0}_{- 9.1}$ \\
 $T$ & 65 & 65 & 80 \\
\end{tabular}
\caption{Table of inferred parameters for the tests in Fig. \ref{prior_noprior_posterior_full_light} using the full light-curve. The inferred parameter is the median value of the posterior probability distribution and the uncertainties quoted are the 68\% confidence intervals. $[1]$ Parameter name; $[2]$ Default simulation by \citealt{pancoast18}: full light-curve and no prior; $[3]$ Full light-curve with a prior on $\tau_{\rm mean}$ centred at $\mu_{\tau} = 3.07$ days and $\sigma_{\tau} = 0.5 \mu_{\tau}$; $[4]$ Full light-curve with a prior on $\tau_{\rm mean}$ centred at $\mu_{\tau} = 3.07$ days and $\sigma_{\tau} = 3 \mu_{\tau}$;  $[5]$ Temperature ($T$) used for each test. More details on $T$ can be found in the Appendix (Appendix~\ref{sec:appendix}.)} 
\label{table_results_light}
\end{table}

\begin{figure*}
\centering
\includegraphics[width=0.99\textwidth]{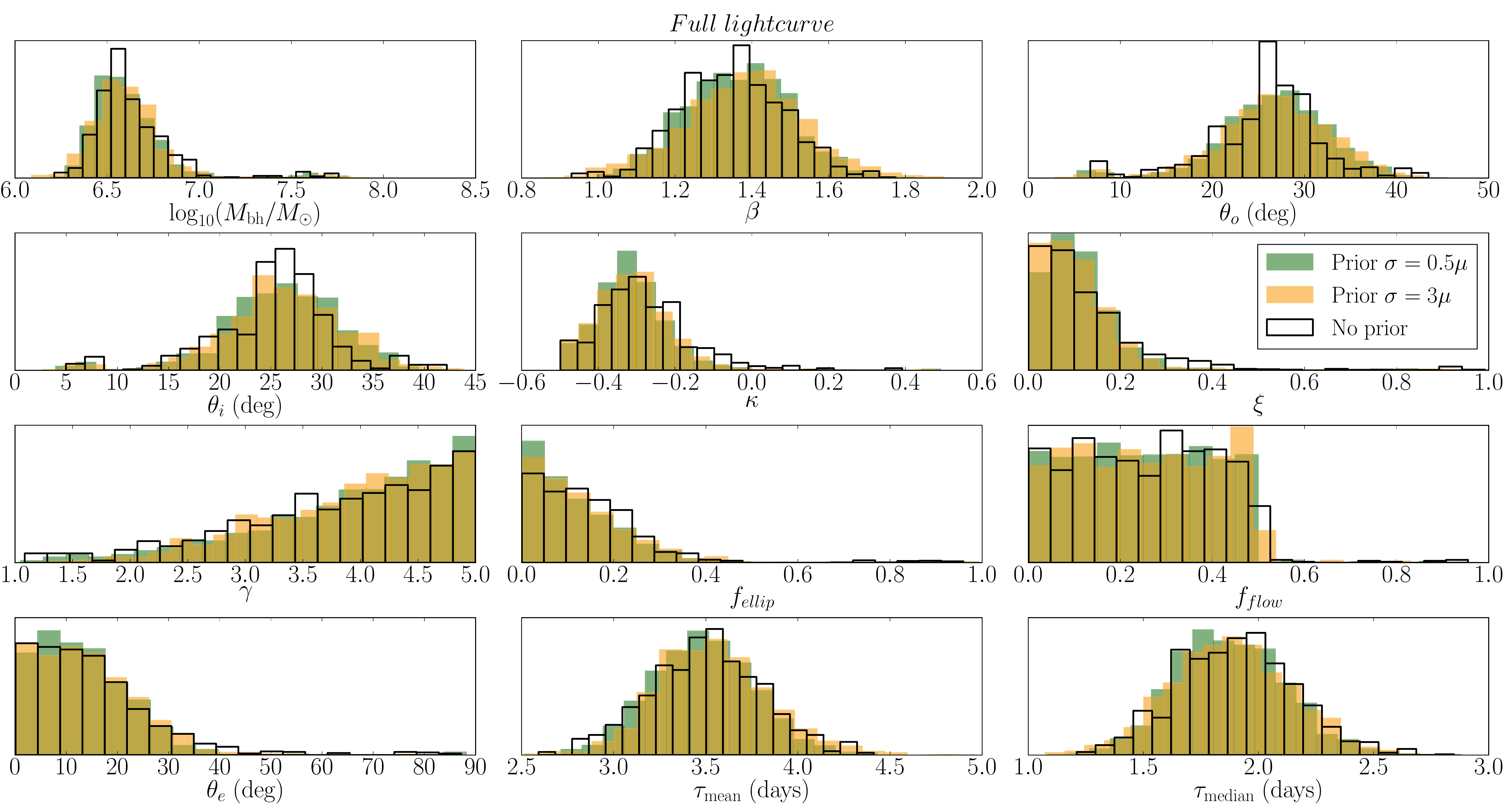}
\caption{Posterior probability distributions for the geometry and dynamics parameters of the BLR of Arp 151 when modelling the full spectroscopic and photometric light-curves. Default modelling of the full light-curve without a Gaussian prior (\citealt{pancoast18}) is shown as the black solid line histogram. The filled histograms show the result of using different Gaussian priors on $\tau_{\rm mean}$. In both cases we use a Gaussian distribution centred at $\mu_{\tau} = $3.07 days but differ on the width ($\sigma_{\tau}$) of the distribution. The green histogram uses $\sigma_{\tau} = 0.5 \times \mu_{\tau}$ and the orange histogram $\sigma_{\tau} = 3 \times \mu_{\tau}$ to represent a well determined value of $\tau_{\rm mean}$ and one with higher associated uncertainty. There is no significant difference in the median and 68\% confidence range of the posterior distributions between using or not using a Gaussian prior on $\tau_{\rm mean}$.}
\label{prior_noprior_posterior_full_light}
\end{figure*}

In the most recent implementation of the model \citep{pancoast18}, the physical scale of the BLR was constrained by the temporal (and scale sensitive) information contained in multi-epoch monitoring data. 
For single-epoch spectra, the temporal information is not available. We must therefore find an alternative method to constrain the physical scale. We do that by setting a prior probability distribution for the mean time delay between the continuum source and the broad line emission, a parameter that in the previous implementation of the model would have been constrained by the temporal information in multi-epoch data. Since the time delay is measured and the light travel speed is known, setting a prior on the mean time delay is equivalent to setting a prior on the size of the broad line region. In practice, we constrain the time delay by adding a Gaussian prior on the mean time delay parameter ($\tau_{\rm mean}$) in the model. The Gaussian prior centre ($\mu_{\tau}$) is set at a fixed mean time delay and we define a characteristic width for the Gaussian prior ($\sigma_{\tau}$) that describes our uncertainty in the prior. The Gaussian prior $\mu_{\tau}$ and $\sigma_{\tau}$ are free parameters and can be defined (by setting meaningful constraints) for each individual AGN modelled. A prime method to set these constraints is to use the R$_{\rm BLR}$ - L$_{\rm AGN}$ relation (\citealt{kaspi00}, \citealt{bentz09}, \citealt{bentz13}). The R$_{\rm BLR}$ - L$_{\rm AGN}$ relation is a robust observational result that originated from reverberation-mapping studies. It provides an established method to constrain the effective size of the BLR, R$_{\rm BLR}$, from the AGN continuum luminosity, L$_{\rm AGN}$, measured from a single-epoch spectrum. For the single-epoch modelling we carry out in this paper, we always define a Gaussian prior for the time delay. Several options for the choice of the time delay prior parameters will be investigated in this work. In particular, a time delay prior derived from the R$_{\rm BLR}$ - L$_{\rm AGN}$ relation, which is the method of choice for the general AGN population, will be discussed in Section \ref{sec:RL}.

To simulate the variable ionising continuum light-curve we generate a continuous model of the AGN continuum light-curve using Gaussian processes, as described by \cite{pancoast14a}. In past implementations of the BLR model, the AGN continuum model would be determined by interpolating between the points of the observed continuum light-curve. In our modified version of the model, we use one continuum flux value and an associated uncertainty estimate as constraint on the continuum light curve. Based on the measured continuum flux and its uncertainty and by extrapolating backwards in time, artificial AGN continuum light-curves are generated. These artificial light-curves sample a broad range of possible continuum light-curves and are formally required for the model to calculate a continuum flux corresponding to the time delay between the continuum emission and each region of the BLR.
The range of possible continuum light curves is determined by three main hyper-parameters: $\mu_{\rm cont}$, the long-term mean flux value of the light curve; $\sigma_{\rm cont}$, the long-term standard deviation of the light curve and $\tau_{\rm cont}$, the typical time-scale for variations (see \citealt{pancoast14a} for more details on these parameters). An important point to highlight is that since our single continuum data point does not constrain the continuum light curve parameters, our modified version of the code explores the continuum hyper-parameter space to find all the possible realistic solutions during the analysis of the single-epoch spectrum. Since we are not able to constrain the continuum light curve we marginalise the posterior probability distributions over the continuum hyper-parameters. We set flat prior probability distributions for $\mu_{\rm cont}$, $\sigma_{\rm cont}$ and $\tau_{\rm cont}$ and run several tests to determine that the inferred parameters do not depend on the prior range assumed.
\subsection{Rationale for model verification}
After modifying the BLR modelling code to model single-epoch spectra, following the method described in the previous section, our goal is to test the model's performance using real observed spectra. We select the AGN Arp 151 (also known as Mrk 40) for which three reverberation mapping campaigns have already been carried out (\citealt{pancoast18} and references therein). 
We extract single-epoch spectra from one of the multi-epoch monitoring datasets (see Section~\ref{sec:data}) and model them individually using our modified version of the model. Each single-epoch spectrum that is modelled results in a predicted set of geometry and dynamics parameters for the BLR. 

The full monitoring Arp 151 dataset has been modelled with previous implementations of the model to determine the geometry and dynamics of the BLR (\citealt{brewer11}, \citealt{pancoast14b}, \citealt{pancoast18}). The existence of a monitoring dataset and the associated modelling results are essential in order to evaluate the present model's performance as we can compare the BLR parameters inferred from single-epoch modelling with the BLR parameters obtained from the full monitoring dataset modelling (\citealt{pancoast14b}, \citealt{pancoast18}). This setup allows us to judge and quantify, in the absence of timing information, the effectiveness of a single spectrum in constraining the parameters of our underlying physical model. In Section~\ref{sec:timing} we discuss in detail the effect of including or not the timing information.

Additionally we validate the model using simulated single epoch spectra. The results of these simulations can be found in Appendix \ref{sec:appendix_sim}.
\subsection{Data}
\label{sec:data}
The data we use in this work is part of the Lick AGN Monitoring Project (LAMP) 2008 (\citealt{bentz09}, \citealt{walsh09}) multi-epoch monitoring dataset for Arp 151. LAMP is a long term project to carry out reverberation-mapping programs on a selected sample of AGN. The dataset comprises photometry and spectroscopy monitoring of Arp 151 that covers the spectral region of the broad H$\beta$ emission line ($\lambda$ [4300 : 7100] \AA). The spectra were decomposed into individual contributions from the AGN continuum, host galaxy stellar emission, narrow emission lines and Fe II emission as described by \cite{barth13} and \cite{pancoast18}. 

The advantage of using Arp 151 is that data from the full monitoring campaign are available, which include the continuum light-curve and the H$\beta$ line spectral information.
The final decomposed AGN continuum and spectra we use in this work to test the single-epoch BLR modelling approach is similar to the ones used by \cite{pancoast18} in their analysis. In this work we will use the multi-epoch monitoring dataset for initial tests and will then extract specific single-epoch spectra for our single-epoch modelling.
Our model results will be compared with the ones by \cite{pancoast18}. \cite{pancoast18} test two approaches to modelling the Arp 151 data: fitting the full H$\beta$ line profile and excluding the red wing of the H$\beta$ line profile. Their final inferred BLR parameters quoted in their Table 2 are determined from combining the posterior distributions obtained from these two approaches. In this paper we fit the full H$\beta$ emission line profile and will therefore compare our single epoch spectra modelling with the full light-curve modelling of \cite{pancoast18} shown as the blue histograms in their Fig. 3 (but not quoted in their Table 2).  
\section{Results}
\label{sec:results}
In this section we describe the tests carried out to evaluate the performance of our single-epoch BLR model. Our goal is to determine how the inferred parameters from single-epoch spectra modelling compare with modelling of the full multi-epoch monitoring dataset. 
We carried out a set of tests using either the full multi-epoch monitoring dataset or single epoch spectra of Arp 151 extracted from the multi-epoch monitoring. 
For comparison we use as baseline reference the modelling result of \cite{pancoast18}, obtained by modelling the LAMP 2008 full multi-epoch monitoring dataset of Arp 151 and the full H$\beta$ line profile described in Section \ref{sec:data}. From here on, we will refer to their result as the `Full light-curve' modelling. The inferred parameter values for the `Full light-curve' modelling are quoted in the second columns of Table~\ref{table_results_light} and Table~\ref{table_results}.
\subsection{The effect of adding a Gaussian prior on the mean time delay}
\label{sec:single_epoch}
We first test our main modification to the model, described in Section~\ref{sec:modifications}, which is the addition of a Gaussian prior on the mean time delay ($\tau_{\rm mean}$). The test is carried out by using our modified version of the model but with the full multi-epoch monitoring dataset as input. This ensures that we isolate the effect of the Gaussian prior on the inferred parameters. 
As a starting point we need to choose the parameters for the $\tau_{\rm mean}$ Gaussian prior: the mean $\mu_{\tau}$ and the standard deviation $\sigma_{\tau}$. In practice, the Gaussian prior represents a best measurement of $\tau_{\rm mean}$ with an associated measurement uncertainty. For these first tests we decide to match $\mu_{\tau}$ to the inferred $\tau_{\rm mean}$ value based on a previous multi-epoch modelling of Arp 151.
At the start of our tests the final results of \cite{pancoast18} on Arp 151 were not yet available. We therefore used the inferred $\tau_{\rm mean}$ from an earlier model of the Arp 151 LAMP 2008 data \citep{pancoast14b} as our assumed $\mu_{\tau}$. Due to the computational time required to run our models, we did not repeat our analysis using the value of $\tau_{\rm mean}$ found by \cite{pancoast18}. However, our assumed confidence ranges in  $\tau_{\rm mean}$ encompass the \cite{pancoast18} $\tau_{\rm mean}$ inferred value.

\begin{figure*}
\centering
\includegraphics[width=0.85\textwidth]{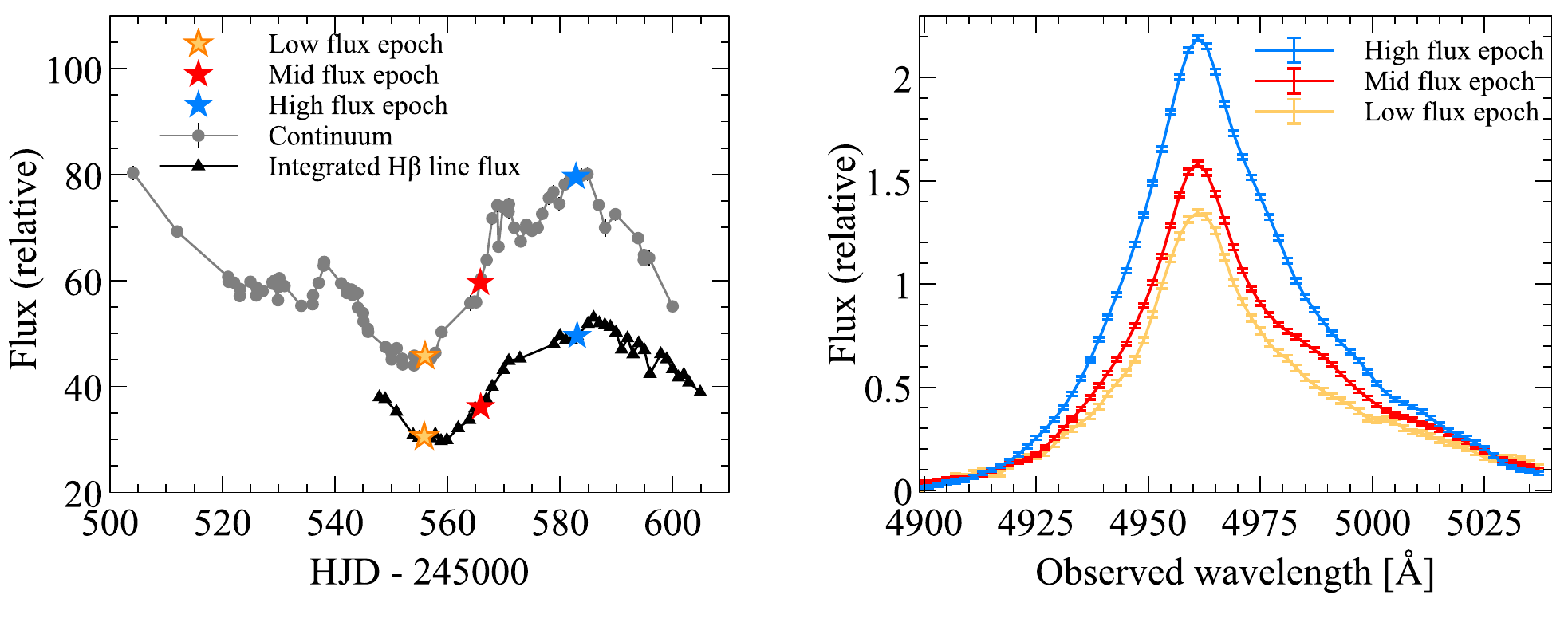}
\caption{Diagrams showing the LAMP 2008 optical light-curves of Arp 151 and the epochs chosen for our study. Left: Arp151 continuum light-curve (grey circles) and integrated broad H$\beta$ emission line flux light-curve (black triangles). The three single epochs are highlighted by the coloured stars. Right: Spectra covering the broad H$\beta$ emission line corresponding to each of the epochs selected in the light-curve. The line profiles shown were those obtained from spectral decomposition.}
\label{3epochs}
\end{figure*}

\begin{figure*}
\centering
\includegraphics[width=0.4\textwidth]{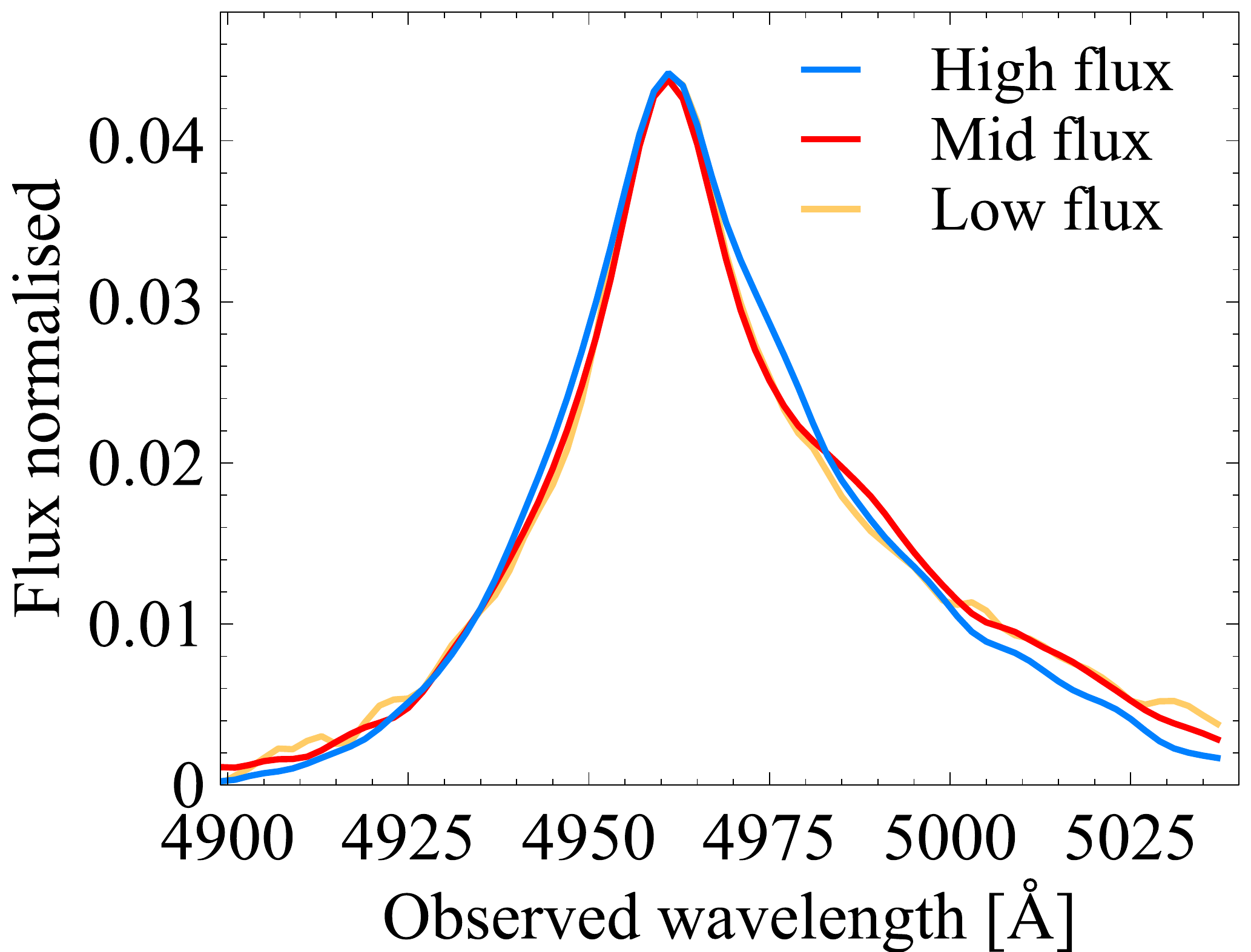}\hspace{0.6cm}
\includegraphics[width=0.4\textwidth]{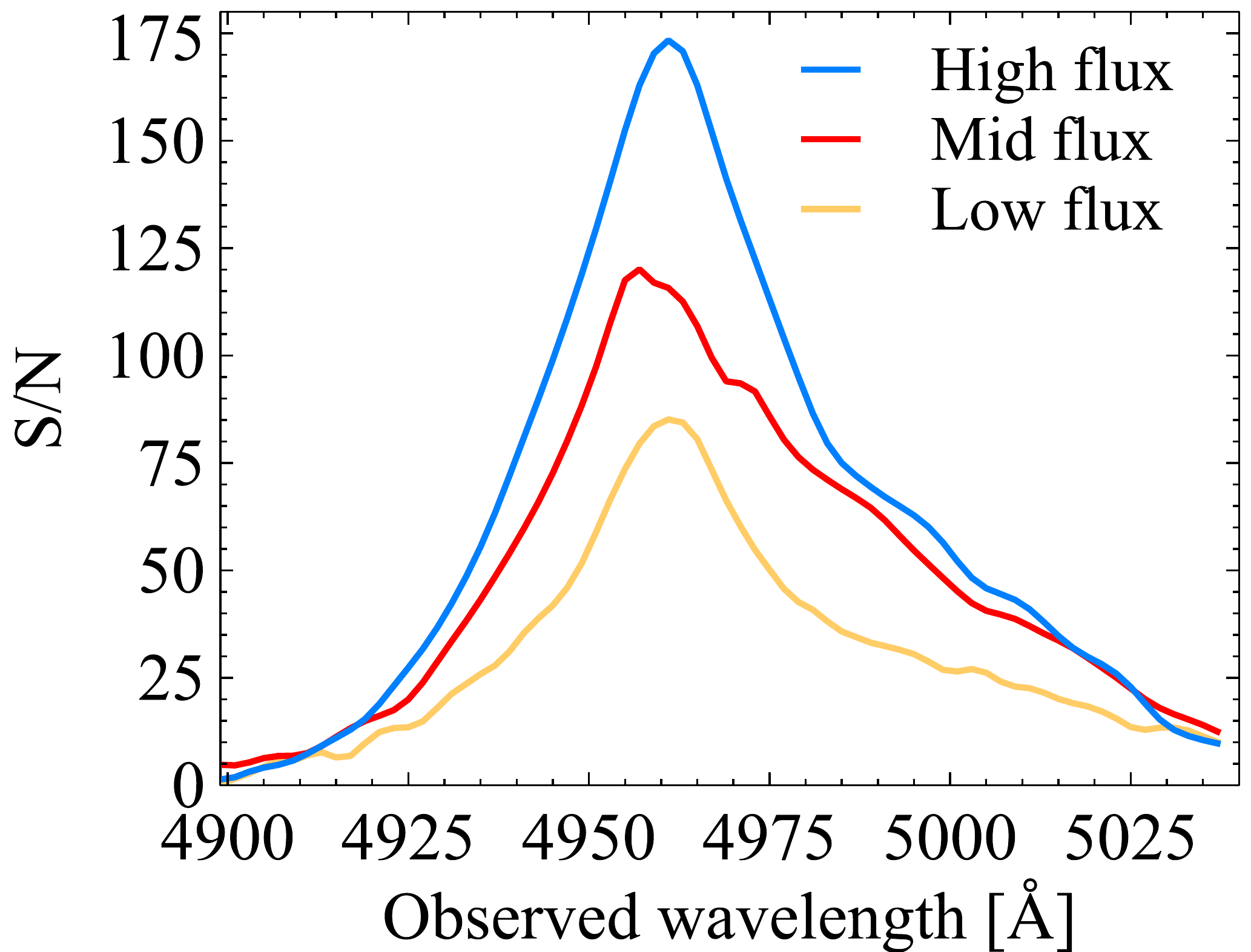}
\caption{Comparison of the continuum-subtracted H$\beta$ spectral line profiles and the signal-to-noise ratio for the three epochs colour coded as in Fig.~\ref{3epochs}. Left panel: Line profile shapes normalised to the integrated line flux to ease the comparison between line shapes. Right: Signal-to-noise ratio of the continuum subtracted profiles.}
\label{3epochs_norm}
\end{figure*}
\cite{pancoast14b} infer that $\tau_{\rm mean}$ = 3.07$^{+0.25}_{-0.2}$ days. 
We set the $\tau_{\rm mean}$ Gaussian prior to be centred at this value: $\mu_{\tau} = 3.07$. We set $\sigma_{\tau}$ to be 0.5 and 3 times the mean time delay ($\mu_{\tau}$) in the prior, i.e. 1.54 and 9.21 days respectively. These uncertainties correspond to a 68\% confidence range of 3.08 and 18.4 days, which are conservatively larger than the inferred $\tau_{\rm mean}$ 68\% confidence range of 0.45 days, determined by \cite{pancoast14b}. Our approach is equivalent to increasing the uncertainty on the prior value of $\tau_{\rm mean}$ so that we do not start with stringent assumptions. 

We expect our modified model to give a similar set of inferred parameters as in \cite{pancoast18} since for this test we are using exactly the same dataset to constrain the model, the only difference is that in our modified model we use a Gaussian prior on $\tau_{\rm mean}$.

The output of the BLR modelling is the posterior probability distribution function for each of the parameters describing the geometry and dynamics of the BLR. For most of the tests we show the posterior probability distributions for the parameters obtained from our model. 
In Fig.~\ref{prior_noprior_posterior_full_light} we compare the posterior distributions for Arp 151 with and without a prior on $\tau_{\rm mean}$. The results for the full light-curve baseline model from \cite{pancoast18} (with no prior on $\tau_{\rm mean}$), are shown as the solid black line histograms. These histograms correspond to the blue histograms in Fig. 3 of \cite{pancoast18}. The filled histograms show the result of using a prior centred at $\mu_{\tau} =   $3.07 days with $\sigma_{\tau} = 0.5 \times \mu_{\tau}$ (green) and a prior centred at $\mu_{\tau} = $3.07 days but with a larger uncertainty $\sigma_{\tau} = 3 \times \mu_{\tau}$ (orange). In Table~\ref{table_results_light} we show the inferred parameter values for each test. The inferred parameter is the median value of the posterior probability distribution and the uncertainties quoted are the 68\% confidence intervals.
The diagram and the table show that adding the Gaussian prior with varying widths does not affect the posterior distributions for the parameters. Therefore we can conclude that adding a Gaussian prior on $\tau_{\rm mean}$ does not affect significantly the inferred parameter values.
\subsection{Three single-epoch spectra and the effect of the flux level}
\label{sec:three_spec}
In this section we explore modelling the BLR geometry and dynamics of Arp 151 using a single-epoch of the light-curve (i.e. a spectrum) and show how the results depend on the chosen epoch. 

\begin{figure*}
\centering
\includegraphics[width=0.999\textwidth]{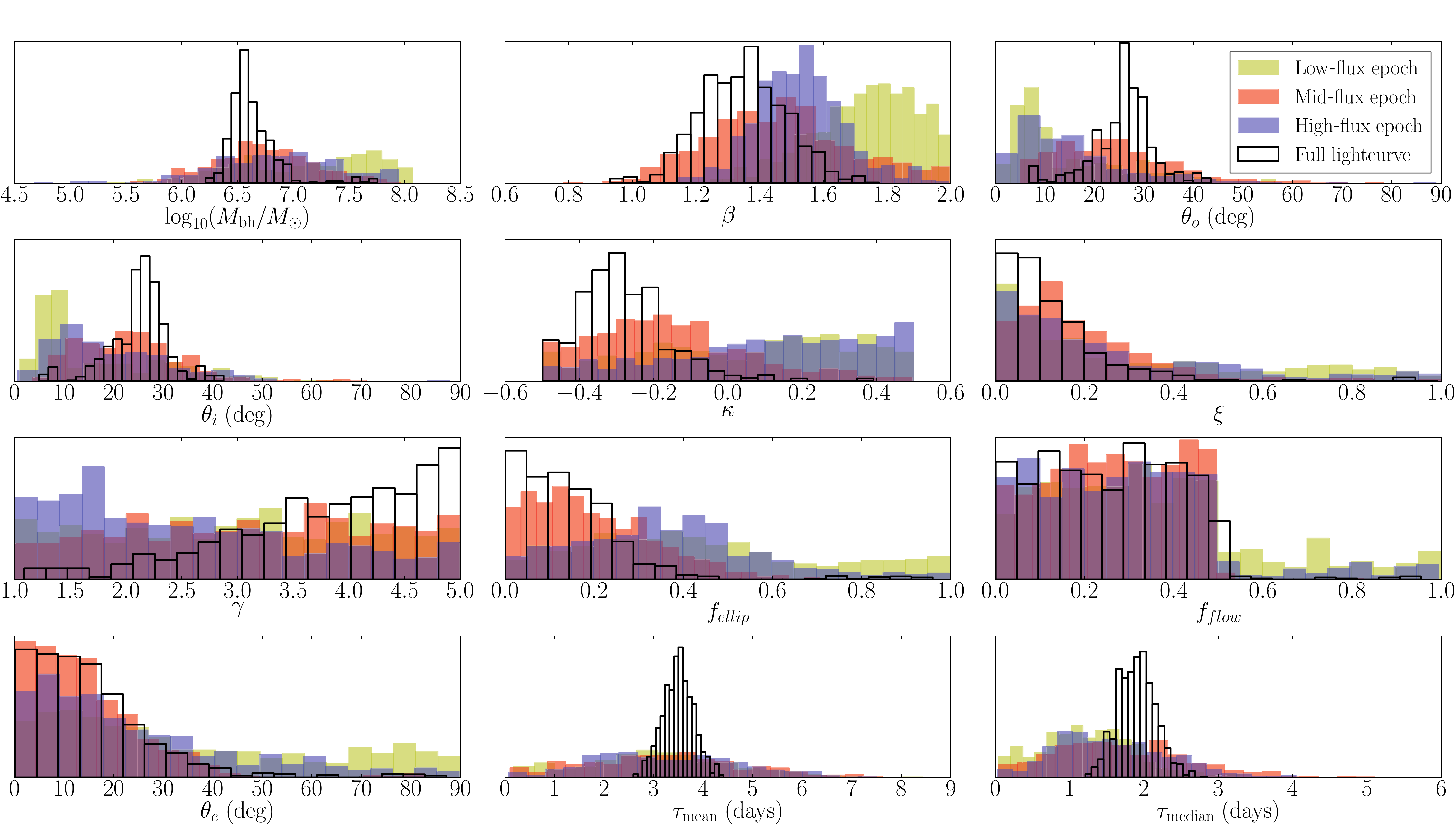}
\caption{Posterior probability distributions for the geometry and dynamics parameters of the BLR of Arp 151 when modelling three single epochs independently. Default modelling of the full light-curve without a Gaussian prior (\citealt{pancoast18}) is shown as the black solid line histogram. The filled histograms show the result for each of the epochs as labelled, using a Gaussian prior on $\tau_{\rm mean}$ centred at $\mu_{\tau} = 3.07$ days and with $\sigma_{\tau} = 0.5 \mu_{\tau}$.}
\label{3epochs_posterior}
\end{figure*}

We start by selecting three epochs (i.e. three time instants) in the Arp 151 multi-epoch monitoring light curve. These epochs are representative in terms of the continuum mean flux state of the source and include an epoch at relatively low-flux `Low-flux epoch', an epoch at high-flux `High-flux epoch' and an epoch with intermediate flux `Mid-flux epoch', identified from the continuum light-curve. In Fig.~\ref{3epochs} we show the three selected epochs overlaid on the full continuum light-curve and the line integrated flux light-curve from the monitoring campaign. The line integrated flux light-curve is obtained by integrating the observed spectrum between 4900 - 5050 \AA\ for each individual epoch. On the right panel we show the corresponding spectrally decomposed emission line profile measured at the specific epoch. The spectral profiles are selected from the H$\beta$ line flux light-curve of Fig.~\ref{3epochs} (black triangles). For each of the above mentioned epochs we choose the emission-line epoch that is closest in time to the continuum epoch. 
In practice, the modified model only needs a continuum value and an associated uncertainty. 
Using the mean of the two continuum values that encompass the epoch of the line profile measurement does not affect the outcome of our test. This is because the absolute continuum flux value does not affect the outcome of the model, as the continuum flux is normalised. There are two free parameters in the model (a multiplying constant and an additive constant) that re-scale the continuum flux (e.g. \citealt{pancoast11}), as described in Section~\ref{sec:model_par}. 

The shape of the line profiles for the three epochs can be more easily compared in Fig.~\ref{3epochs_norm}. The left panel shows each line profile normalised to its total integrated flux. The right panel shows the signal-to-noise ratio for each epoch. 

To set the physical scale we once again implement a Gaussian prior on $\tau_{\rm mean}$ centred at the inferred value found by \cite{pancoast14b} when modelling the full light-curve, i.e. 3.07 days, and an uncertainty of $\sigma_{\tau} = 1.54$ days. As in the previous section, we deliberately use a conservative width for the Gaussian prior which corresponds to a wider 68\% confidence range (3.08 days) than that found by \cite{pancoast14b} (0.45 days) for the inferred $\tau_{\rm mean}$. We model each of the three epochs independently, the only common assumption is a similar prior for the mean time delay. We use our continuum model to extrapolate the continuum light curve backwards in time. Since the continuum acts as the ionising source and it is located at a certain distance from the BLR, we need to simulate a past continuum flux history responsible for producing the broad line emission observed at the present time.
This extrapolated continuum light-curve does not necessarily reproduce what has been observed in the multi-epoch monitoring campaign. We are simulating a typical situation in AGN studies where we only have access to observations at one epoch by performing our analysis based on a wealth of simulated light curves, as explained in Section~\ref{sec:modifications}. A single epoch does not have enough information to determine what is the real continuum light-curve for Arp 151. As we will show below, we can still infer some of the BLR parameters even when there is no information on the real continuum light curve. 

\begin{figure*}
\centering
\includegraphics[width=0.34\textwidth]{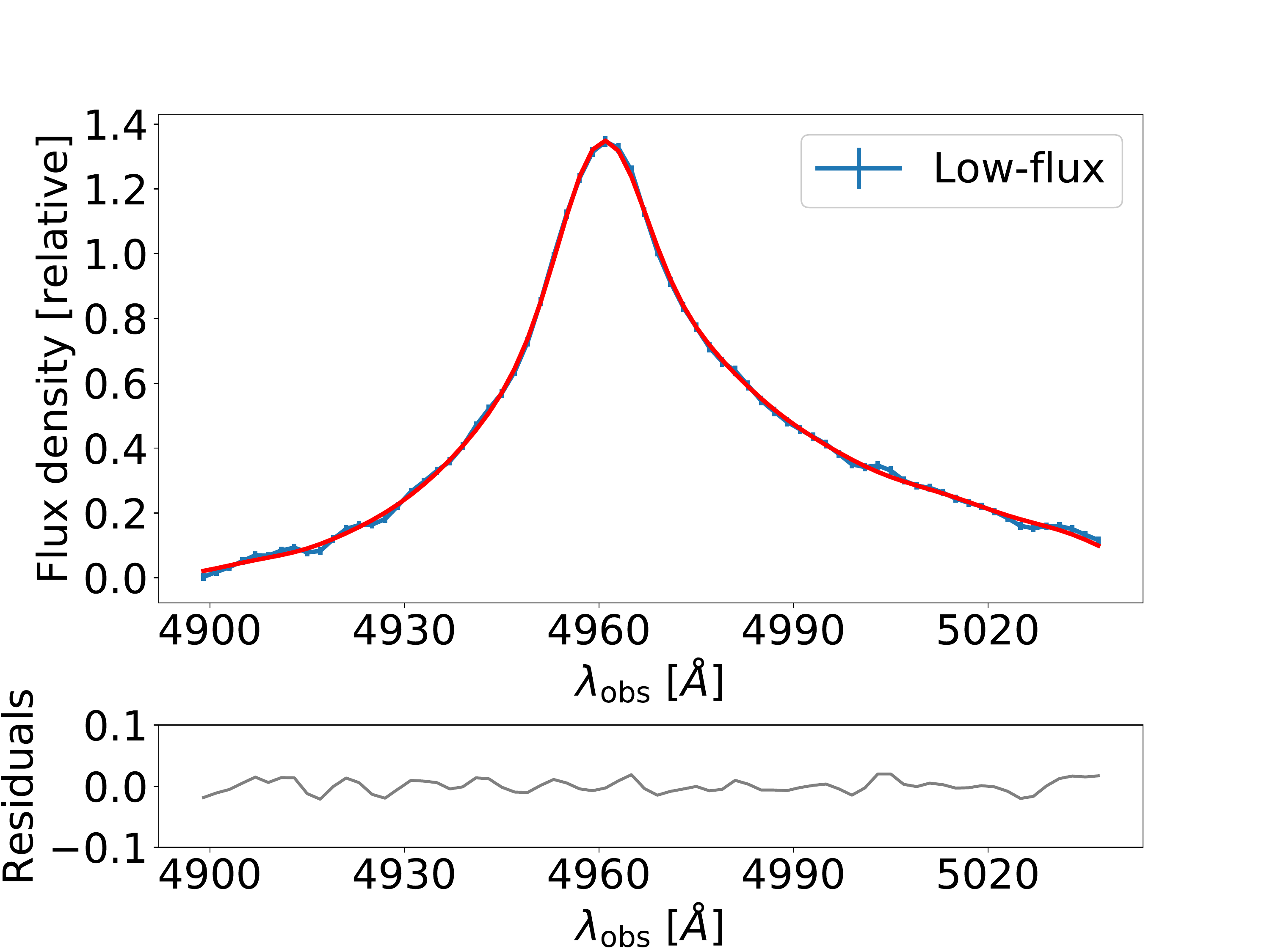}\hspace{-0.35cm}
\includegraphics[width=0.34\textwidth]{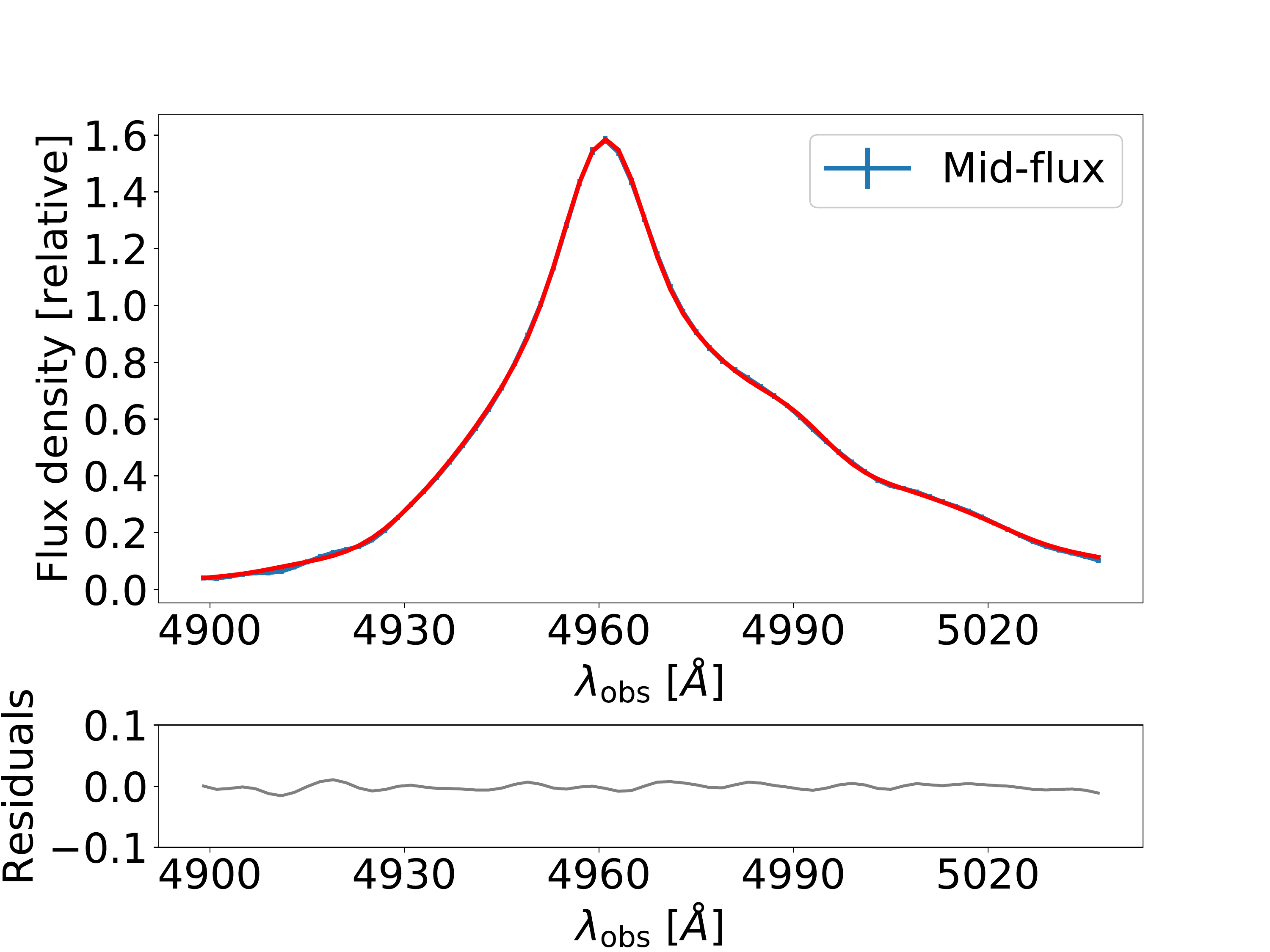}\hspace{-0.35cm}
\includegraphics[width=0.34\textwidth]{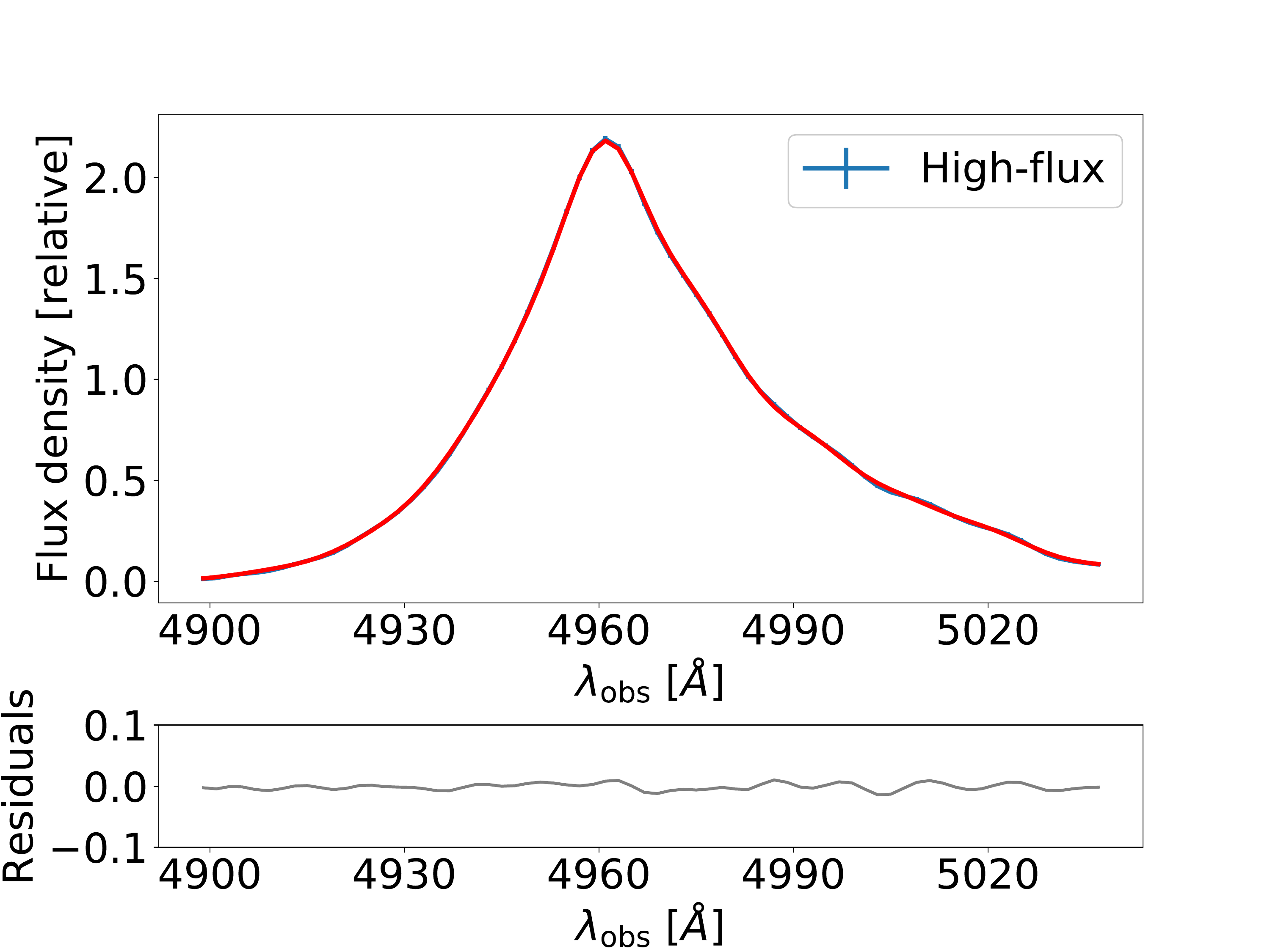}\hspace{-0.35cm}
\caption{Comparison between the input observed spectrum and a representative model line profile drawn from the posterior probability distribution for each epoch. The data is in blue, the model in red and the residuals are shown in grey in the bottom panels.}
\label{data_model_comparison}
\end{figure*}

The results of the modelling for each epoch are shown in Fig.~\ref{3epochs_posterior}. As a comparison, the posterior distribution obtained by \cite{pancoast18} with the full light-curve modelling is shown as the black solid line histogram. The posterior distribution for each parameter is consistent within the 68\% confidence range among epochs but also broader in general than what was found using the full light-curve. This is expected if there is less information in a single epoch spectrum to derive the geometry and dynamics of the BLR. However, the inferred values of the parameters for each of the epochs and the full light-curve do agree within the 68\% confidence range, which indicates: 1) that some of the structural and dynamical information about the BLR is included in a single epoch; 2) that the BLR geometry and dynamics information is not strongly dependent at which flux level of a light-curve a particular epoch was observed. This suggests that provided that a good estimate for the mean time lag is available, some of the geometry parameters can be recovered using single epoch spectra. The inferred parameters and their uncertainties are shown as cases $[2]$, $[3]$ and $[4]$ in Table~\ref{table_results} and in Fig~\ref{inferred_parameters}. In Fig.~\ref{data_model_comparison} we also show a representative comparison between the model, data and respective residuals for each epoch. We show the input spectrum in blue and an example of a model line profile drawn from the posterior probability distribution in red. The residuals (data$-$model) are shown in the bottom panel. The model generated profiles shown here are representative of the solutions found in the posterior probability distribution.

The mid-flux epoch appears to be the epoch that more closely approaches the results for the full lightcurve, possibly because the particular line shape observed at that epoch allows us to decrease the uncertainties on the inferred parameters. The low-flux epoch gives the most dissimilar results in terms of constraints on the parameters, in particular predicting higher values for the $\beta$ parameter, which describes the shape of the spatial distribution of the BLR emission. This is the only parameter for which the inferred value is not within the 68\% confidence range of the full light-curve inferred $\beta$. The low-flux epoch also predicts lower values for the angles than when using the full light-curve but with a wide distribution, which makes the inferred values still consistent within the uncertainties. The higher inferred $\beta$ is not the effect of assuming a prior on the mean time delay because in Fig.~\ref{prior_noprior_posterior_full_light} this is not observed. It is likely that high or low flux states may highlight significantly different portions of the BLR and therefore $\beta$ will describe the portion of BLR highlighted and not the full BLR distribution. For example, at a specific epoch a different portion of the BLR may be emitting, allowing the model to access information that was not present in the spectra of the other epochs.

\begin{figure*}
\centering
\includegraphics[width=0.49\textwidth]{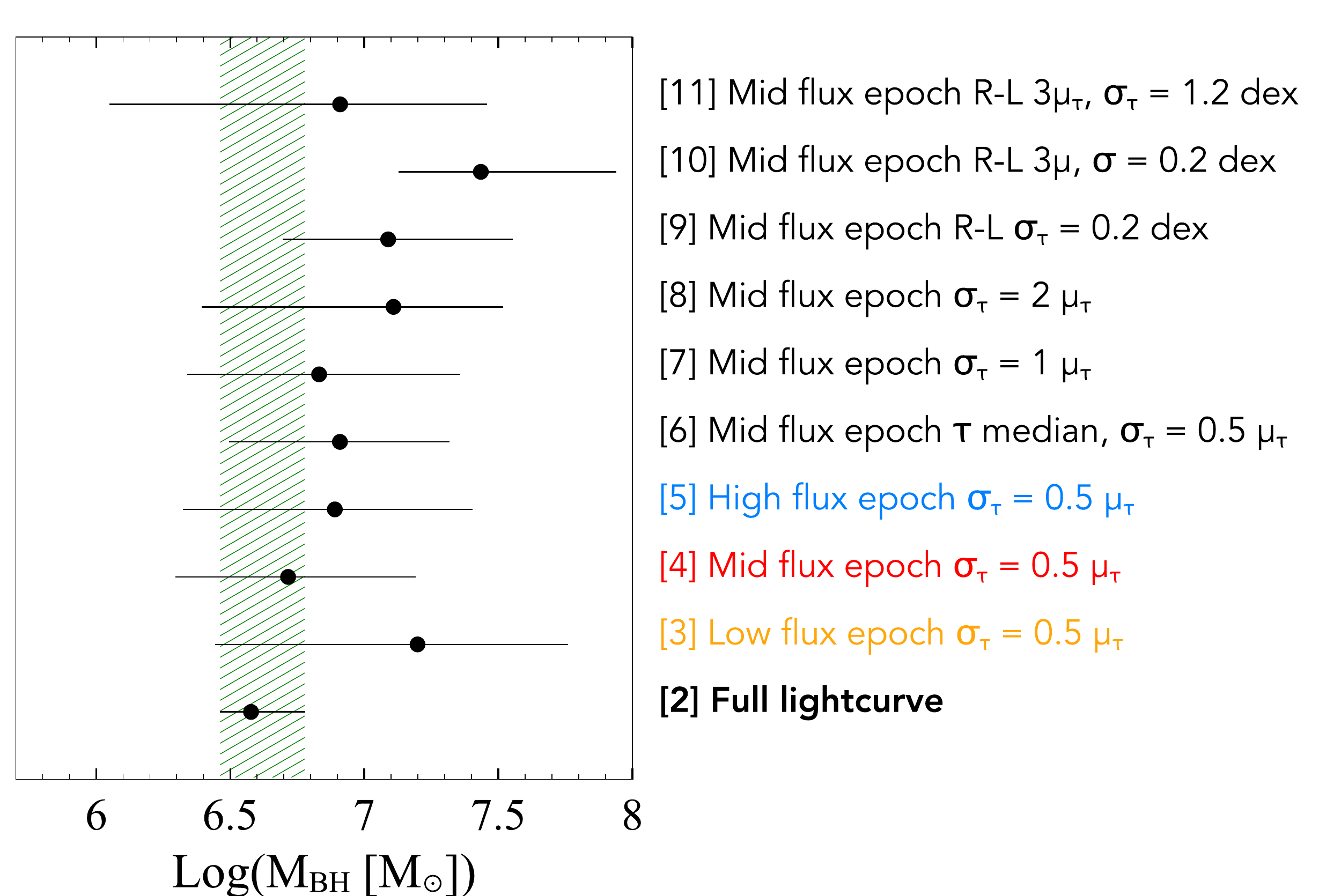}
\includegraphics[width=0.49\textwidth]{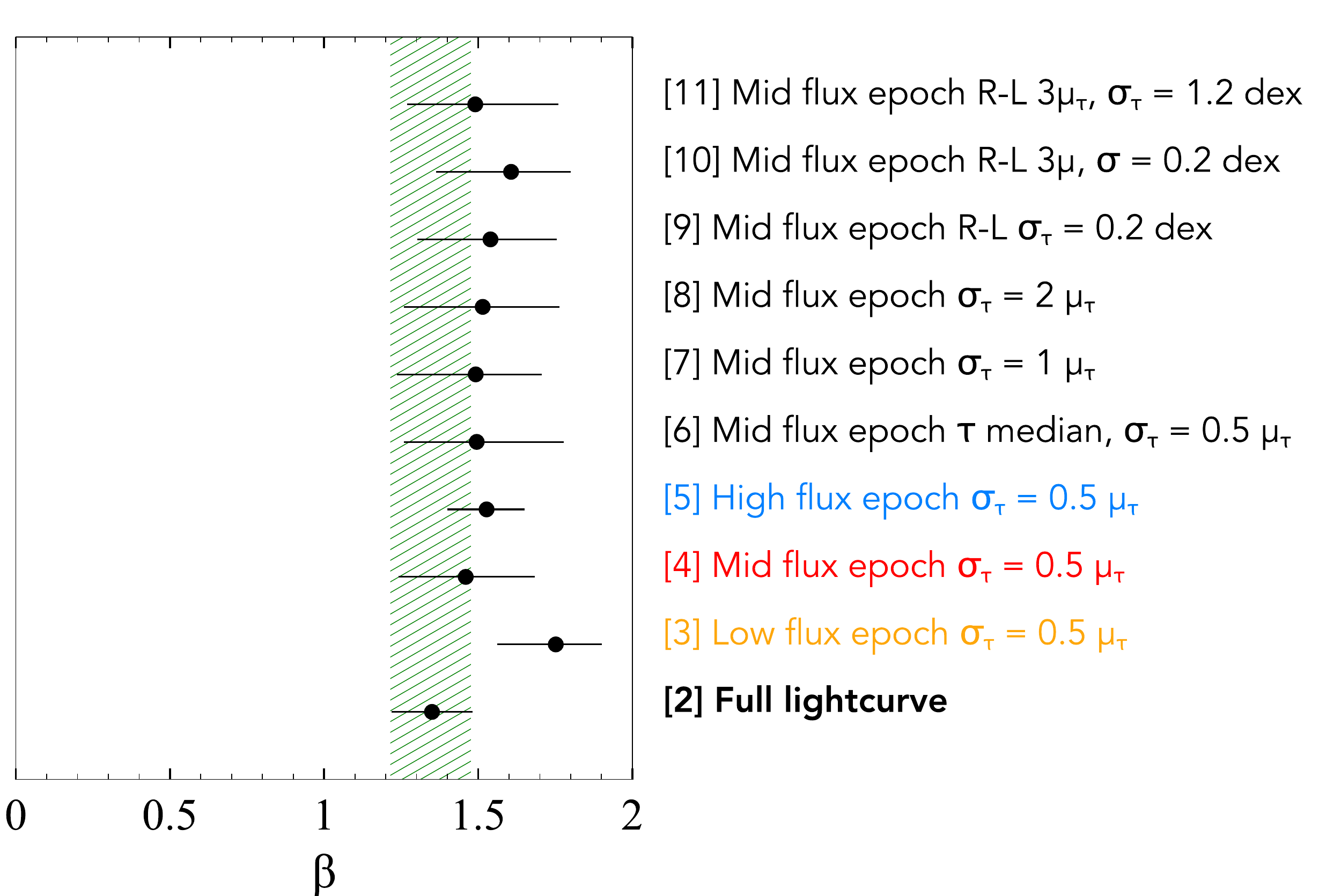}\\
\includegraphics[width=0.49\textwidth]{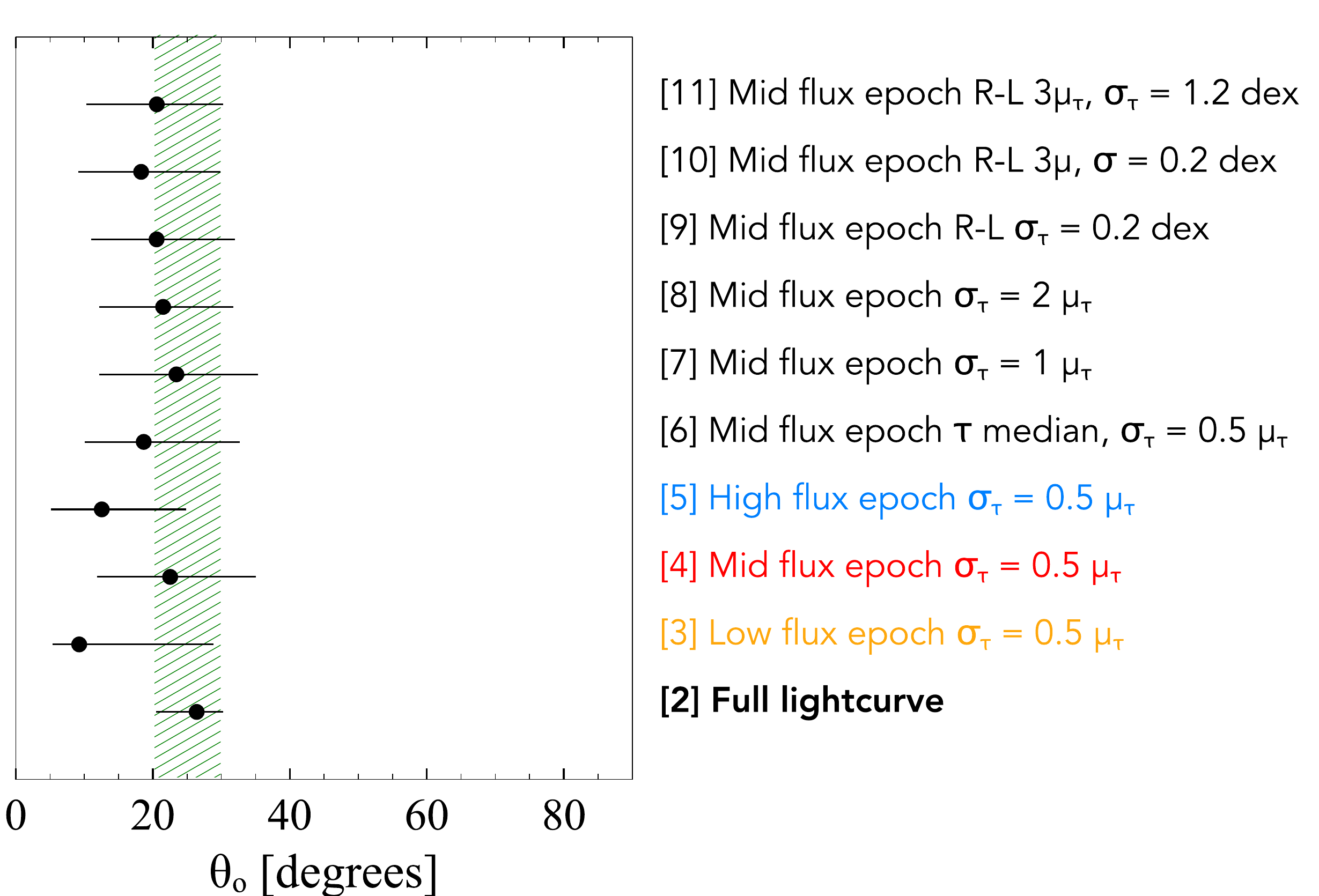}
\includegraphics[width=0.49\textwidth]{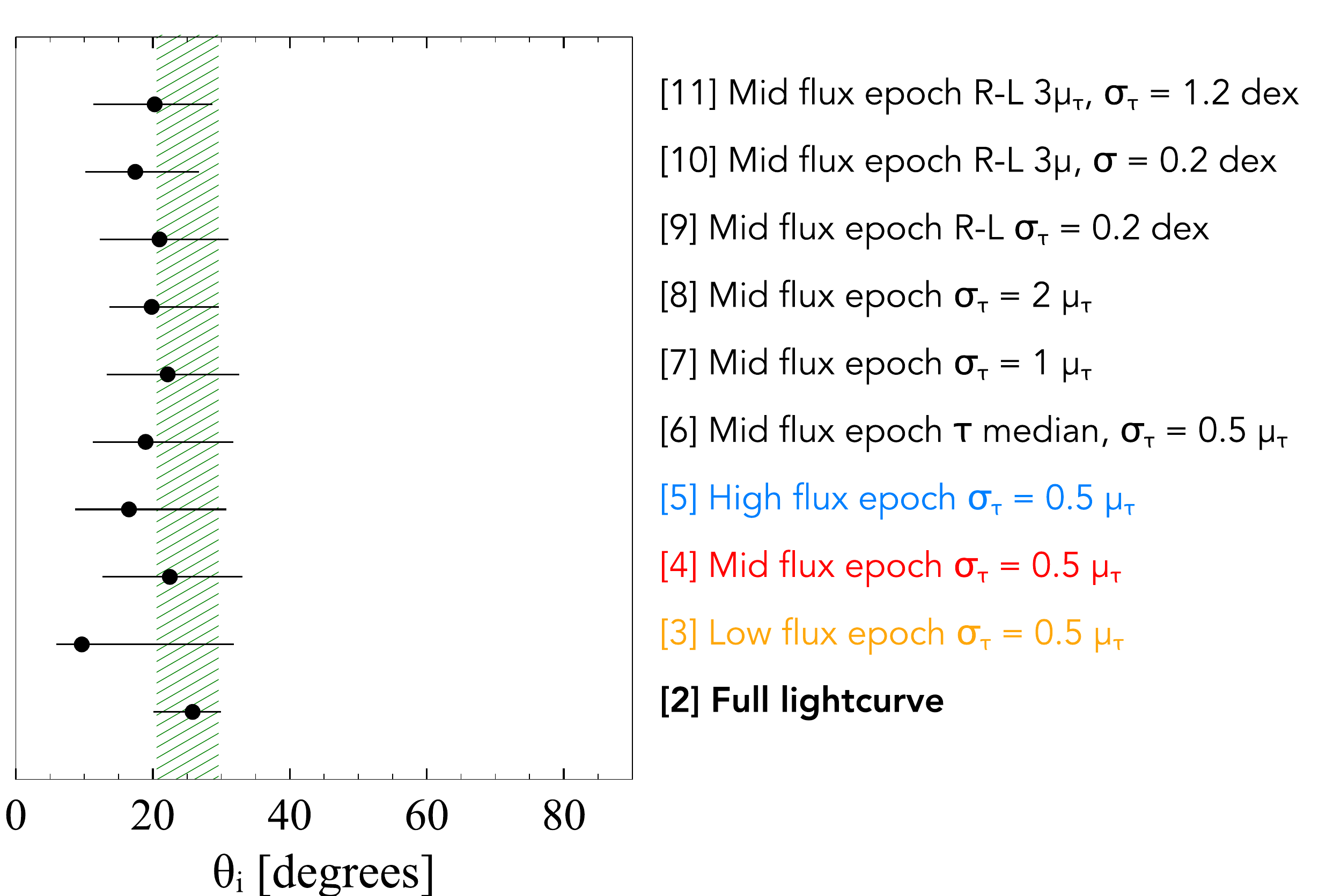}\\
\includegraphics[width=0.49\textwidth]{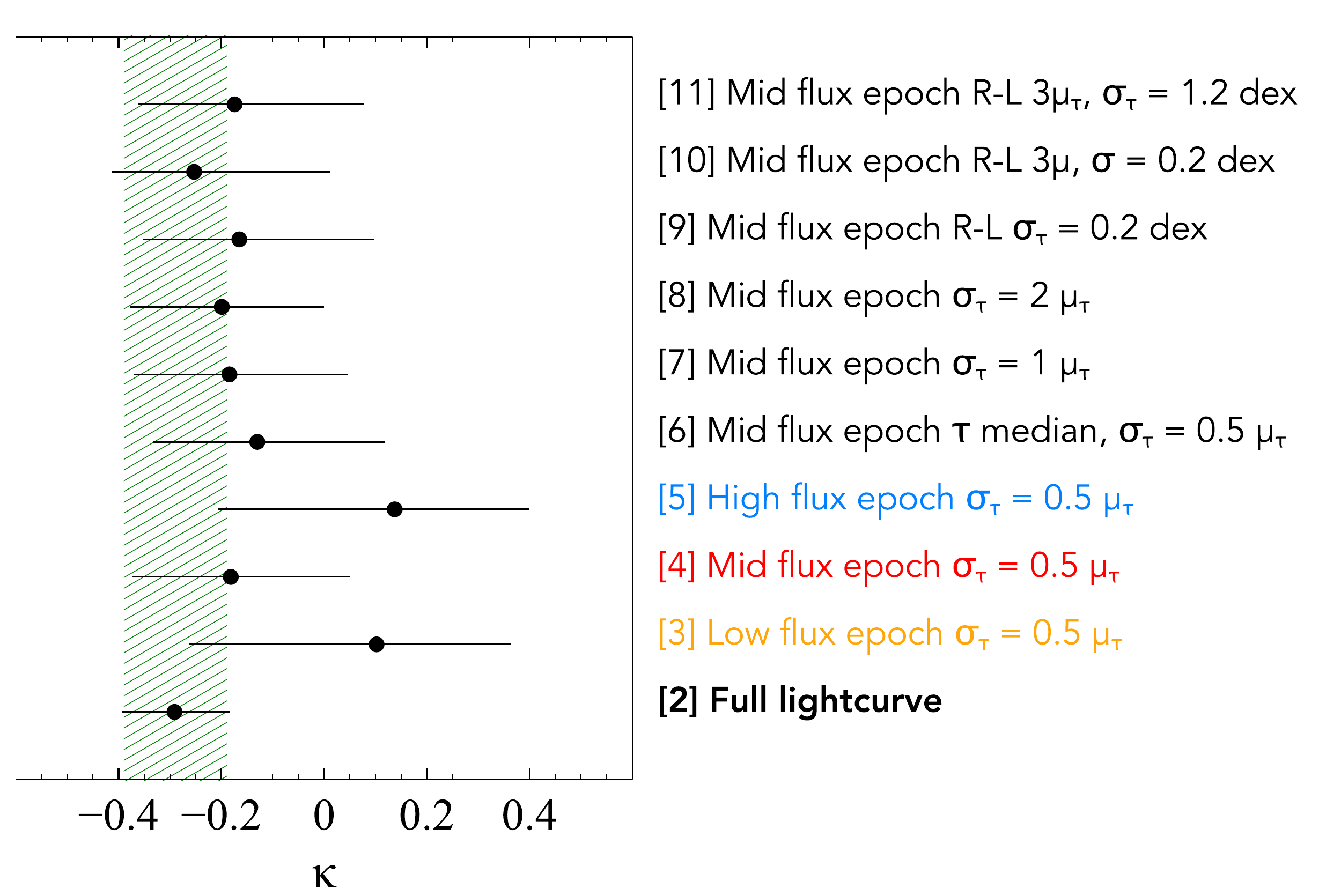}
\includegraphics[width=0.49\textwidth]{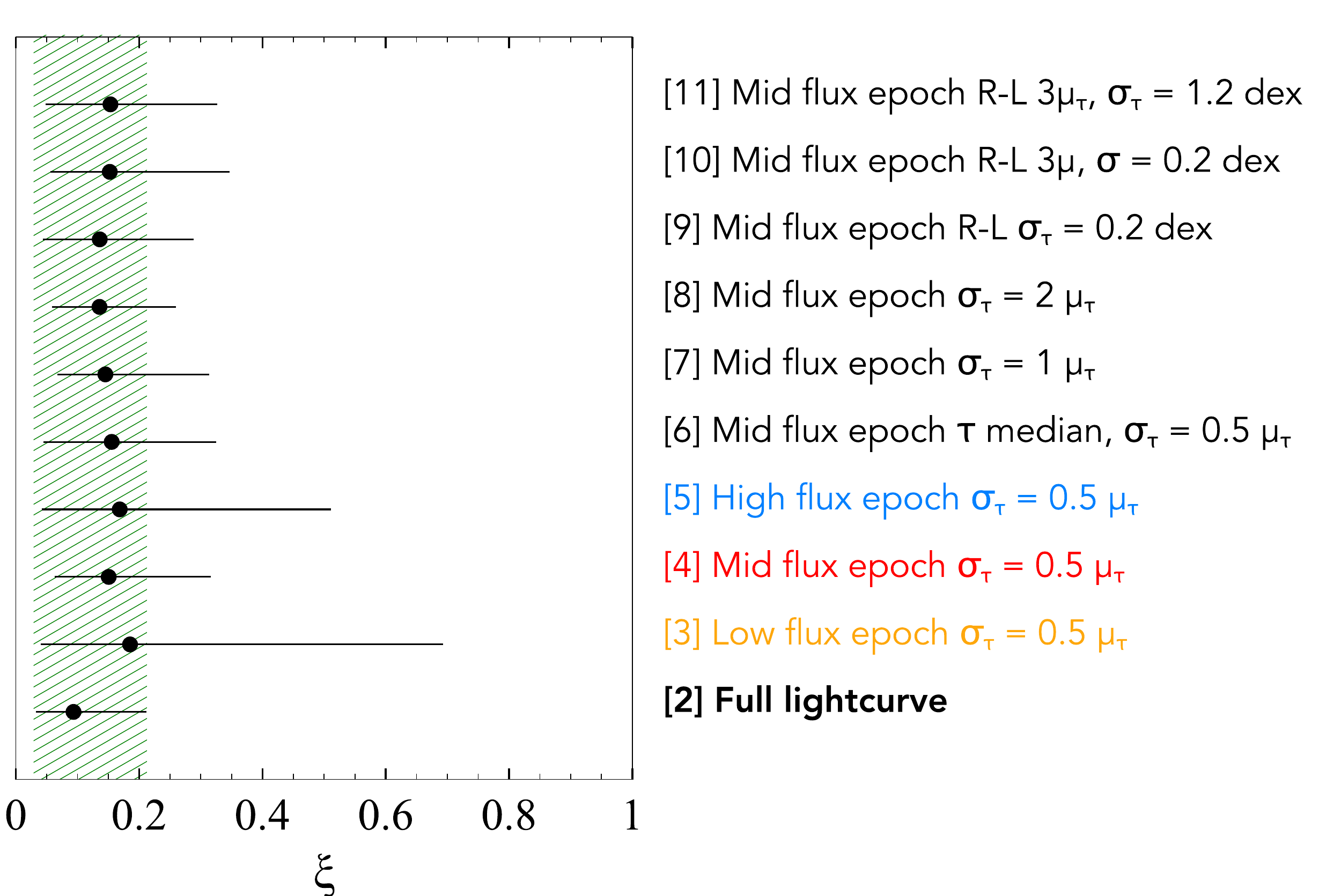}\\
\caption{Inferred values for the parameters and their respective 68\% confidence regions for each of the tests carried out in this work. The shaded green vertical region is the 68\% confidence region for the inferred parameters, determined from the full light-curve modelling result in \citealt{pancoast18}. The inferred parameter value is determined from the median of the parameter posterior probability distribution. Scale in the y-axis is arbitrary for drawing purposes. Each test is labeled with a number referring to the relevant column of inferred parameters in Table~\ref{table_results}.}
\end{figure*}
\begin{figure*}\ContinuedFloat
\centering
\includegraphics[width=0.49\textwidth]{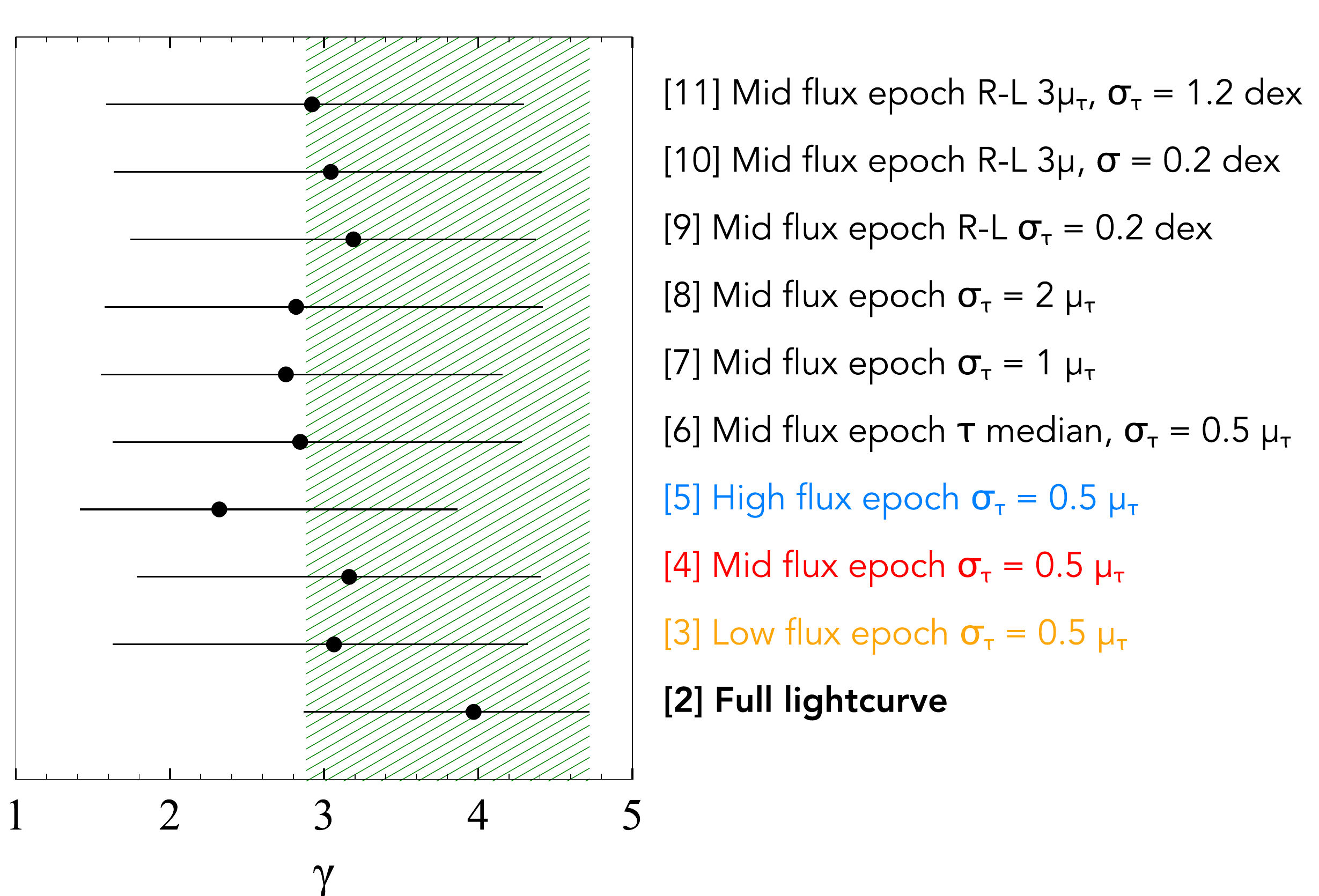}
\includegraphics[width=0.49\textwidth]{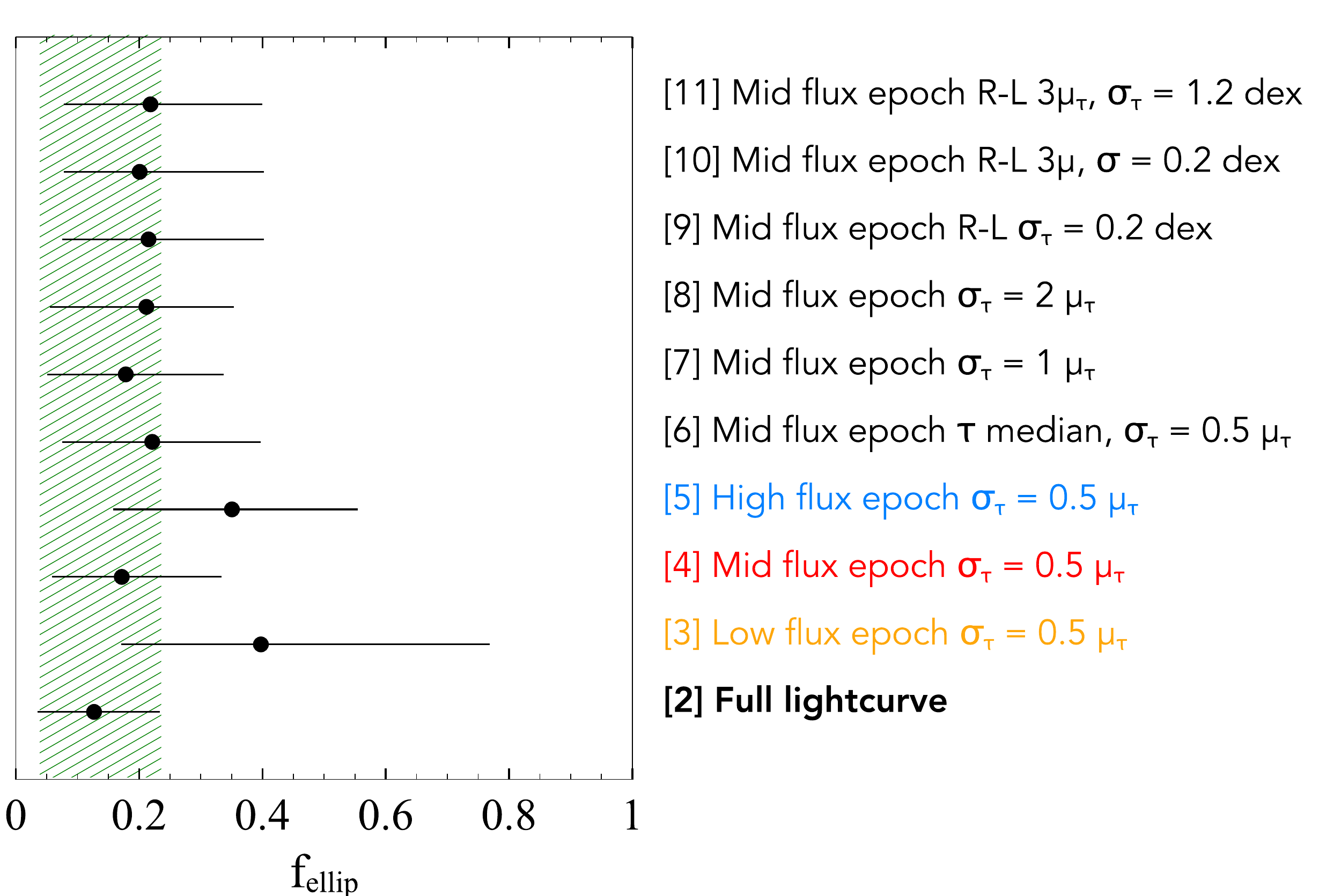}\\
\includegraphics[width=0.49\textwidth]{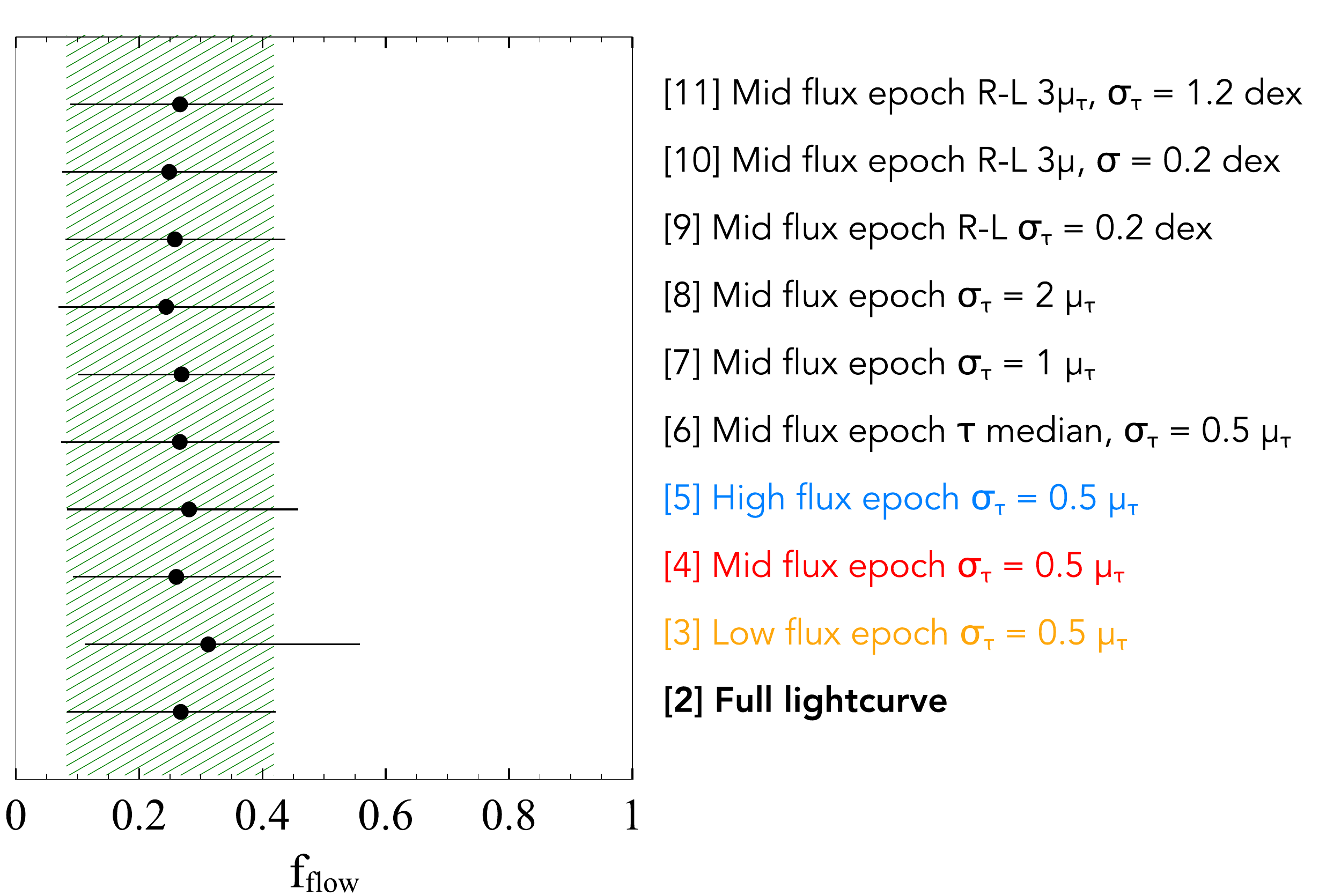}
\includegraphics[width=0.49\textwidth]{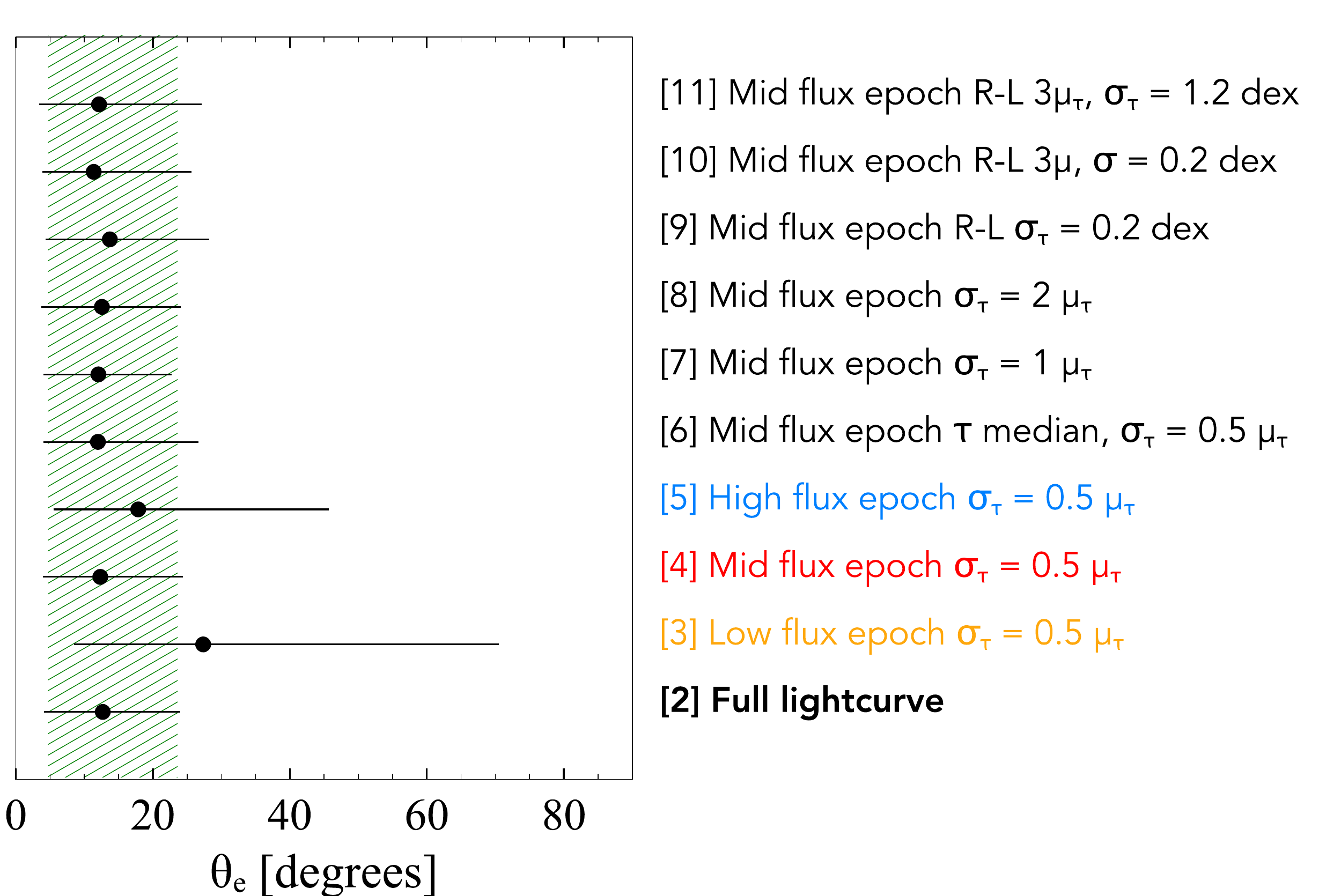}
\caption{Continued.}
\label{inferred_parameters}
\end{figure*}
\subsection{Using a prior on $\tau_{\rm mean}$ or $\tau_{\rm median}$}
Our ultimate goal is to apply the single epoch BLR modelling to other AGN and therefore in the following tests we will continue to use a single-epoch spectrum for the BLR modelling, as in Section~\ref{sec:three_spec}. We choose the mid-flux epoch of Arp 151 as an example, as its inferred parameters are closest to the full-lightcurve inferred parameters.

We test if there is a difference between the modelling results when using a prior on the mean ($\tau_{\rm mean}$) or median ($\tau_{\rm median}$) time delay. We find that there is no significant difference between these two approaches, as described next.

\cite{pancoast14b} found that, for Arp 151, the mean time delay determined by the model, $\tau_{\rm mean}$, matched the values quoted by \cite{bentz09} for the reverberation mapping cross-correlation time lag and therefore should give a better indication of the BLR size. The time lags quoted by \cite{bentz09} were used to calculate the BLR radius for the radius-luminosity relation. However, for one of the modelled AGN by \cite{pancoast14b}, SBS 1116+583A, the time lag from reverberation mapping is closer to the median time delay ($\tau_{\rm median}$) obtained from the model. For the purpose of testing if there is a difference in constraining $\tau_{\rm mean}$ or $\tau_{\rm median}$, we will now use a Gaussian prior on $\tau_{\rm median}$ as opposed to $\tau_{\rm mean}$ on single epoch spectra modelling. We use the mid-flux epoch of Arp 151 and set a Gaussian prior on $\tau_{\rm median}$, centred at the inferred value of 1.75 days from BLR modelling of the full light-curve \citep{pancoast14b}. We assume $\sigma_{\tau} = 0.5 \times \mu_{\tau}$, i.e. $\sigma_{\tau} = 0.875$ days. 

\begin{figure}
\centering
\includegraphics[width=0.4\textwidth]{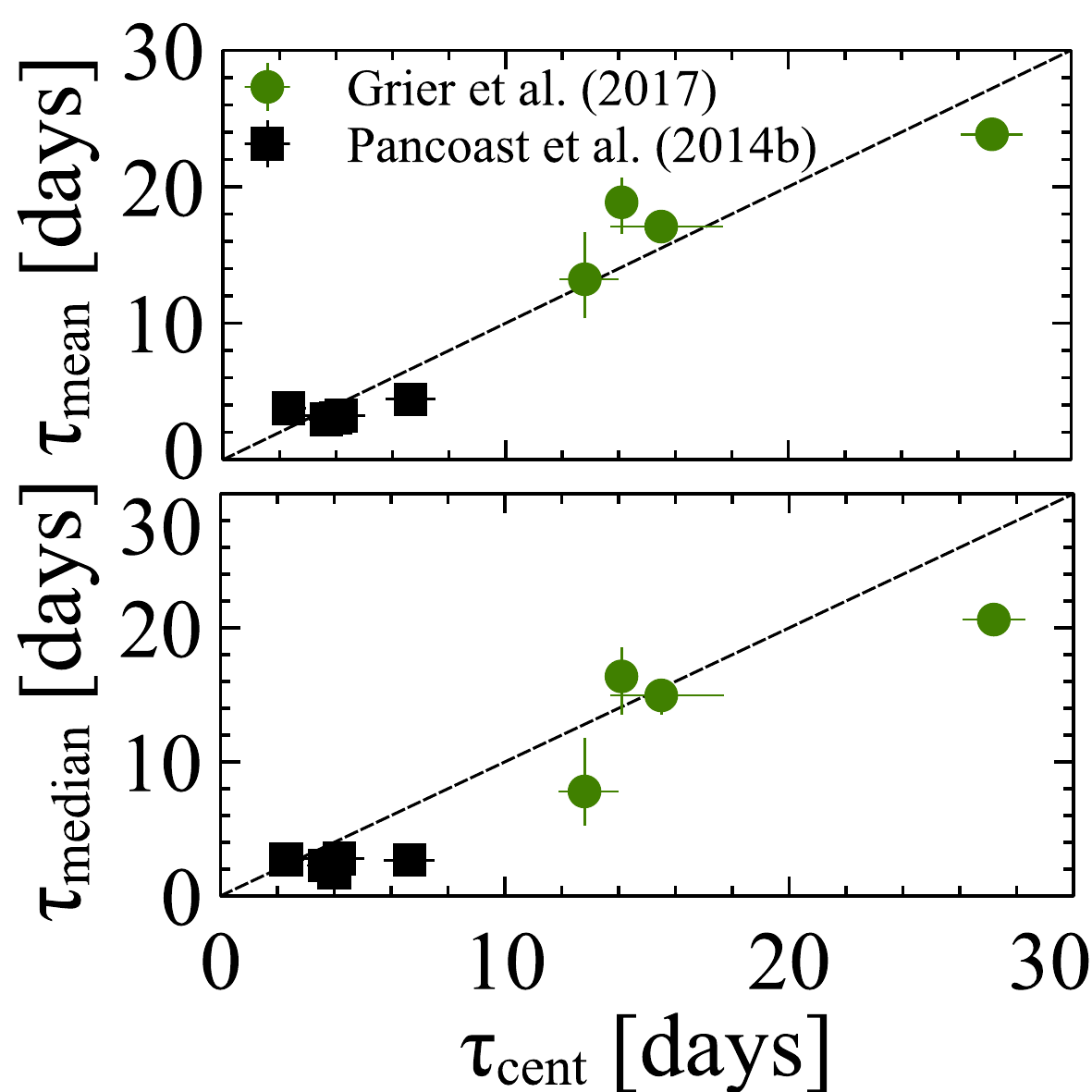}
\caption{Model-determined $\tau_{\rm mean}$ (top panel) and $\tau_{\rm median}$ (bottom panel) calculated by \citealt{pancoast14b} (black squares) and \citealt{grier17} (green circles) for a sample of AGN with reverberation mapping data. Mean and median time delays are shown as a function of the reverberation mapping cross correlation time delay $\tau_{\rm cent}$ determined by \citealt{bentz09} and \citealt{grier12}. The dashed line shows the 1:1 relation.}
\label{tau_median_mean_comp}
\end{figure}

We find that the posterior distribution for the parameters is consistent, within the 68\% confidence ranges, between all three cases shown. A figure showing the posterior distributions can be found in Fig.~\ref{tau_median_mean} in the Appendix~\ref{sec:appendix_extra_fig}.   
However, a prior on $\tau_{\rm mean}$ (red histogram) appears to produce a better match to the full light-curve results for M$_{\rm BH}$, $\beta$, $\theta_{o}$ and $\theta_{i}$ as can be seen from Fig.~\ref{inferred_parameters}, and Table~\ref{table_results} cases $[4]$ and $[6]$. Additionally, $\tau_{\rm mean}$ appears to be closer to the reverberation mapping time delay measured for most sources (\citealt{pancoast14b}, \citealt{grier17}). In Fig.~\ref{tau_median_mean_comp} we show $\tau_{\rm mean}$ and $\tau_{\rm median}$ as a function of the cross-correlation time delay for all the sources modelled by \cite{pancoast14b} and \cite{grier17}. It has been shown that for Arp 151 in particular, $\tau_{\rm mean}$ is the best indicator of the cross-correlation time delay. This could be due to the fact that the mean time delay or mean radius is more sensitive to long tails in the particle radial distribution and therefore more sensitive to $\beta$. We will therefore continue to use a Gaussian prior on $\tau_{\rm mean}$ to explore a possible extension of this study to a more general set of sources.
\subsection{Changing the mean and confidence range of the Gaussian prior}
In this section we are interested in testing what is the effect of increasing the $\tau_{\rm mean}$ Gaussian prior distribution width ($\sigma_{\tau}$). This test represents a scenario in which $\tau_{\rm mean}$ may not be well known and one wants to increase the uncertainty associated with that measurement.

\subsubsection*{Width of the Gaussian prior}

We use different $\sigma_{\tau}$ ($0.5 \times \mu_{\tau}$, $1 \times \mu_{\tau}$, $2 \times \mu_{\tau}$) for the Gaussian prior, where $\mu_{\tau}$ is the Gaussian mean. We set $\mu_{\tau} = 3.07$ days as in the previous test.  We find that with a fixed value for $\mu_{\tau}$, the uncertainty associated with the $\tau_{\rm mean}$ prior does not significantly impact the inferred parameter values as they are still consistent within the 68\% confidence ranges. For M$_{\rm BH}$ and the time delays ($\tau_{\rm mean}$ and $\tau_{\rm median}$), the width of the Gaussian prior naturally influences the precision of the inferred values, as the final posterior distributions are wider. For the remaining parameters, the prior width does not seem to have a significant effect, as the mean posterior values and uncertainties appear to be mostly constrained by the input spectrum and not by the width of the prior. This can be more clearly seen from the inferred parameters and associated uncertainties for this test, shown in columns 4, 7 and 8 of Table \ref{table_results} and in Fig~\ref{inferred_parameters}. The posterior probability distributions are shown in Fig.~\ref{tau_diff_sigma_posterior} in Appendix~\ref{sec:appendix_extra_fig}.
\subsubsection*{Mean value of the Gaussian prior}
We also test if changing the mean value of the Gaussian prior ($\mu_{\tau}$) will affect the inferred parameters. We assume a case where $\mu_{\tau} = 5.21$ days, the value corresponding to the R$_{\rm BLR}$ - L$_{\rm AGN}$ expected value for Arp 151 which is higher than the value of $\tau_{\rm mean} = 3.07$ days found by \cite{pancoast14b}. We assume $\sigma_{\tau} = 2 \times \mu_{\tau} = 2 \times 3.07$ as for the test in the previous section. This simulates a situation where the mean value for the Gaussian prior is offset from the real $\tau_{\rm mean}$, but still within the 68\% confidence range of the real $\tau_{\rm mean}$. We find that offsetting $\mu_{\tau}$ from the real $\tau_{\rm mean}$ does not strongly impact the inferred parameters. The posterior probability distributions for this case compared with the full light curve result and the test in the previous section are shown in Fig.~\ref{tau_diff_sigma_mu_posterior} in the Appendix~\ref{sec:appendix_extra_fig}. 
\subsection{Using the radius-luminosity relation to constrain single-epoch spectra modelling}
\label{sec:RL}
In this section we extend the scope of our previous tests and describe how to model the BLR of the more general AGN population. In the previous tests we used prior knowledge of the $\tau_{\rm mean}$ of Arp 151 to set the physical scale for the model. For the general AGN population such information is not available, however, we will show that the AGN radius-luminosity relation derived for the H$\beta$ line can be used to set the needed constraints in the model.

As mentioned in Section \ref{sec:modifications}, the radius-luminosity (R$_{\rm BLR}$ - L$_{\rm AGN}$) relation (\citealt{kaspi00}, \citealt{bentz09}, \citealt{bentz13}) establishes a correlation between the characteristic size of the BLR, R$_{\rm BLR}$,  and the AGN continuum luminosity, L$_{\rm AGN}$. This relation was determined based on monitoring data in reverberation mapping studies, using the time delay between the AGN continuum and the broad line region response to derive a characteristic size of the BLR. 
The R$_{\rm BLR}$ - L$_{\rm AGN}$ relation can be used to determine the broad line region size of a much larger sample of AGN for which meaningful monitoring data have not been, or are difficult to obtain. The community has used this relation to extend BLR size and black hole mass measurements to a wider population of AGN for which only single-epoch spectra exist (e.g. \citealt{vestergaard02}, \citealt{mclure&jarvis02}, \citealt{vestergaard&peterson06}, \citealt{shen11}, \citealt{kozlowski17}). 

As we showed in our tests, some knowledge of the mean time delay together with BLR modelling can produce meaningful constraints on the BLR geometry and dynamics parameters, even when single epochs are used. 
Many known AGN have single epoch measurements that provide spectroscopy covering the H$\beta$ line and the continuum. The R - L relation can be used to determine the effective size of the variable BLR (parameterised by R$_{\rm BLR}$) from the AGN continuum luminosity, $\lambda L_{\lambda}$, \citep{bentz13}:

\begin{equation}
\log\left[ \frac{R_{\rm BLR}}{\rm 1\,lt{-}day} \right]=1.527^{+0.031}_{-0.031} + 0.533^{+0.035}_{-0.033}\,\rm log\left[\frac{\lambda L_{\lambda}}{10^{44} erg\,s^{-1}} \right]
\label{eq:RL}
\end{equation}

This relation has been derived based on data on the H$\beta$ broad line emission and has a scatter of typically 0.2 dex \cite{bentz13}. We should highlight that single-epoch spectra include starlight emission flux from the galaxy. To determine the AGN continuum it is necessary to remove the galaxy starlight contribution via spectral decomposition or, better yet, via surface brightness modelling of images, when available (e.g. \citealt{bentz13}). For more luminous AGN the galaxy starlight contribution will be less important than for lower luminosity AGN.

For the purpose of our work, since R$_{\rm BLR}$ measured in light-days is a proxy for the mean time delay, $\tau_{\rm mean}$, the R - L relation can provide the necessary constraints on $\tau_{\rm mean}$ based on the AGN continuum luminosity. As we have shown in the previous sections, information on $\tau_{\rm mean}$ is the only constraint needed to extend our analysis of the BLR geometry and dynamical modelling to a large sample of AGN with single-epoch data.

To test this method, we simulate the conditions that would be available for a typical AGN which is not a multi-epoch monitoring target: a spectrum containing information on the H$\beta$ line profile, the AGN continuum luminosity at 5100 \AA\ and the R-L relation. We use a single epoch of Arp 151 as a test since we know the intrinsic BLR parameters from modelling the full light-curve. We choose the mid-flux epoch and use the AGN 5100 \AA\ continuum luminosity for Arp 151 measured by \cite{bentz13} after accounting for the galaxy's starlight contribution: $\lambda L_{\lambda \rm , AGN} = 3.02\times 10^{42}$ erg s$^{-1}$. Using this continuum value and the radius-luminosity (R-L) relation quoted by \cite{bentz13} (Equation~\ref{eq:RL}), we obtain an estimated R$_{\rm BLR} = 5.21$ light-days. The radius determined is different from the cross correlation time delay value measured directly by \cite{bentz13} (3.99 days) and used to calculate the R-L relation itself. This is because Arp 151 lies slightly below the R-L relation. Nevertheless we will set $\mu_{\tau} = 5.21$ days since this simulates a situation without prior knowledge on the reverberation mapping results. 

\begin{figure*}
\centering
\includegraphics[width=1.0\textwidth]{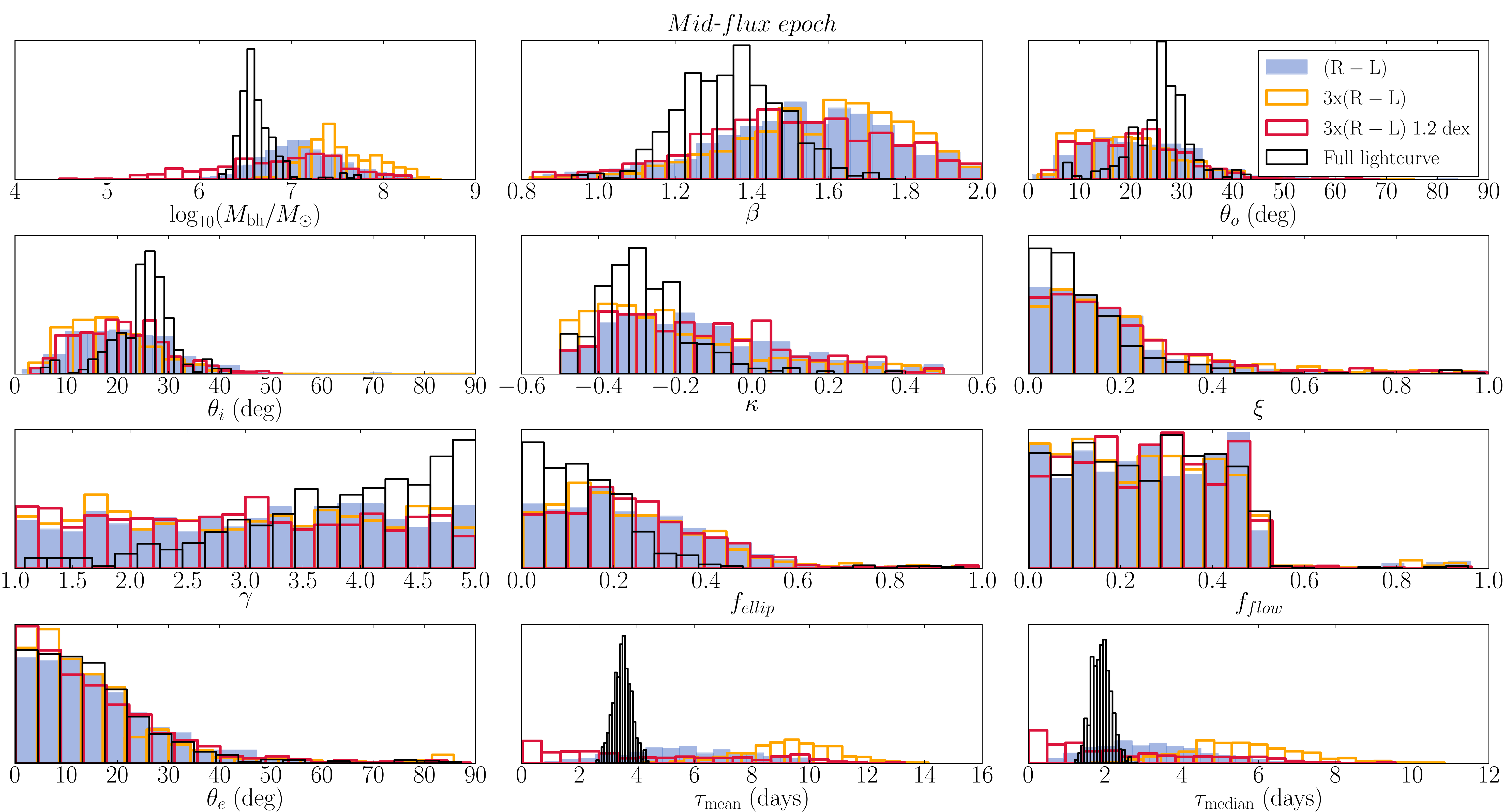}
\caption{Posterior probability distributions for the BLR geometry and dynamics parameters determined using the mid-flux epoch. Default modelling of the full light-curve without a Gaussian prior (\citealt{pancoast18}) is shown as the black solid line histogram. Blue: Gaussian prior on $\tau_{\rm mean}$ centred at the value found from the radius-luminosity relation and with $\sigma_{\tau} = 0.2$ dex. Yellow: Gaussian prior on $\tau_{\rm mean}$ centred at 3$\times$ the time delay predicted by the R-L relation and with $\sigma_{\tau} = 0.2$ dex. Red: Gaussian prior on $\tau_{\rm mean}$ centred at 3$\times$ the time delay predicted by the R-L relation and with $\sigma_{\tau} = 1.2$ dex. }
\label{tau_diff_mean_value_posterior}
\end{figure*}

\cite{bentz13} find a spread of $\sim$0.2 dex in the logarithm of R$_{\rm BLR}$. To mimic these findings we set a Gaussian prior in the logarithm of $\tau_{\rm mean}$, centred at $\mu_{\tau} = 5.21$ days and with $\sigma_{\tau} = 0.2$ dex, and perform the BLR modelling using the Arp 151 mid-flux epoch spectrum. 
In addition to the default parameters described above, we also test a case where we use a different mean value ($\mu_{\tau}$) for the Gaussian prior. This is to simulate a situation where the R-L relation predicts a mean time delay that is too high (in our tests by a factor of 3) compared with the intrinsic unknown value. In the latter test we assume two possible values for $\sigma_{\tau}$, the default of 0.2 dex and a significantly higher uncertainty of 1.2 dex.

The results are shown in Fig.~\ref{tau_diff_mean_value_posterior}. For the case where we use the R-L relation results directly (blue histogram) the posterior distributions are consistent with the results for the full light-curve. This is an important result considering that the R-L relation itself overestimates $R_{\rm BLR}$ for Arp 151. The inferred geometry and dynamics parameters ($\beta$, $\theta_{o}$, $\theta_{i}$, $\kappa$, $\xi$, $\gamma$, $f_{\rm ellip}$, $f_{\rm flow}$ and $\theta_{e}$) are consistent within the uncertainties with what has been determined from the full light-curve modelling, as can be seen in columns 9, 10 and 11 of Table~\ref{table_results} and in Fig.~\ref{inferred_parameters}. The black hole mass is more strongly correlated with the time delay. The worst result is obtained when the $R_{\rm BLR}$ is overestimated by a factor of $\times 3$ while maintaining a narrow uncertainty of $\sigma_{\tau} = 0.2$ dex on the prior (yellow histogram). This setup does not allow the model to significantly explore the low black hole mass region which results in an overestimated $M_{\rm BH}$. 
The last case where $\sigma_{\tau} = 1.2$ dex (red histogram) results in a broader posterior distribution and larger final uncertainty in the inferred black hole mass. For this case the broad distribution encompasses the M$_{\rm BH}$ estimate based on the full light curve results.

Our test case of $\mu_{\tau}$ equal 3 times the R-L inferred value is somewhat extreme, considering that the intrinsic dispersion in the R-L relation is 0.2 dex (\citealt{bentz13}). However our goal was to show the model's behaviour when tested with extreme cases. Considering this, it is remarkable that the inferred parameters for this test are consistent within the 68\% confidence range with the result from the full light-curve, as can be seen in Fig.~\ref{inferred_parameters}. The test we carried out for this extreme case may also be helpful in light of recent findings of samples of AGN that lie below the R-L relation (e.g. \citealt{grier17b}, \citealt{du18}). 

From these tests we can infer that some of the information on the BLR geometry and dynamics parameters is contained in the spectrum and does not completely depend on the assumed prior on the time delay to set the parameters. We want to highlight that all of the three tests mentioned above provided equally good constraints on all the geometry and dynamics parameters. This means that even though the prior on the mean time delay changes in these tests, the model is able to extract the necessary information from the broad line profile. An exception to this is the black hole mass which is more sensitive to the time delay assumed. If one assumes a too narrow confidence range on $\tau_{\rm{mean}}$, and by chance the R-L determined $R_{\rm BLR}$ is offset from the intrinsic M$_{\rm BH}$ value, this will result in a relatively narrow posterior distribution for the M$_{\rm BH}$ parameter but the mean value of the posterior distribution will be offset. The best approach to a more general application of this method is to set a wide prior that does not overly restrict the model, to take into account scenarios where the R-L determined $R_{\rm BLR}$ is uncertain or poorly constrained. 

\section{Discussion}
\label{sec:discussion}
\subsection{Constraining the geometry and dynamics of the BLR from single-epoch spectra}
In this section we discuss the constraints that can be derived for the BLR geometry and dynamics parameters from modelling of single-epoch spectra, the effect of the signal-to-noise ratio and the particular features that may make Arp 151 a good test case.

In Table~\ref{table_results} we show the inferred values for the parameters and their respective 68\% confidence ranges obtained for each of our different tests. Each inferred parameter posterior probability distribution is determined by marginalising over the remaining model parameters. The inferred parameter value is calculated from the median of the parameter posterior probability distribution. The distribution of the inferred parameters for each test can be visually compared in Fig.~\ref{inferred_parameters}. The black circles indicate the inferred values for the parameters and their respective 68\% confidence ranges. The shaded green vertical bars show the 68\% confidence range region for the full light-curve modelling inferred parameters (labelled as [1] in Fig.~\ref{inferred_parameters} and in Table~\ref{table_results}). Both Fig.~\ref{inferred_parameters} and Table~\ref{table_results} will be used as reference for the discussion that follows in this section.

The BLR parameters can be divided into three groups, based on our analysis for the case of Arp 151: 1) parameters that are well constrained by single-epoch spectra; 2) parameters that depend on the epoch; 3) parameters that cannot be constrained by single-epoch spectra. We discuss each of these groups below.

\subsubsection{BLR parameters constrained by single-epoch spectra}
Some of the inferred parameters show a consistent trend towards the full-lightcurve result, independently of the epoch chosen: $\theta_{o}$, $\theta_{i}$, $\xi$, $f_{\rm flow}$, $\theta_{e}$. 
M$_{\rm BH}$ also does not depend strongly on the epoch. However, it is more closely connected with $\tau_{\rm mean}$, and therefore can be biased if too stringent confidence ranges are imposed on the $\tau_{\rm mean}$ Gaussian prior, as in test number [9]. Fig.~\ref{mbh_tau_mean} shows the two dimensional posterior probability distribution for M$_{\rm BH}$ and $\tau_{\rm mean}$ obtained for the mid-flux epoch to illustrate the dependence of the black hole mass on the mean time delay.

The three angles $\theta_{o}$, $\theta_{i}$ and $\theta_{e}$ are consistent within the 68\% confidence range with the inferred values from the full light-curve modelling. We find $\theta_{o}$ = (\Remark{low_opening}, \Remark{mid_opening}, \Remark{high_opening}) and $\theta_i$ = (\Remark{low_inc}, \Remark{mid_inc}, \Remark{high_inc}) for the low-flux, mid-flux and high-flux epochs, respectively. While the median values seem to be underestimated for two of the epochs with respect to the value found from the full light-curve modelling ($\theta_{o} = 26.4^{+ 3.9}_{- 5.9}$ and $\theta_i = 25.8^{+ 4.2}_{- 5.7}$), the fact that all three values are consistent within their 68\% confidence region suggests that the perceived BLR angles do not vary strongly between epochs. The uncertainties associated with our inferred $\theta_{o}$, $\theta_{i}$ are only a factor of $\sim$2 higher than those from the full light-curve modelling. For $\theta_{e}$ we find $\theta_{e}$ = (\Remark{low_thetae}, \Remark{mid_thetae}, \Remark{high_thetae}) for the low-flux, mid-flux and high-flux epochs, respectively. For the low-flux epoch, $\theta_{e}$ is not well constrained by the data as can be seen from its broad 68\% confidence range. However, $\theta_{e}$ for the mid-flux epoch is remarkably well constrained, and has a similar median value and similar uncertainties to the inferred full-lightcurve $\theta_{e} = 12.7^{+11.3}_{- 8.6}$. 

In Fig.~\ref{inc_opening_mbh} we show the line profile shapes generated by the model, for a similar set of parameters as those inferred from the full light curve modelling of Arp 151 and listed in column 2 of Table~\ref{table_results}. {This figure illustrates how the inclination angle and opening angle affect the line profile. The asymmetry between the blue and red peaks observed in the line profiles is due to a preference for the far side of the BLR to emit more ($\kappa = -0.29$) combined with the inflowing gas orbits ($f_{\rm ellip} = 0.13$ and $f_{\rm flow} = 0.27$), which makes the blue-shifted emission more prominent than the redshifted emission. The effect of an almost opaque mid-plane ($\xi = 0.09$), is mostly seen at low inclinations, due to the presence of the inflowing gas orbits. An example is the $\theta_{i} = 0$ orbit with $\theta_{o}$ = 15$^{\circ}$. For that case the mid-plane of the BLR is in the plane of the sky, which means that mostly only foreground gas is observed. As there is a strong component of inflowing gas orbits, by obscuring the background gas, it means that most of the blue-shifted component of the line-of-sight gas velocity is removed, skewing the line towards the red.

\begin{figure}
\centering
\includegraphics[width=0.49\textwidth]{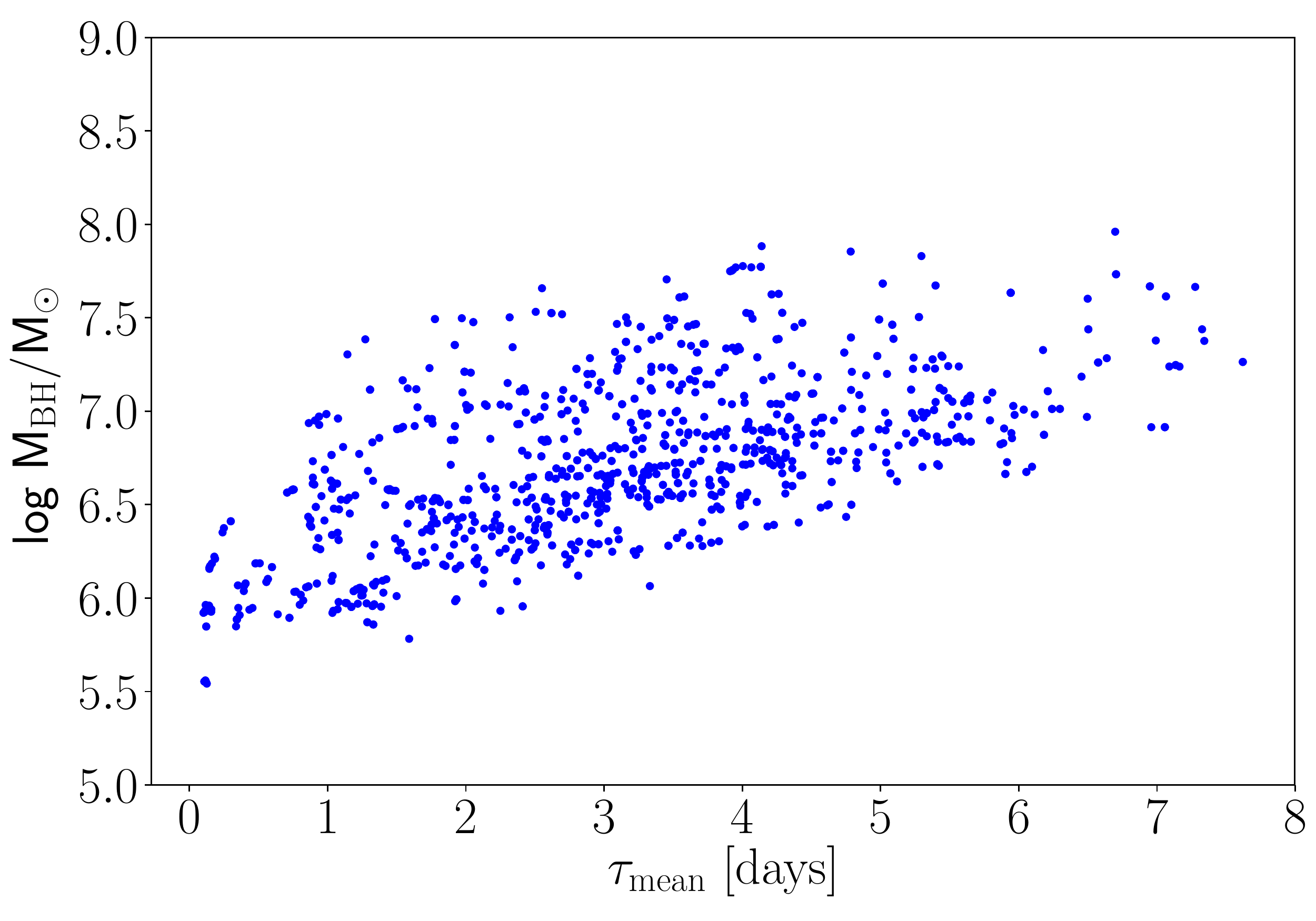}
\caption{Figure illustrating the dependence between the black hole mass ($M_{\rm BH}$) and the mean time delay ($\tau_{\rm mean}$) for the mid-flux epoch of Arp 151. The values for the parameters were taken from the posterior probability distribution.}
\label{mbh_tau_mean}
\end{figure}

\begin{figure*}
\centering
\includegraphics[width=0.9\textwidth]{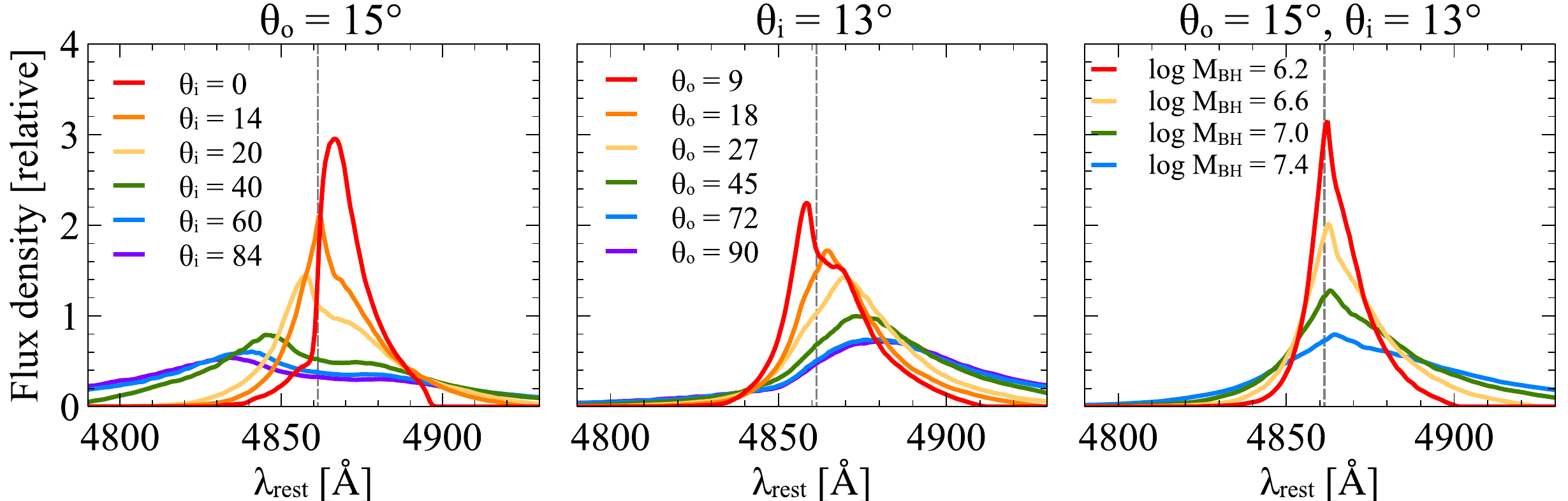}
\caption{Simulated broad line profiles for different inclinations and opening angles for a case similar to Arp 151. The remaining parameters are set to the inferred values for Arp 151 quoted in column 2 of Table~\ref{table_results}. The vertical dashed line indicates the central H$\beta$ rest-frame wavelength. The left panel shows the effect of changing the inclination angle for a fixed opening angle of $\theta_{o}$ = 15 degrees. The central panel shows the effect of changing the opening angle for a fixed inclination of $\theta_{i}$ = 13 degrees. The right panel shows the effect of changing the black hole mass for fixed inclination ($\theta_{i}$ = 13 degrees) and opening angle ($\theta_{o}$ = 15 degrees). }
\label{inc_opening_mbh}
\end{figure*}

\begin{figure}
\centering
\includegraphics[width=0.49\textwidth]{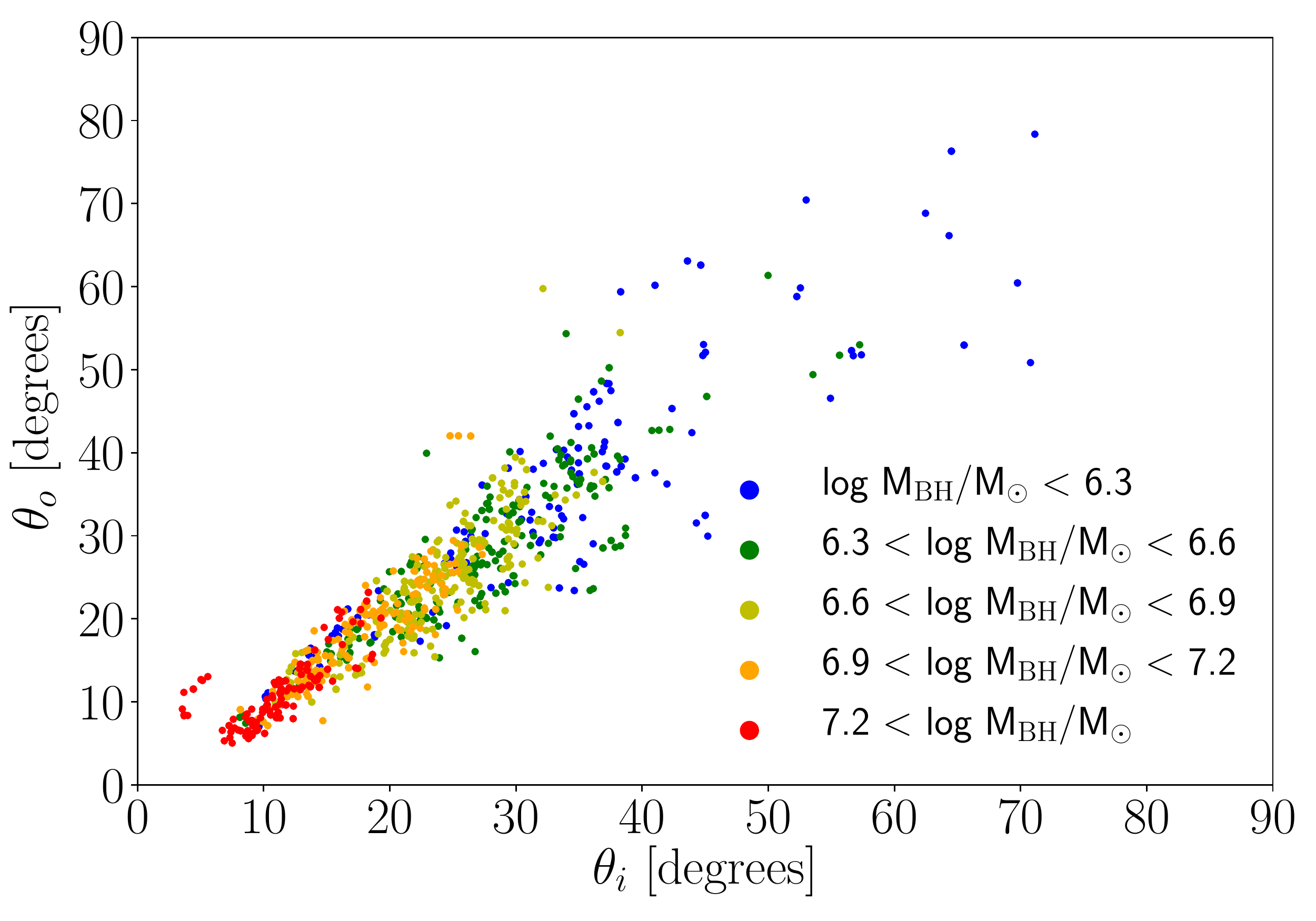}
\caption{Figure illustrating the dependence between the inclination angle, opening angle and black hole mass for Arp 151. The values for the parameters were taken from the posterior probability distribution. The coloured points represent different black hole mass bins.}
\label{corner_plot_angles}
\end{figure}

\begin{figure*}
\centering
\includegraphics[width=0.7\textwidth]{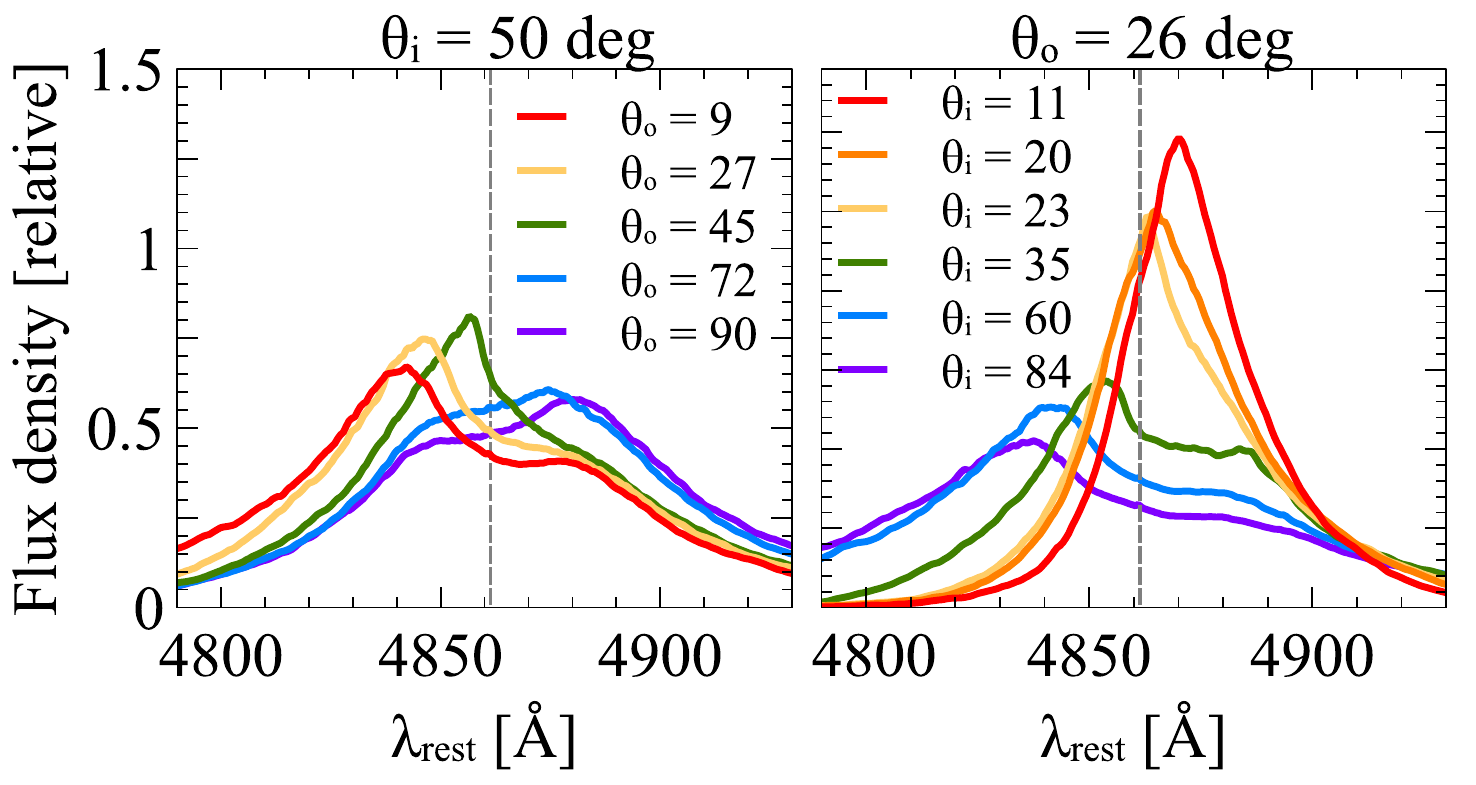}
\caption{Simulated broad line profiles for different inclinations and opening angles for a case similar to Arp 151. The remaining parameters are set to the inferred values for Arp 151 quoted in column 2 of Table~\ref{table_results}. The vertical dashed line indicates the central H$\beta$ rest-frame wavelength. The left panel shows the effect of changing the opening angle for a fixed inclination of $\theta_{i}$ = 50 degrees, illustrating why high inclination angles are not considered a good fit of the observational data. The right panel shows the effect of changing the inclination angle for a fixed opening angle ($\theta_{o}$ = 26 degrees), corresponding to the inferred value of $\theta_{o}$ from the full light curve modelling of Arp 151. The right panel illustrates the preferred inclination range of $\theta_{i} \sim$ 10 - 30 degrees found by the model.}
\label{inc_opening_high}
\end{figure*}

\begin{figure*}
\centering
\includegraphics[width=0.7\textwidth]{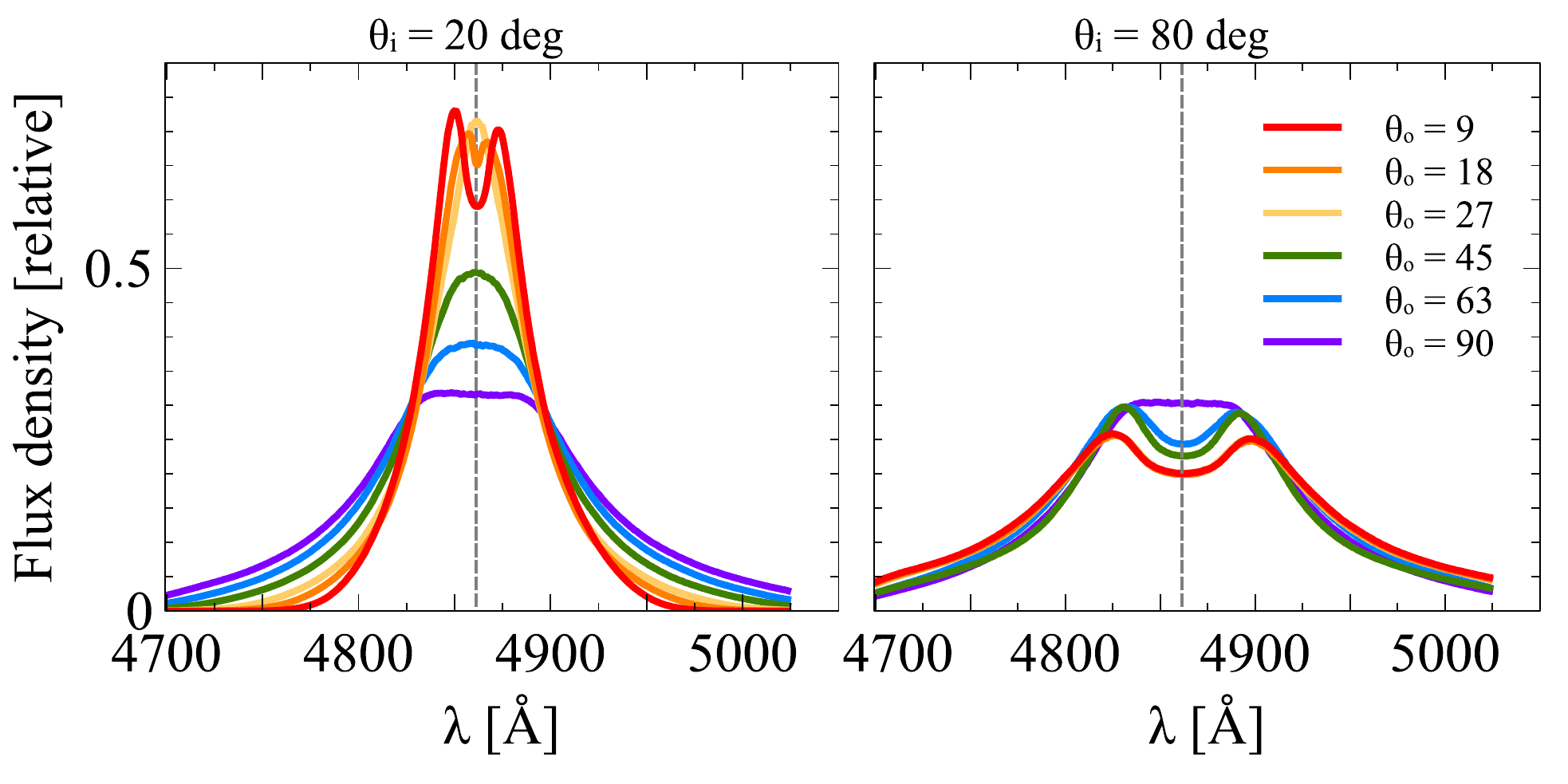}
\caption{Simulated broad line profiles for different inclinations and opening angles. The vertical dashed line indicates the central H$\beta$ rest-frame wavelength. To better illustrate the effect of changing the angles, the line profiles were generated assuming $f_{\rm ellip} = 1$, $\gamma = 1$, $\xi = 1$ and $\kappa = 0$}.
\label{sphere_disc}
\end{figure*}

Similar to what was found in previous modelling of the full light curves (\citealt{pancoast14b}, \citealt{grier17}, \citealt{pancoast18}), we find a correlation between the inclination angle and the opening angle. In Fig.~\ref{corner_plot_angles} we show the two dimensional posterior distributions for Arp 151, showing the correlation between $\theta_{i}$ and $\theta_{o}$ colour coded as a function of black hole mass.  As can be seen in Fig.~\ref{inc_opening_mbh} and Fig.~\ref{corner_plot_angles}, an increase in inclination angle and opening angle, can be compensated by a decrease in black hole mass, so that small variations in these three parameters can produce various line profiles that are consistent. This results in several possible solutions for the angles that spread as a function of black hole mass. This dependence is mostly driven by the fact that larger black hole masses will increase the line of sight velocity of the gas and therefore increase the line profile width. The inclination affects the line-of-sight velocity via a factor of sin($\theta_{i}$), broadening the line profiles as the inclination increases.

There is also the question of why the opening angle and inclination angle are similar for Arp 151 (and in fact for most of the AGN modelled via this technique - see \citealt{grier17} for a discussion). In the simulations we show in Appendix \ref{sec:appendix_sim}, the model is able to recover the real $\theta_{i}$ and $\theta_{o}$ for different setups, including $\theta_{i} \sim \theta_{o}$, $\theta_{o} > \theta_{i}$ and $\theta_{o} < \theta_{i}$. This shows that the model does not drive the trend $\theta_{o} \sim \theta_{i}$. The fact that the angles found for real AGN are similar is mostly due to the transition that occurs in the line profile when $\theta_{o} \sim \theta_{i}$. In a simplified way, a BLR with an inclination $\theta_{i}$ and opening angle of $\theta_{o}$ can be imagined as a collection of thin discs with inclinations between $\theta_{i} - \theta_{o}$ and $\theta_{i} + \theta_{o}$. When $\theta_{o}$ is of the order of, or higher than $\theta_{i}$, a significant fraction of orbits close to the plane of the sky will be included. These orbits have velocities close to the rest-frame velocity of the AGN and therefore will transform the line profiles into `box-like' profiles. For $\theta_{o} \gg \theta_{i}$, the line profiles will tend to have flatter tops. On the other hand, for $\theta_{o} \ll \theta_{i}$, the line profiles will tend to show double peaks. If the inclination is low, as one would expect for type 1 AGN, a combination of low inclination and lower opening angles will result in the gas having a limited range in line of sight velocities creating line profiles that have more compact wings. Both the very boxy profiles and the compact wing profiles as generated by the model tend to not be a good representation of the observed line profiles for the sub-sample of AGN analysed with the \cite{pancoast14a} code. In Fig~\ref{inc_opening_high} we show different line profiles for the Arp 151 parameters to illustrate part of the $\theta_{i}$ vs $\theta_{o}$ parameter space for this AGN {and the arguments outlined above.

As the line profiles for Arp 151 have a significant contribution from parameters that cause asymmetry, we produced line profiles using a simple structure for the BLR to isolate the effect of the inclination and opening angles. In Fig.~\ref{sphere_disc} we show line profiles generated assuming that there was no inflow or outflow ($f_{\rm ellip} = 1$), a uniform distribution of particles ($\gamma = 1$), a transparent mid-plane ($\xi = 1$) and isotropic emission ($\kappa = 0$). As can be seen from the figure, changing the inclination or opening angles make the line top flat or double-peaked and affect the width and shape of the tail of the emission line. The panel on the right shows that when the opening angle is lower than the inclination angle, and there are not significant asymmetries in the BLR, the profiles are difficult to distinguish, which means that for those cases the model may only be able to determine an upper limit for the opening angle.

\begin{figure*}
\centering
\includegraphics[width=0.99\textwidth]{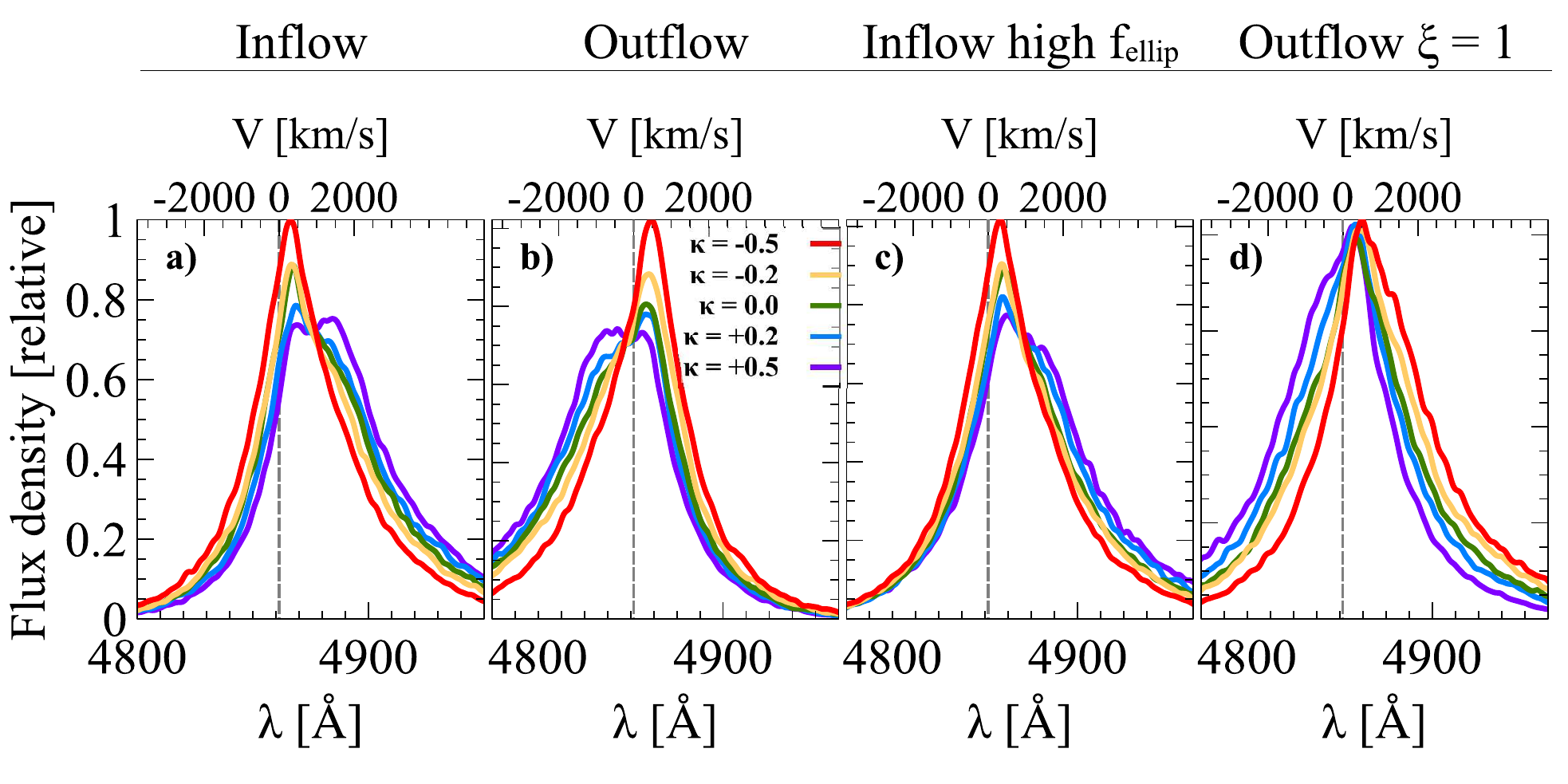}
\caption{Model-generated line profiles as a function of geometry and dynamics parameters, illustrating how the BLR structure and kinematics affect the shape of the emission line in a single-epoch spectrum. The parameters $f_{\rm flow}$, $f_{\rm ellip}$, $\kappa$ and $\xi$ are different in the four panels, otherwise the parameters used to generate the line profiles are similar to those inferred from the full light curve result. The solid coloured lines represent the line profiles generated for different values of $\kappa$. Panel a) and b) show an example of the effect of inflowing and outflowing gas orbits, with $f_{\rm flow} = 0.27$ and $f_{\rm flow} = 0.73$ respectively, $f_{\rm ellip} = 0.13$ and obscuration in the mid-plane of the BLR ($\xi = 0.09$). Panel c) shows a similar configuration to panel a) with the only difference being a higher value of $f_{\rm ellip} = 0.9$, i.e. less particles in unbound orbits. Panel d) shows the effect of removing the obscuration in the mid-plane of the BLR ($\xi = 1$), for a fully outflowing BLR ($f_{\rm ellip} = 0.0$). The vertical line indicates the rest-frame of the H$\beta$ emission line.}
\label{line_shapes_kappa}
\end{figure*}

The parameter $f_{\rm flow}$, which describes the tendency of BLR particles to be in inflowing or outflowing orbits is also consistent ($f_{\rm flow} < 0.5$) for all the tests we carried out. A change in $f_{\rm flow} < 0.5$ to $f_{\rm flow} > 0.5$ causes a strong asymmetry in the spectral shape of the broad line. This is why this parameter is well determined from single-epoch data, as the information provided to the model comes from the line profile.
In Fig.~\ref{line_shapes_kappa} we show examples of line profiles generated by our model for different combination of parameters, to illustrate possible configurations of the BLR and their influence on the line profile. Panels a) and b) of Fig.~\ref{line_shapes_kappa} show examples of line profiles where inflowing or outflowing orbits dominate (with $f_{\rm flow} = 0.27$ and $f_{\rm flow} = 0.73$ respectively and $f_{\rm ellip} = 0.13$). The signature of inflowing/outflowing orbits are asymmetric tails towards lower or higher wavelengths in the line profile. The tails are less pronounced if $f_{\rm ellip}$ is higher, which means that a smaller percentage of orbits are in inflowing or outflowing configurations. An example of this is shown in panel c) of Fig.~\ref{line_shapes_kappa}, which was generated assuming $f_{\rm flow} = 0.27$ and $f_{\rm ellip} = 0.9$.

We infer values of $\xi =$ (\Remark{low_xi}, \Remark{mid_xi}, \Remark{high_xi}) for the low-flux, mid-flux and high-flux epochs, respectively. The median values of this parameter are similar to the result from the full light-curve modelling: $\xi = 0.09^{+0.12}_{-0.06}$ with associated uncertainties that are a factor of $\sim$ 1.5 to 3.5 times higher. For the low flux epoch, the values of $\xi$ are still consistent within the uncertainties with the full light curve, however, the 68\% confidence region is very broad, which indicates that $\xi$ is not significantly constrained for this epoch. The $\xi$ parameter controls the BLR mid-plane transparency and is responsible for breaking the symmetry between the back and the front side of the BLR with respect to its mid-plane. This can be seen in Fig.~\ref{line_shapes_kappa}. Panel d) shows a case where $\xi = 1$, i.e. the BLR mid-plane is completely transparent, and it is fully outflowing ($f_{\rm ellip} = 0.0$). For this case, the line profile is close to symmetric for changes in $\kappa$. A similar line profile will be obtained for an inflowing BLR with anisotropic emission towards the ionising source ($\kappa < 0$) and an outflowing BLR with anisotropic emission away from the ionising source ($\kappa > 0$). A value of $\xi \neq 1$ such as observed in Arp 151, will break the symmetry. This can be seen in panels a) to c) of Fig.~\ref{line_shapes_kappa} that assume $\xi = 0.09$. 
Such behaviour could explain why $\xi$ is constrained from single-epoch spectra.

Focusing on the three single epochs selected for Arp 151 (columns 3 - 5 in Table~\ref{table_results}), we can see that the geometry and dynamics parameters mentioned above ($\theta_{o}$, $\theta_{i}$, $\xi$, $f_{\rm flow}$ and $\theta_{e}$) can be determined with uncertainties that are comparable with those obtained using the full light-curve, or up to a factor of 3.5 times higher, depending on the epoch. Such uncertainties are noteworthy considering that a single-epoch spectrum requires significantly less observing time than multi-epoch monitoring data.

We note that due to the probabilistic approach of our model, the parameter space is searched for all possible solutions (in terms of sets of parameters) that reproduce the line profile observed. Therefore, multiple solutions and their relative probabilities will be recorded and that information is part of the posterior probability distribution. The shape of the posterior distribution, that we characterise by the median value and 68\% confidence range, will contain the information about how well a parameter can be constrained from the data. In Fig.~\ref{corner_plot} in the Appendix~\ref{sec:appendix_extra_fig} we show the two dimensional posterior probability distributions, or corner plots for a selected set of our parameters for the mid-flux epoch.
\subsubsection{BLR parameters dependent on the epoch}
The inferred values of some BLR parameters depend more strongly on the chosen epoch than others. The parameters that depend strongly on the epoch will show different distributions for each of the epochs but consistent distributions between mid-flux epoch [3] and the remaining tests. These are: $\beta$, $\kappa$ and $f_{\rm ellip}$. 

The parameter $\beta$, which controls the shape of the radial BLR cloud distribution, shows similar median values and uncertainties of up to a factor of 2 higher than the full light-curve result ($\beta$ = $1.35^{+0.13}_{-0.13}$) for all the tests with the exception of the low-flux epoch. We find $\beta =$ (\Remark{low_beta}, \Remark{mid_beta}, \Remark{high_beta}) for the low-flux, mid-flux and high-flux epochs, respectively. The low-flux epoch predicts a higher $\beta$ than the full light-curve modelling. This may indicate that different portions of the BLR will react depending on a combination between the flux level and the prior continuum history. The line from the low-flux epoch may only be emitted from the regions closest to the black hole, which would result in a cloud distribution that decays more steeply with increasing radius. This is the so-called breathing effect: in low flux states it is expected that the BLR is smaller, more compact, and intrinsically more responsive (\citealt{netzer&maoz90}, \citealt{korista&goad04}, \citealt{cackett&horne06}).
In this case, one would indeed expect a $\beta$ value that is high, in line with the relatively high $\beta$ value inferred from the modelling of the low-flux epoch compared to the other epochs. It may also depend on the previous continuum history and show effects of hysteresis: even if the flux is the same, the line profile may vary depending on whether the source was previously in a higher or lower flux state (e.g. \citealt{perry94}).

The $\kappa$ parameter is related to a preferential BLR near-side or far-side emission relative to the ionising source, as seen from the observer. All tests produce similar results for this parameter with the exception of the low and high-flux epochs, for which $\kappa$ cannot be constrained. We find $\kappa$ =  (\Remark{low_kappa}, \Remark{mid_kappa}, \Remark{high_kappa}) for the low-flux, mid-flux and high-flux epochs, respectively. The full light-curve result is $\kappa = -0.29^{+0.11}_{-0.10}$. The 68\% confidence ranges for the low and high-flux epochs are consistent with both negative and positive values, indicating that the ability to constrain $\kappa $ may depend on the shape of the line profile which varies from epoch to epoch. 
In Fig.~\ref{line_shapes_kappa} we show an example of the effect of $\kappa$ on the line profile. This parameter has a strong effect on the prominence of the tail of the line profile when inflowing or outflowing orbits are present, as can be seen in panels a) and b). However, the effect of $\kappa$ is less noticeable when $f_{\rm ellip}$ is large, i.e. only a small fraction of orbits are in inflowing or outflowing configurations, as in panel c) calculated with $f_{\rm ellip} = 0.9$. It is difficult to distinguish the effect of small changes in $f_{\rm ellip}$ and $\kappa$ in the line profile, for specific BLR configurations such as the one in Arp 151 (exemplified in Fig.~\ref{line_shapes_kappa}). This explains why $\kappa$ and $f_{\rm ellip}$ are difficult to constrain independently for every single epoch.

The parameter f$_{\rm ellip}$ shows a somewhat similar behaviour to $\kappa$, as mentioned above. It may depend on the epoch as the inferred parameters only marginally agree within the uncertainties. For the low-flux epoch f$_{\rm ellip}$ is mostly unconstrained. We find $f_{\rm ellip}$ = \Remark{low_fellip}, \Remark{mid_fellip}, \Remark{high_fellip} for the low-flux, mid-flux and high-flux epochs, respectively. The full light-curve result is $f_{\rm ellip}$ = $0.13^{+0.11}_{-0.09}$. The expression 1$-$f$_{\rm ellip}$ determines the fraction of particle orbits around the radial inflowing or outflowing escape velocity. It is interesting that $f_{\rm flow}$, the parameter that determines if the 1-f$_{\rm ellip}$ particles are mostly close to inflowing or outflowing velocities, seems to be well determined and consistent for single epochs. If indeed f$_{\rm ellip}$ depends on the epoch, one interpretation is that the tendency for inflow or outflow does not change between epochs, but that the fraction of particles that can be in those inflows or outflows does change.
\subsubsection{Unconstrained BLR parameters}
In our single-epoch spectra modelling, $\gamma$, the particle concentration, is almost completely unconstrained by the data for every test that we carried out. The parameter $\gamma$ does not have a strong effect on the broad line spectral profile which is consistent with our findings of non-existent constraints on this parameter. From our tests we can conclude that single-epoch spectra may not be able to constrain $\gamma$, and is unlikely to be useful in BLR population studies.
\subsubsection{The effect of signal-to-noise ratio}
The Arp 151 data we use in this work have high signal-to-noise ratio (S/N). The three epochs selected have S/N $\sim$ 85 - 175 at the peak of the H$\beta$ line. Fig.~\ref{3epochs_norm}, right panel, shows the S/N for each of the epochs after subtracting the continuum flux. The low-flux epoch that has the lowest S/N is also the epoch for which the posterior probability distribution functions for the parameters are broader, as can be seen from the 68\% confidence ranges in Fig.~\ref{inferred_parameters}. This indicates that the parameters are more loosely constrained and sometimes not constrained at all as shown in the previous sections. From the tests we did here, the mid-flux epoch is the one that provides inferred parameter values closest to the full light-curve modelling. However, the mid-flux epoch is not the one with the highest S/N. This suggests that successful modelling is not only a function of S/N but that it depends on other effects such as the line profile shape, discussed in the previous sections. Still, all the results of our tests have been obtained with relatively high quality spectra (S/N $>$ 85) when compared with all the single-epoch spectra available in the literature. This should be taken into consideration when selecting targets to extend the single epoch BLR modelling to a larger sample of AGN. 
\subsubsection{Arp 151 as a test case}
Particular features of Arp 151 dataset make it a good test case. In addition to the high signal-to-noise mentioned above, the H$\beta$ line profile in Arp 151 has the advantage of being reasonably well separated from other spectral features. In some AGN the wings of the H$\beta$ emission line are blended with other emission lines such as Fe II and He II, making it more difficult to separate broad and narrow emission features. This should be taken into account when modelling a larger sample of AGN. In future work we will test the model's performance for AGN in which the H$\beta$ line profile is not so clearly isolated.
\subsection{The effect of using monitoring data}
\label{sec:timing}
In this section we discuss what would be the expected contribution to the model if timing information and spectral information (i.e. a monitoring dataset) were to be used to constrain the model, as opposed to single-epoch spectra.

The first effect is the number of broad line spectral profiles that are used as input to the model. In our model we select single-epochs and therefore only one broad emission spectral profile is used as input. As a comparison, the input data used in the full light curve modelling of \cite{pancoast18} contains 43 spectral profiles (a full spectral dataset). As can be seen from the results in this work, 43 spectra provide stronger constraints on the BLR parameters than a single spectrum. This is of course expected since the model tries to reproduce all the input spectra by searching the parameter space. It is also expected, as we see in our work, that not all of the parameters can be constrained from a single spectrum. The parameters that can be constrained using a single epoch spectrum will typically have larger uncertainties than when using the full monitoring data set.

The second effect is the timing information. \cite{pancoast14a} describe simulations using two different datasets to constrain the model: 1) flux light curves, comprised of the continuum flux light curve and the integrated H$\beta$ line flux light curve; 2) continuum and integrated H$\beta$ line flux light curves and a mean spectrum of the broad H$\beta$ line. The first dataset contains only the light curve information and no spectral information. They find that with this dataset, only the time delays and mean radius of the BLR are constrained by the model. All the remaining parameters are mostly unconstrained. This shows that the light curves mostly contain information on the time delays and size of the BLR.
The second dataset simulated by \cite{pancoast14a} still contains the timing information from the light curves but adds the information contained in one spectrum which in the study by \cite{pancoast14a} is the mean spectrum. They find that the parameters inferred by the model are fully consistent with those determined from the modelling of the monitoring data, but with larger uncertainties. Since this second dataset contains the light curves and therefore timing information, they are able to constrain the time delays, radius of the BLR and black hole mass albeit with larger uncertainties than for the monitoring dataset case. Notably, the mean time delay and mean BLR radius have uncertainties that are ten times larger than those determined from modelling of the full monitoring dataset (light curves + spectra), indicating that the full spectral dataset has a major role in accurately constraining the time delays.

In our model we go one step beyond the second dataset referred to above. We exclude the light curves and therefore all the timing data completely. Our findings are similar to what \cite{pancoast14a} find using the second dataset: the parameters inferred are fully consistent with those determined when modelling the monitoring dataset but with larger uncertainties. The difference between using a single-epoch spectrum and the second dataset of \cite{pancoast14a} is the lack of timing information. As discussed above, the timing information determines the ability (or inability) of the model to constrain the time delays, radius of the BLR and to some extent the black hole mass. In this work we test the performance of the model when the timing information is replaced with information on the prior probability distribution of the mean time delay. Our findings are in agreement with what \cite{pancoast14a} found using the second dataset, in that our results are consistent with those from the monitoring dataset modelling but with larger uncertainties, with the exception of the $\beta$ parameter which appears to change with the epoch. The other major difference is that without the light curves we are not able to independently determine the time delays and radius of the BLR. These parameters and the black hole mass will be influenced by our selection of the prior probability distribution for the mean time delay.
\subsection{Black hole mass}
The black hole mass parameter is more sensitive to the prior constraint on $\tau_{\rm mean}$, than other parameters, as can be seen from the first panel of Fig.~\ref{inferred_parameters} and Fig.~\ref{mbh_tau_mean}. 
Focusing on the result for the three different epochs (low-flux, mid-flux and high-flux) we can see that the black hole mass obtained for each of the epochs is consistent with the one derived from the full light-curve modelling. The uncertainties, as expected, are larger than those obtained with the full light-curve modelling, of the order of 1 dex for each individual epoch. This uncertainty is similar to the inferred maximal statistical uncertainty estimated for mass determinations based on single-epoch spectra and black hole mass scaling relations \citep{vestergaard&peterson06}. We find $\log_{10}(M_{\rm BH}/M_\odot)$ = \Remark{low_Mbh},  \Remark{mid_Mbh}, \Remark{high_Mbh} for the low-flux, mid-flux and high-flux epochs, respectively. The full light-curve result is $\log_{10}(M_{\rm BH}/M_\odot)$ = $6.58^{+0.20}_{-0.12}$. Our median black hole mass value is overestimated with respect to the full light-curve result. However, the analysis of the full light-curves of Arp 151 from 2011 and 2015 provides higher inferred median M$_{\rm BH}$ values as well (log(M$_{\rm BH}/M_\odot) = 6.93^{+0.33}_{-0.16}$ and log(M$_{\rm BH}/M_\odot) = 6.92^{+0.50}_{-0.23}$ respectively) \citep{pancoast18}. Our inferred median $M_{\rm BH}$ from single-epoch modelling is in better agreement with the 2011 and 2015 estimates.
In general, when we assume a prior for the time delay, the inferred $M_{\rm BH}$ values are consistent, within the uncertainties, with the result of the full light-curve modelling. This is valid unless the mean value for that prior is significantly different (in our case by a factor of $\times$ 3) from the real time delay and the prior assumes a narrow uncertainty. For future modelling of single-epoch spectra, the best approach is to use the R$-$L determined value for R$_{\rm BLR}$ but assume a conservative $\sigma_{\tau}$ for the prior, so that the prior does not limit the parameter space search in case R$_{\rm BLR}$ is over or under estimated. However, we highlight that the strength of our model is in determining the BLR geometry and dynamics parameters. Even with an overestimated mean value for the prior on $\tau_{\rm mean}$ the inferred parameters (with the exception of M$_{\rm BH}$) are remarkably consistent.
\subsection{Future applications of the model}
The fact that some of the parameters are being constrained by single epoch data is an encouraging result. It indicates that it is possible to extend BLR modelling to other AGN for which only single epoch spectra and time lag estimates (possibly derived from the radius-luminosity relation) are available. 
In follow up papers we will utilise these new results to explore the potential of applying this modelling to a wider sample of AGN, with and without monitoring data and spanning a wide range of redshifts. We will also use the same approach to model additional broad lines such as Mg II and C IV in particular for high redshift AGN. Our ultimate goal is to constrain the typical BLR geometry and dynamical parameters for improved insight on the physics of the AGN central engine.

The BLR modelling in this work relies on an underlying physical prescription for the BLR. The free parameters in the model add flexibility to this prescription in that a variety of BLR configurations can be simulated by the model. However there is an intrinsic limitation to the physics that can be described by our model, and physical processes that are beyond the prescription assumed will not be captured by the model. Future applications of the model will be to include further physical processes that are known to be important in the BLR, such as photoionisation physics, radiation pressure or winds. Comparing and implementing other underlying physical models for the BLR is also important, since there are currently several physical interpretations for the BLR that have been suggested to explain the current observational data (e.g. \citealt{goad12}, \citealt{elvis17}).
\section{Conclusions}
\label{sec:conclusions}
The broad line region model of \cite{pancoast14a} has been successfully used to model monitoring data to determine the BLR geometry and dynamics of a small sample of local AGN. However, due to the observational resources required, data from monitoring campaigns are only available for a small sub-sample of the known AGN. In this work we modify the BLR model of \cite{pancoast14a} for application to single-epoch AGN spectra with the major modification being the addition of a prior probability distribution for the BLR size. We test our modified model to determine if it can be used to obtain meaningful constraints on the BLR parameters when single-epoch spectra (i.e. one single spectrum) are used.

We test our modified model on the AGN Arp 151, for which full light-curve broad line region modelling has been carried out by \cite{pancoast14b} and \cite{pancoast18}. The tests are performed by extracting single-epoch spectra from Arp 151's full light-curve and analysing them independently. This setup tests for the first time the performance of the model when applied to data that does not contain timing information.
We find that a large fraction of the BLR geometry and dynamics parameters inferred from single epoch spectra modelling are consistent within the uncertainties with the full light-curve modelling result. 
Our study identifies which BLR geometry and dynamics parameters are weakly dependent on the epoch, and therefore are the best candidates to be constrained by single-epoch spectra. Notably, these best candidate parameters: (the opening angle $\theta_{o}$, inclination angle $\theta_{i}$, the transparency of the BLR mid-plane $\xi$, the inflow or outflow orbits $f_{\rm flow}$ and the rotation angle between the circular and inflow/outflow orbits $\theta_{e}$) can be determined from single-epoch spectra with uncertainties that are comparable with those obtained using the full light-curve, or up to a factor of $\sim$ 3.5 times higher, depending on the epoch selected. Our tests indicate that BLR modelling using single-epoch spectra and an underlying physical model can provide constraints on the BLR structure. Considering the wealth of available single-epoch AGN observations, our approach can potentially be used to determine the overall AGN population trends in the geometry and dynamics of the BLR. We will further explore this opportunity in future work.
\section*{Acknowledgements}
The authors would like to thank the anonymous referee for a thorough review of the manuscript that improved the quality and clarity of the paper. The authors also thank Bradley Peterson for helpful discussions and comments on an early version of this work. The authors acknowledge financial support from the Independent Research Fund Denmark via grant no. DFF-4002-00275 (PI: Vestergaard). AP is supported by NASA through Einstein Postdoctoral Fellowship grant number PF5-160141 awarded by the Chandra X-ray Center, which is operated by the Smithsonian Astrophysical Observatory for NASA under contract NAS8-03060. Research by A.J.B. is supported by NSF grant AST-1412693.

\bibliographystyle{mnras}
\bibliography{AGN}

\begin{landscape}
\mbox{}\vfill
\begin{table}
\vspace{3cm}
\begin{tabular}{c | c | c | c | c | c | c | c | c | c | c}
 & &  \multicolumn{3}{ c }{Single epochs} &  \multicolumn{3}{ c }{Mid-flux epoch} &   \multicolumn{3}{ c }{(R-L)} \\
  \cmidrule(lr){3-5}\cmidrule(lr){6-8}\cmidrule(lr){9-11}
 Parameter & Full light. & Low-flux & Mid-flux & High-flux & $\tau_{\rm median}$ & $\sigma_{\tau} = 1\mu_{\tau}$ & $\sigma_{\tau} = 2\mu_{\tau}$ & $\sigma_{\tau} = 0.2$ dex & $3\mu_{\tau}$ $\sigma_{\tau} = 0.2$ dex & $3\mu_{\tau}$ $\sigma_{\tau} = 1.2$ dex\\
  $[1]$ & $[2]$ & $[3]$ & $[4]$ & $[5]$ & $[6]$ & $[7]$ & $[8]$ & $[9]$ & $[10]$ & $[11]$\\
 \hline
$\tau_{\rm mean}$  (days)  &   $3.43^{+0.29}_{-0.32}$ &   $3.01^{+1.47}_{-1.56}$ &   $3.21^{+1.54}_{-1.72}$ &   $3.09^{+1.65}_{-1.26}$ &   $3.61^{+1.61}_{-1.73}$ &   $3.98^{+3.10}_{-2.49}$ &   $6.42^{+3.04}_{-4.23}$ &   $5.72^{+2.52}_{-2.29}$ &   $9.33^{+1.55}_{-1.76}$ &   $3.22^{+5.53}_{-2.60}$ \\ 
$\tau_{\rm median}$  (days)  &   $1.86^{+0.23}_{-0.24}$ &   $1.21^{+0.60}_{-0.64}$ &   $1.60^{+0.96}_{-0.85}$ &   $1.48^{+0.86}_{-0.64}$ &   $1.82^{+0.80}_{-0.81}$ &   $2.07^{+1.74}_{-1.32}$ &   $3.58^{+2.10}_{-2.51}$ &   $3.08^{+1.62}_{-1.37}$ &   $5.50^{+1.63}_{-1.31}$ &   $1.84^{+3.15}_{-1.51}$ \\ 
$\beta$ &   $1.35^{+0.13}_{-0.13}$ &   $1.75^{+0.15}_{-0.19}$ &   $1.46^{+0.22}_{-0.22}$ &   $1.53^{+0.12}_{-0.13}$ &   $1.50^{+0.28}_{-0.24}$ &   $1.49^{+0.21}_{-0.26}$ &   $1.51^{+0.25}_{-0.26}$ &   $1.54^{+0.22}_{-0.24}$ &   $1.61^{+0.19}_{-0.24}$ &   $1.49^{+0.27}_{-0.22}$ \\ 
$\theta_o$ (degrees) &   $26.4^{+ 3.9}_{- 5.9}$ &   $ 9.3^{+19.6}_{- 3.9}$ &   $22.5^{+12.5}_{-10.7}$ &   $12.5^{+12.3}_{- 7.4}$ &   $18.7^{+14.0}_{- 8.6}$ &   $23.4^{+11.9}_{-11.3}$ &   $21.5^{+10.2}_{- 9.3}$ &   $20.6^{+11.4}_{- 9.6}$ &   $18.3^{+11.6}_{- 9.2}$ &   $20.6^{+ 9.7}_{-10.3}$ \\ 
$\theta_i$ (degrees) &   $25.8^{+ 4.2}_{- 5.7}$ &   $ 9.6^{+22.2}_{- 3.7}$ &   $22.5^{+10.6}_{- 9.8}$ &   $16.5^{+14.2}_{- 7.8}$ &   $18.9^{+12.8}_{- 7.7}$ &   $22.2^{+10.5}_{- 8.9}$ &   $19.8^{+ 9.8}_{- 6.2}$ &   $21.0^{+10.1}_{- 8.7}$ &   $17.4^{+ 9.3}_{- 7.3}$ &   $20.3^{+ 8.4}_{- 9.0}$ \\ 
$\kappa$ &   $-0.29^{+0.11}_{-0.10}$ &   $0.10^{+0.26}_{-0.37}$ &   $-0.18^{+0.23}_{-0.19}$ &   $0.14^{+0.26}_{-0.34}$ &   $-0.13^{+0.25}_{-0.20}$ &   $-0.18^{+0.23}_{-0.19}$ &   $-0.20^{+0.20}_{-0.18}$ &   $-0.16^{+0.26}_{-0.19}$ &   $-0.25^{+0.26}_{-0.16}$ &   $-0.17^{+0.25}_{-0.19}$ \\ 
$\gamma$ &   $3.97^{+0.75}_{-1.10}$ &   $3.06^{+1.26}_{-1.44}$ &   $3.16^{+1.25}_{-1.38}$ &   $2.32^{+1.54}_{-0.90}$ &   $2.85^{+1.44}_{-1.22}$ &   $2.75^{+1.41}_{-1.20}$ &   $2.82^{+1.60}_{-1.24}$ &   $3.19^{+1.18}_{-1.45}$ &   $3.04^{+1.37}_{-1.41}$ &   $2.92^{+1.37}_{-1.33}$ \\ 
$\xi$ &   $0.09^{+0.12}_{-0.06}$ &   $0.19^{+0.51}_{-0.14}$ &   $0.15^{+0.17}_{-0.09}$ &   $0.17^{+0.34}_{-0.13}$ &   $0.16^{+0.17}_{-0.11}$ &   $0.15^{+0.17}_{-0.08}$ &   $0.14^{+0.12}_{-0.08}$ &   $0.14^{+0.15}_{-0.09}$ &   $0.15^{+0.19}_{-0.10}$ &   $0.15^{+0.17}_{-0.10}$ \\ 
$\log_{10}(M_{\rm BH}/M_\odot)$ &   $6.58^{+0.20}_{-0.12}$ &   $7.20^{+0.56}_{-0.76}$ &   $6.72^{+0.48}_{-0.42}$ &   $6.89^{+0.51}_{-0.57}$ &   $6.91^{+0.41}_{-0.41}$ &   $6.83^{+0.53}_{-0.49}$ &   $7.11^{+0.41}_{-0.72}$ &   $7.09^{+0.47}_{-0.39}$ &   $7.43^{+0.51}_{-0.31}$ &   $6.91^{+0.55}_{-0.86}$ \\ 
$f_{\rm ellip}$ &   $0.13^{+0.11}_{-0.09}$ &   $0.40^{+0.37}_{-0.23}$ &   $0.17^{+0.16}_{-0.11}$ &   $0.35^{+0.20}_{-0.19}$ &   $0.22^{+0.18}_{-0.15}$ &   $0.18^{+0.16}_{-0.13}$ &   $0.21^{+0.14}_{-0.16}$ &   $0.22^{+0.19}_{-0.14}$ &   $0.20^{+0.20}_{-0.12}$ &   $0.22^{+0.18}_{-0.14}$ \\ 
$f_{\rm flow}$ &   $0.27^{+0.15}_{-0.18}$ &   $0.31^{+0.25}_{-0.20}$ &   $0.26^{+0.17}_{-0.17}$ &   $0.28^{+0.18}_{-0.20}$ &   $0.27^{+0.16}_{-0.19}$ &   $0.27^{+0.15}_{-0.17}$ &   $0.24^{+0.18}_{-0.18}$ &   $0.26^{+0.18}_{-0.18}$ &   $0.25^{+0.18}_{-0.17}$ &   $0.27^{+0.17}_{-0.18}$ \\ 
$\theta_e$ (degrees) &   $12.7^{+11.3}_{- 8.6}$ &   $27.3^{+43.2}_{-18.9}$ &   $12.3^{+12.1}_{- 8.3}$ &   $17.9^{+27.8}_{-12.4}$ &   $12.0^{+14.7}_{- 8.0}$ &   $12.1^{+10.7}_{- 8.0}$ &   $12.6^{+11.5}_{- 8.9}$ &   $13.7^{+14.5}_{- 9.4}$ &   $11.4^{+14.3}_{- 7.5}$ &   $12.2^{+15.0}_{- 8.7}$ \\ 

 $T$ & 65 & 1 & 1 & 1 & 3 & 1 & 1 & 3 & 3 & 3\\
\end{tabular}
\caption{Table of inferred parameters for the tests we carried out on Arp 151. The inferred parameter is the median value of the posterior probability distribution and the uncertainties quoted are the 68\% confidence intervals. $[1]$ Parameter name; $[2]$ Default simulation by \citealt{pancoast18}: full light-curve and no prior; $[3]$ Single epoch (low-flux) with a Gaussian prior on $\tau_{\rm mean}$ centred at $\mu_{\tau} = 3.07$ days and $\sigma_{\tau} = 0.5 \mu_{\tau}$; $[4]$ Single epoch (mid-flux) with a Gaussian prior on $\tau_{\rm mean}$ centred at $\mu_{\tau} = 3.07$ days and $\sigma_{\tau} = 0.5 \mu_{\tau}$; $[5]$ Single epoch (high-flux) with a Gaussian prior on $\tau_{\rm mean}$ centred at $\mu_{\tau} = 3.07$ days and $\sigma_{\tau} = 0.5 \mu_{\tau}$; $[6]$ Single epoch (mid-flux) with a Gaussian prior on $\tau_{\rm median}$ centred at $\mu_{\tau} = 1.75$ days and $\sigma_{\tau} = 0.5\mu_{\tau}$; $[7]$ Single epoch (mid-flux) with a Gaussian prior on $\tau_{\rm mean}$ centred at $\mu_{\tau} = 3.07$ days and $\sigma_{\tau} = 1 \mu_{\tau}$; $[8]$ Single epoch (mid-flux) with a Gaussian prior on $\tau_{\rm mean}$ centred at $\mu_{\tau} = 3.07$ days and $\sigma_{\tau} = 3 \mu_{\tau}$; $[9]$ Single epoch (mid-flux) with $\mu_{\tau} = 5.21$ and $\sigma_{\tau} = 0.2$ dex from the R$-$L relation; $[10]$ Single epoch (mid-flux) with $\mu_{\tau} = 15.63$ (3$\times$ R-L predicted value) and $\sigma_{\tau} = 0.2$ dex; $[11]$ Single epoch (mid-flux) with $\mu_{\tau} = 15.63$ (3$\times$ R-L predicted value) and $\sigma_{\tau} = 1.2$ dex; $[12]$ Temperature ($T$) used for each test. More details can be found in Appendix~\ref{sec:appendix}. The parameters are all measured in the rest-frame. } 
\label{table_results}
\end{table}
\end{landscape}

\clearpage

\appendix

\section{Simulations}
\label{sec:appendix_sim}

In this section we describe the simulations we carried out to further validate the single-epoch model. We generate broad emission line profiles using our model, assuming that the BLR is characterised by a defined set of physical parameters. The black hole mass is one of the input parameters and we explore values of log($M_{\rm BH}/M_{\odot}$) = 7.0, 7.5 and 8.0. The continuum light curve is generated as described in Section~\ref{sec:modifications} and the prior probability distributions for the remaining parameters are also the same as those used for the modelling of the Arp 151 line profiles. Since the simulated continuum flux and line flux are rescaled, we do not have an absolute continuum luminosity to use in the R$-$L relation to infer a mean time delay. Instead we centre the mean time delay prior at values representative of those observed for black holes in the mass range we assume, using the AGN black hole mass database \citep{bentz&katz15}. These time delays are the same as those used in the original determination of the R$-$L relation \citep{bentz13}. This is a conservative approach that includes the situation where the assumed central value for the prior mean time delay may be offset from the intrinsic one, as explored for Arp 151. The model is then used to infer the BLR parameters using the simulated broad emission line profile with added noise as input. We vary the noise level for each simulation, to represent cases where the spectrum has a peak S/N $\sim$ 100 - 300, to explore not only the S/N level of Arp 151 but also to test the model's performance for higher quality data.

Due to the large number of free parameters in our model, it is not practical or feasible in a work such as this to simulate all possible combinations of parameters and their corresponding BLR structures. Instead we select combinations of parameters that are representative of the possible BLR geometries and dynamics and allow us to illustrate the performance of the model. Table~\ref{table_sim} summarises the input parameters for each simulation.
We test the model performance for the 9 main BLR parameters ($\beta$, $\theta_{o}$, $\theta_{i}$, $\kappa$, $\xi$, $\gamma$, $f_{\rm ellip}$, $f_{\rm flow}$, $\theta_{e}$) for five simulations. We find that 34 out of 45 parameters are recovered within the 68\% confidence range of the posterior probability distribution and 44 out of 44 within the 95\% confidence range. This is what is expected since the true parameter value should lie within the 68\% confidence range $\sim$68\% of the time and within the 95\% confidence region $\sim$95\% of the time. We describe the setup and results for each simulation below.

For each simulation we show two diagrams. The first shows the posterior probability distributions for the parameters. Overlaid on the posterior distributions are the parameter values used as input to the model (red vertical lines), and the median values of the posterior probability distributions (green vertical lines). The second diagram shows the 68\% confidence regions for the inferred parameters, with the red vertical line indicating the input value of each parameter. The line profile generated for each simulation can be found in Fig~\ref{line_profiles}.
\subsubsection*{Simulation 1}
The first simulation is of a relatively thin BLR seen at an inclination of 20 degrees, with half of the orbits in inflowing configurations, isotropic emission and a transparent mid-plane. The opening angle and inclination angle have the same value. The results are shown in Fig.~\ref{posterior_sim1} and Fig.~\ref{inferred_sim1}. The parameters $\theta_{o}$, $\theta_{i}$ and $\kappa$ are constrained and agree within the 68\% confidence range with the input parameter values. The parameter $\beta$ is only marginally constrained. Due to their broad posterior probability distributions we consider that the remaining parameters are not constrained. In this case the model is not able to identify the inflowing velocity orbits as $f_{\rm flow}$ is unconstrained.

\subsubsection*{Simulation 2}
The second simulation assumes the same angular thickness as the previous case, but an intermediate inclination with half of the orbits in outflowing configurations. The simulated BLR has preferential emission away from the ionising source with some obscuration in the mid-plane. The results are shown in Fig.~\ref{posterior_sim2} and Fig.~\ref{inferred_sim2}. The parameters $\theta_{i}$, $\theta_{o}$ and $\beta$ are constrained, while $\xi$ is only marginally constrained in this simulation.

This case illustrates three important features: 1) we use the median of the posterior probability distribution as the value of the inferred parameter. However, when the input value of the parameter is close to the limit of the parameter range, the median may not be a good indicator of the parameter value but indicate a lower or upper limit. This can be seen for the parameters $\kappa$ and $\theta_{e}$ which agree with the real value at the 95\% confidence level. What we observe is a common feature of the original model as well, and has to do with using the median of the posterior as the inferred value for the parameter. Such feature highlights the importance of analysing the shape of the parameter posterior distribution together with the median and 68\% confidence ranges; 2) Even though the model accurately identifies $f_{\rm flow}$, it is only able to derive an estimate for $f_{\rm ellip}$ that is within the 95\% confidence range of the input value. 3) The model is able to recover the real value of $\theta_{o}$ even when it is significantly lower than $\theta_{i}$.

\subsubsection*{Simulation 3}
The BLR simulated here is close to spherical seen at a 30$^{\circ}$ inclination. The mid-plane is opaque but the emission is isotropic and half of the orbits have velocities drawn from the distribution of outflowing velocities. The results are shown in Fig.~\ref{posterior_sim3} and Fig.~\ref{inferred_sim3}. The model is able to constrain all the parameters. The true values for $\beta$, $\theta_{o}$, $\theta_{i}$, $\kappa$, $\gamma$ and $f_{\rm flow}$ are recovered within the 68\% confidence range. The true values of the parameters $\xi$ and $\theta_{e}$ are recovered within the 95\% confidence range, possibly due to the true values being at the limit of the parameter range. The true value of the parameter $f_{\rm ellip}$ is recovered at the 99\% confidence range. This simulation illustrates the ability of the model to constrain the inclination and opening angles for the case when $\theta_{o} \gg \theta_{i}$.

\subsubsection*{Simulation 4}
In this case we simulate a spherical BLR with only a small percentage of inflowing orbits, with isotropic emission, a transparent mid-plane and a uniform angular distribution of particles. The results are shown in Fig.~\ref{posterior_sim4} and Fig.~\ref{inferred_sim4}. This case is an example when the model is not able to constrain most of the parameters. The only constrained parameters are $\beta$ and $\kappa$. This simulation illustrates a case where $f_{\rm flow}$ cannot be constrained due to the low fraction of particles in inflowing orbits (large $f_{\rm ellip}$).

\subsubsection*{Simulation 5}
The simulated BLR is a close to face on, thick BLR with a significant percentage of inflowing orbits. The results are shown in Fig.~\ref{posterior_sim5} and Fig.~\ref{inferred_sim5}. For this case all of the inferred parameters recover the true value within the 68\% confidence region. However, the posterior probability distributions for some of the parameters are broad. We find that $\theta_{o}$, $\theta_{i}$, $f_{\rm ellip}$ and $f_{\rm flow}$ are constrained by the model, the parameters $\beta$, $\kappa$, $\xi$ and $\theta_{e}$ are marginally constrained and $\gamma$ is unconstrained.

\begin{table}
\begin{tabular}{c | c | c | c | c | c}
Parameter & Sim 1 & Sim 2 & Sim 3 & Sim 4 & Sim 5 \\
 \hline
$\beta$ &  1.0 & 1.5  & 1.0 & 1.0  & 0.8 \\ 
$\theta_o$ (degrees) & 20 & 20 & 80 & 80 &30\\ 
$\theta_i$ (degrees) &  20 &  40 & 30 & 30 &15\\ 
$\kappa$ &  0 &  0.5 & 0 & 0 & 0.2\\ 
$\gamma$ &  1 &  4 & 4 & 1 & 2\\ 
$\xi$ & 1 & 0.5 & 0 & 1 & 0.2\\ 
$f_{\rm ellip}$ & 0.5 & 0.5 & 0.5 & 0.9 & 0.2\\ 
$f_{\rm flow}$ & 0.2  & 0.8 & 0.8 & 0.2 & 0.25\\ 
$\theta_e$ (degrees) & 30 & 0 & 0 & 30 & 20\\ 
$\log_{10}(M_{\rm BH}/M_\odot)$ & 8 &  7 & 7 & 7 & 7.5\\ 
$\tau_{\rm mean}$ (days) & 27 & 7 & 7 & 3.5 & 3.5 \\
\hline
\end{tabular}
\caption{Table of input parameters for the simulations.} 
\label{table_sim}
\end{table}

\subsubsection*{Overall findings}
The simulations show that it is important to analyse the shape of the posterior probability distributions, as the median value may not always be a good measurement of the true parameter value. This is also valid for the original model and when modelling monitoring data as the median of the posterior probability distribution is used to infer the parameter value (e.g. \citealt{pancoast14b}). We find that in general the angles tend to be constrained, as well as the dominant outflow or inflow configuration for the orbits. The parameter $f_{\rm ellip}$ tends to have broad posterior probability distributions, which indicates that the model has difficulty in constraining the exact fraction of particles in inflowing/outflowing orbits. Better results on $f_{\rm ellip}$ tend to be obtained when $f_{\rm ellip} \lesssim 0.5$, which corresponds to more than 50\% of the orbits with velocities drawn from the inflowing or outflowing velocity distribution. 
When $f_{\rm ellip}$ is high, meaning that the fraction of particles in inflowing or outflowing orbits is small, the model has difficulty constraining the parameters that are related to $f_{\rm ellip}$, i.e. $f_{\rm flow}$ and $\theta_{e}$, likely due to a small contribution of this type of orbits to the line profile shape.
As for Arp 151, $\gamma$ is not constrained for most of the simulations thereby highlighting a typical limitation of this model. Interestingly, the parameter $\beta$ is constrained or marginally constrained for all our simulations. This is in line with the findings for Arp 151, where we showed that although $\beta$ may be constrained from a single-epoch spectrum, its value may change with the epoch and not be a good indication of the average value of $\beta$ for a full monitoring campaign.

\begin{figure}
\centering
\includegraphics[width=0.3\textwidth]{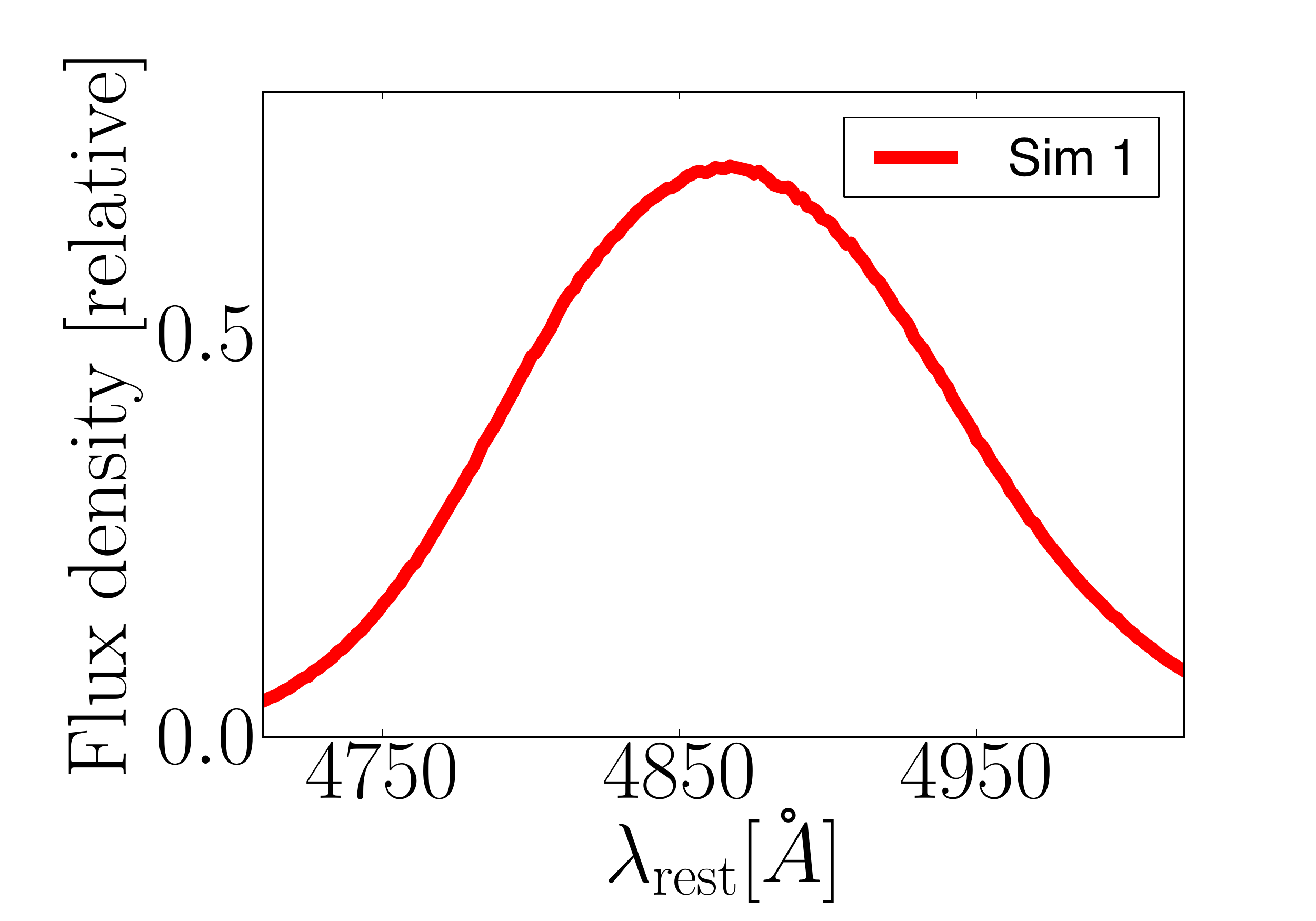}\\[0.1cm]
\includegraphics[width=0.3\textwidth]{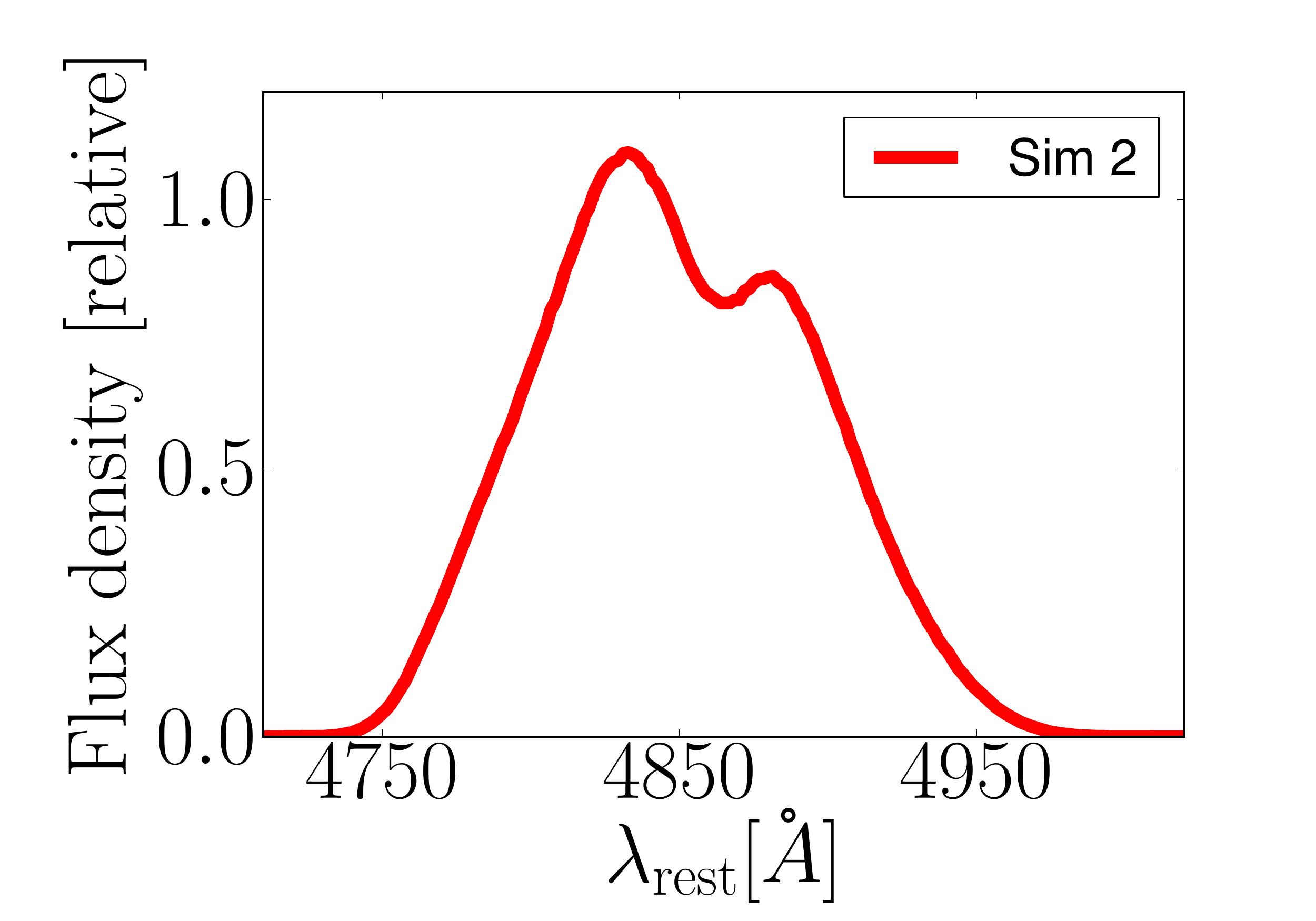}\\[0.1cm]
\includegraphics[width=0.3\textwidth]{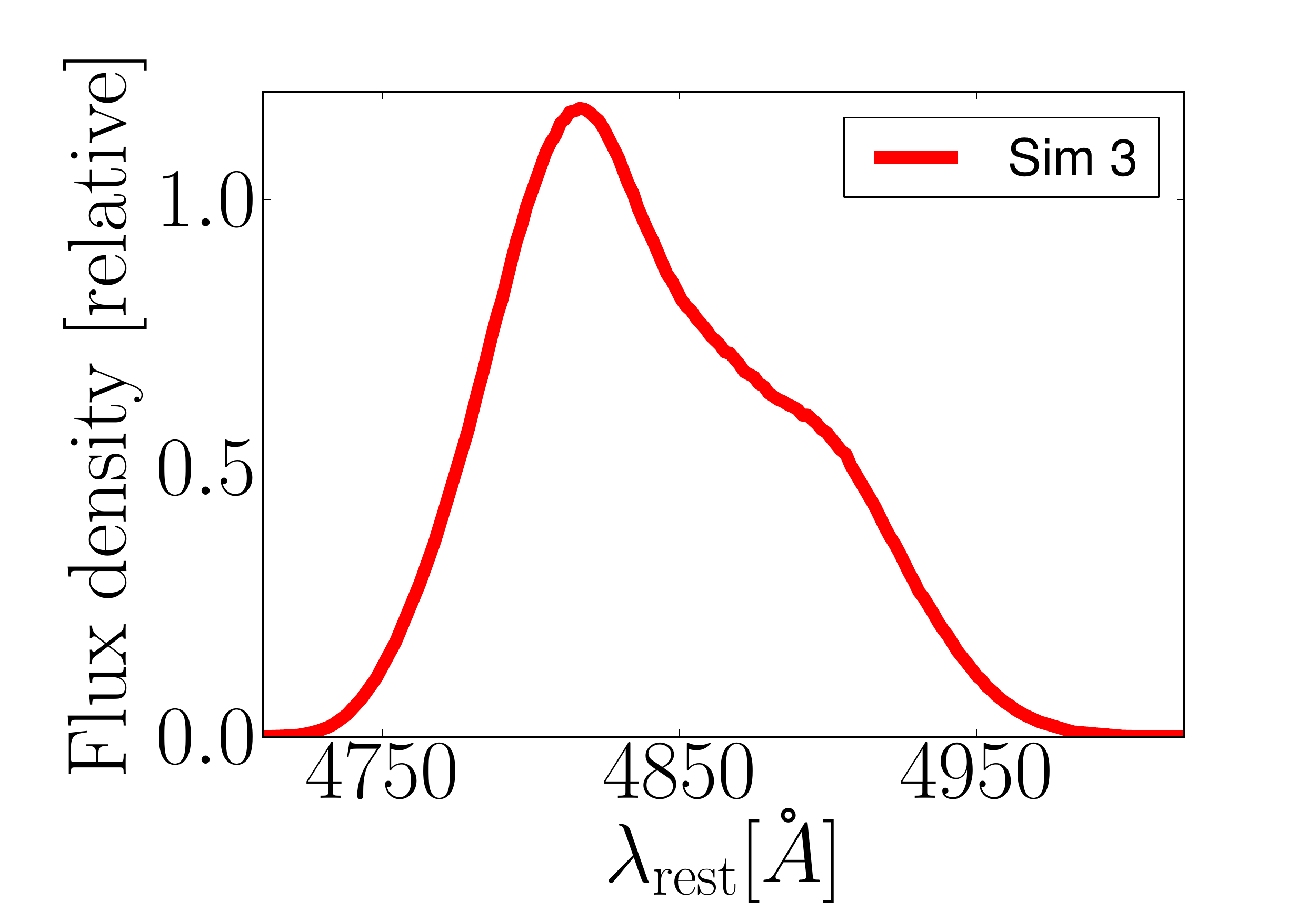}\\[0.1cm]
\includegraphics[width=0.3\textwidth]{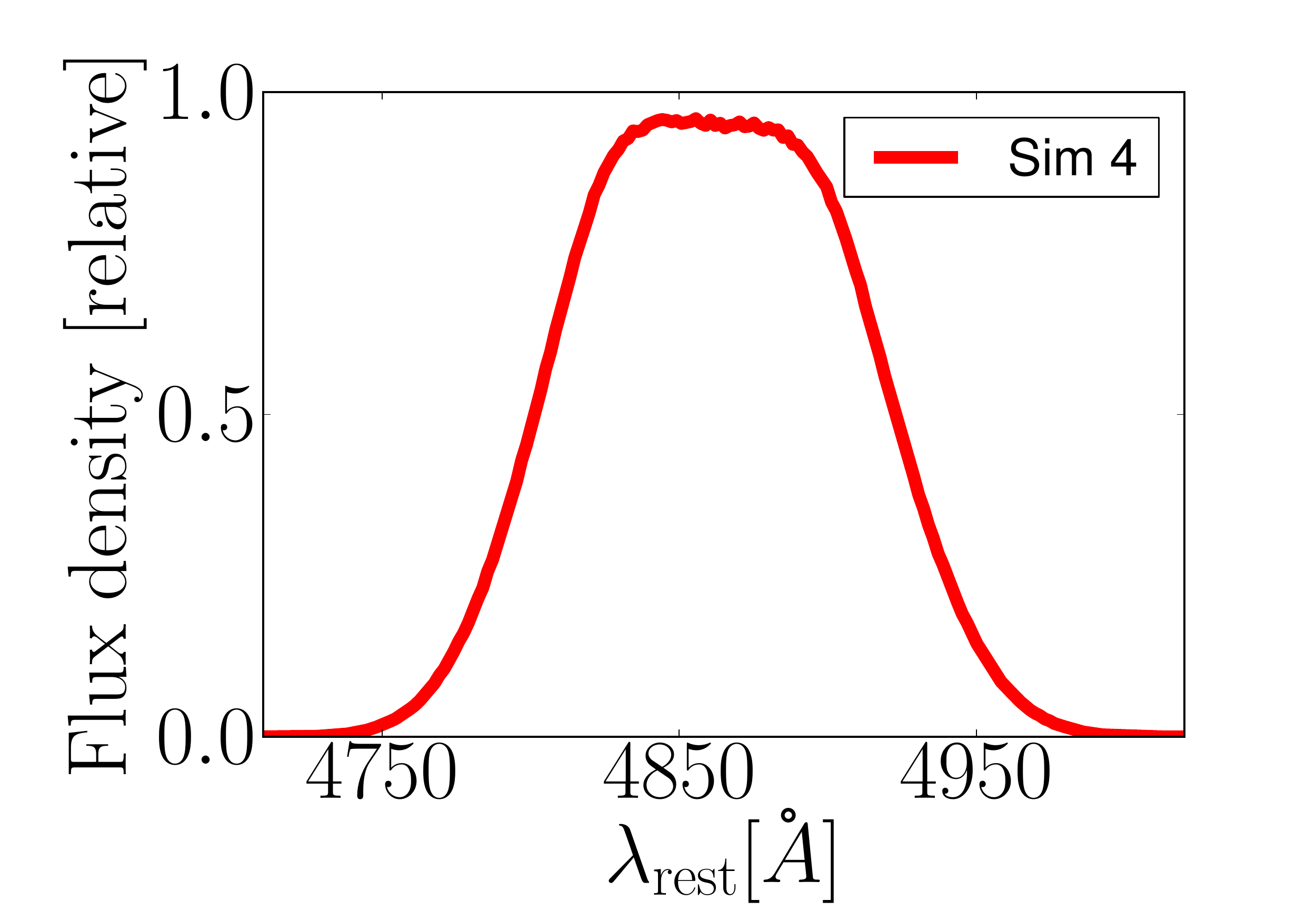}\\[0.1cm]
\includegraphics[width=0.3\textwidth]{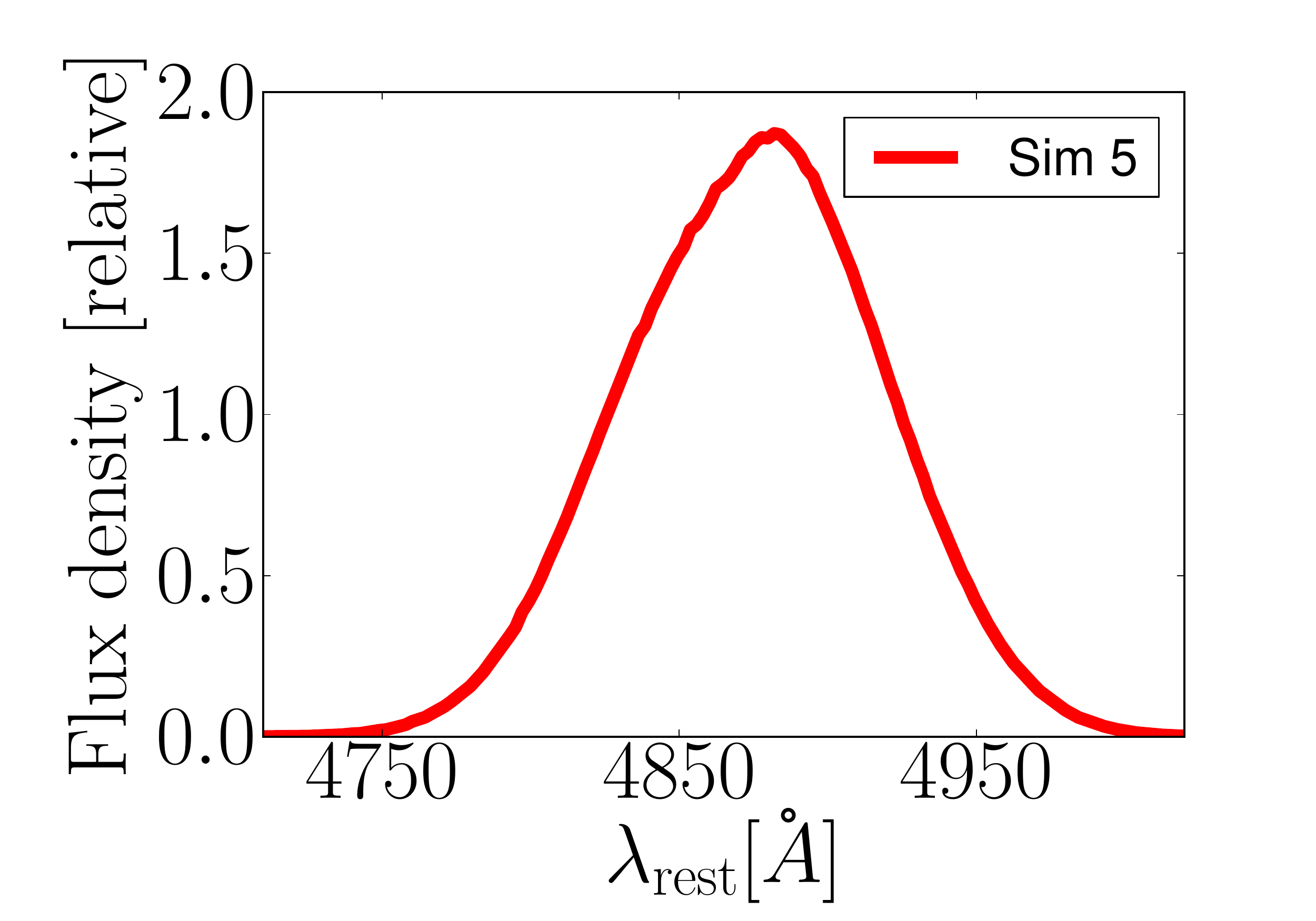}\\[0.1cm]
\caption{Line profiles generated for each of the simulations.}
\label{line_profiles}
\end{figure}

\begin{figure*}
\centering
\includegraphics[width=0.9\textwidth]{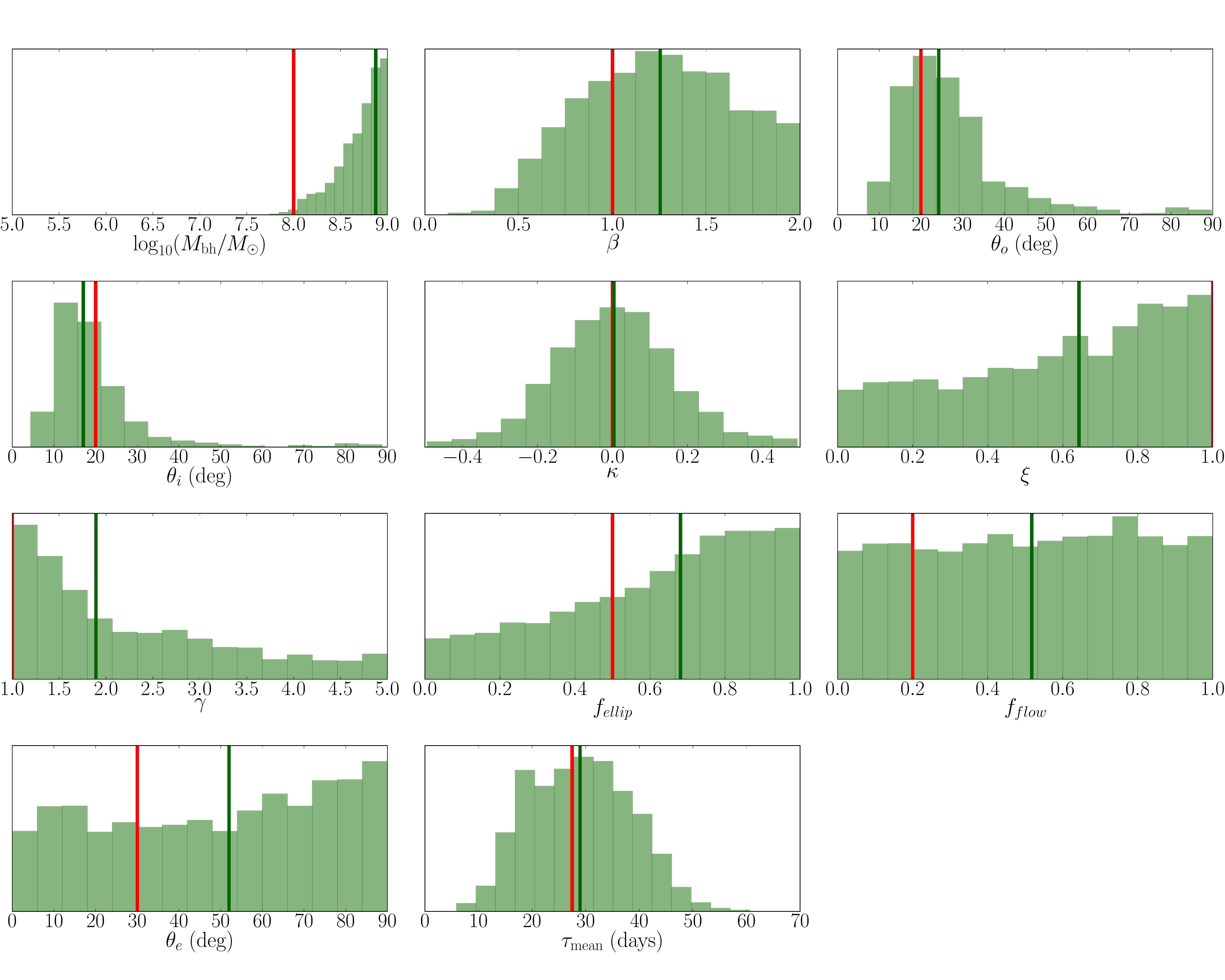}
\caption{Histograms showing the posterior probability distributions for the parameters in single-epoch BLR modelling simulations. The red vertical lines show the input parameter values used in the simulations and the green vertical line shows the median value of the posterior probability distribution. The results shown are for Simulation 1.}
\label{posterior_sim1}
\end{figure*}

\begin{figure*}
\centering
\includegraphics[width=0.8\textwidth]{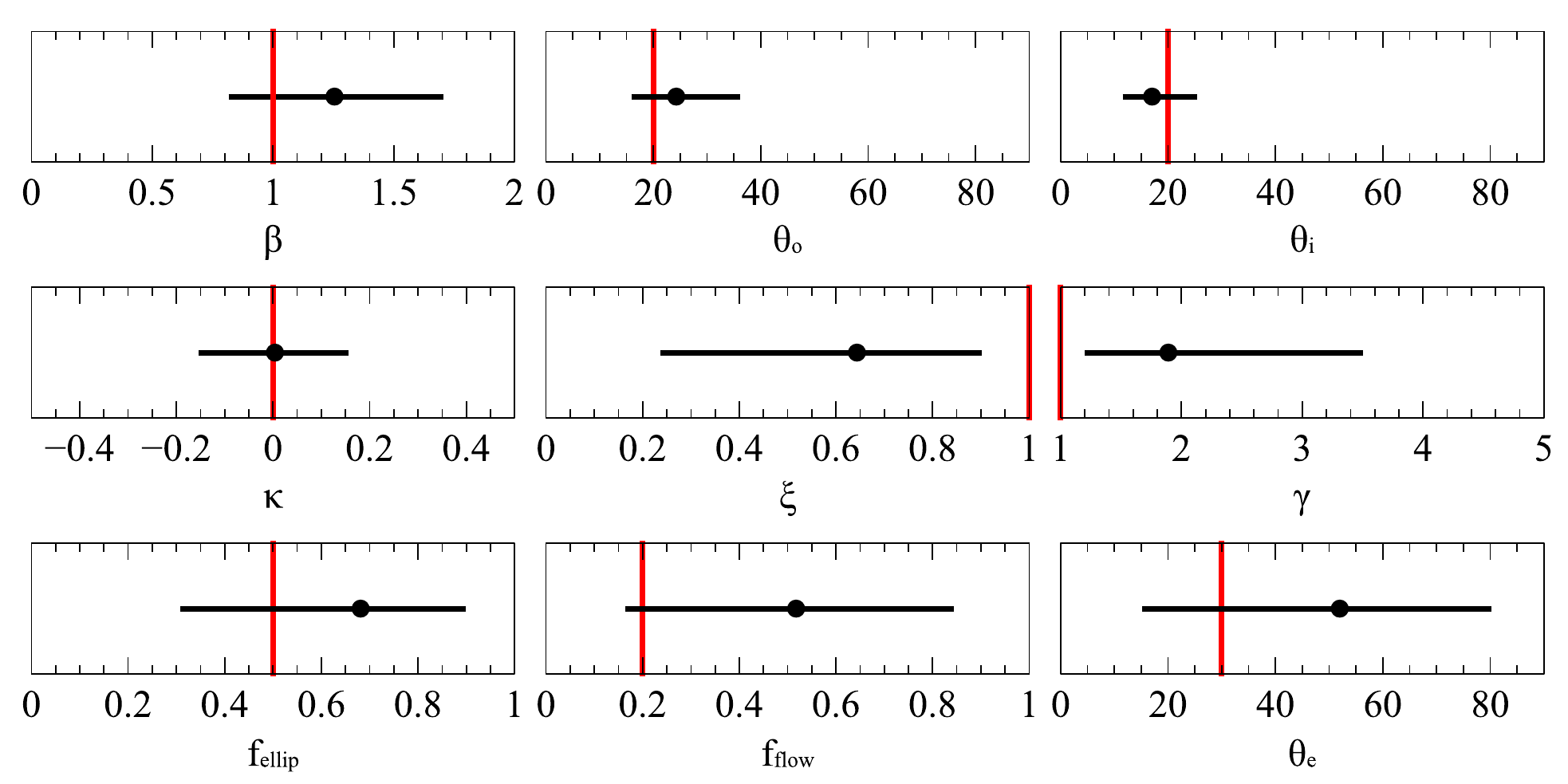}
\caption{Visual comparison of the inferred parameters with the input model parameters. The filled circles are the inferred values for the parameters (i.e. median value of the posterior probability distribution), the black horizontal line is the 68\% confidence region. The green vertical lines indicate the input value of the parameter in the simulation. The results shown are for Simulation 1.}
\label{inferred_sim1}
\end{figure*}

\begin{figure*}
\centering
\includegraphics[width=0.9\textwidth]{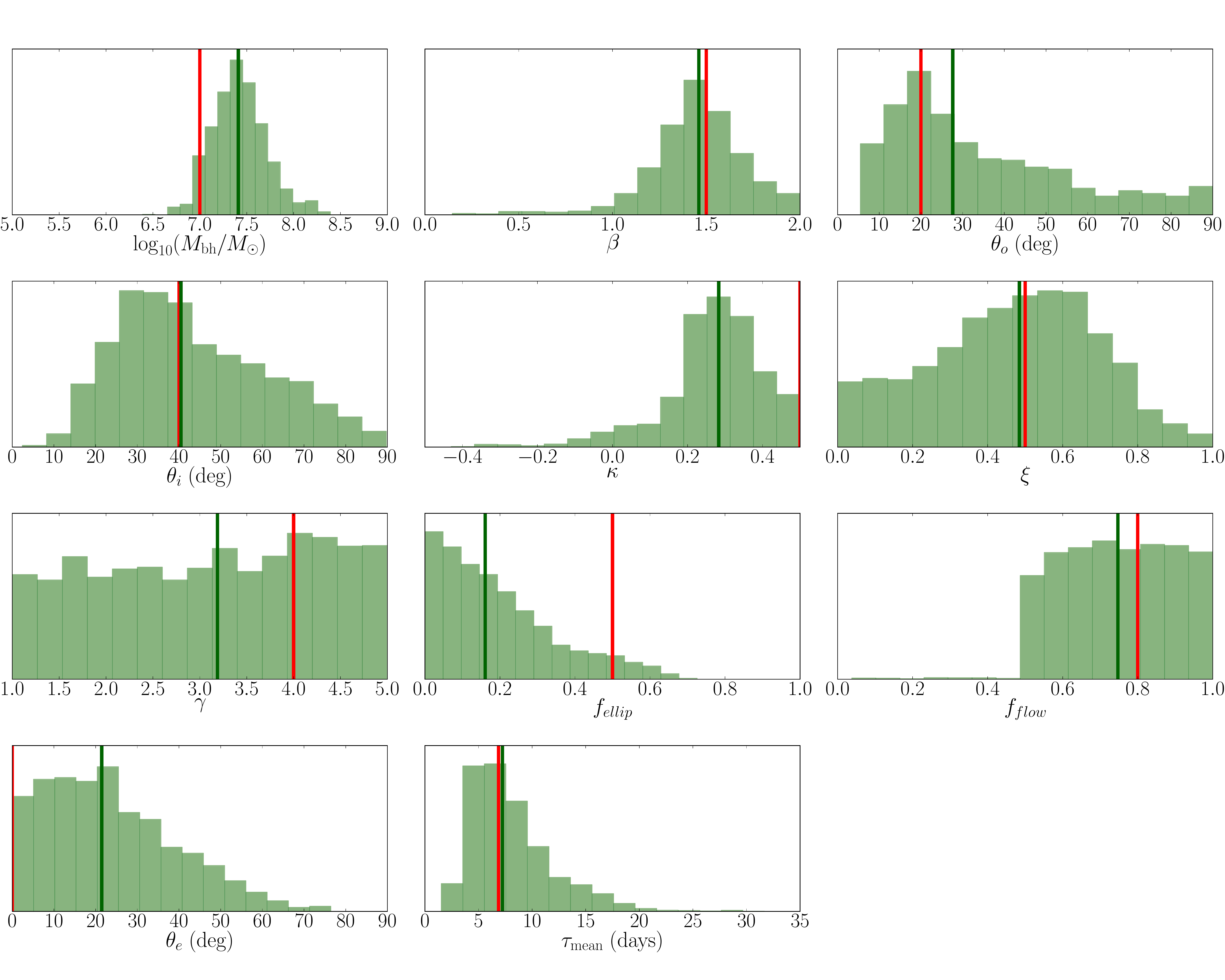}
\caption{Same as Fig.~\ref{posterior_sim1} but for different input parameters - see text. The results shown are for Simulation 2.}
\label{posterior_sim2}
\end{figure*}

\begin{figure*}
\centering
\includegraphics[width=0.8\textwidth]{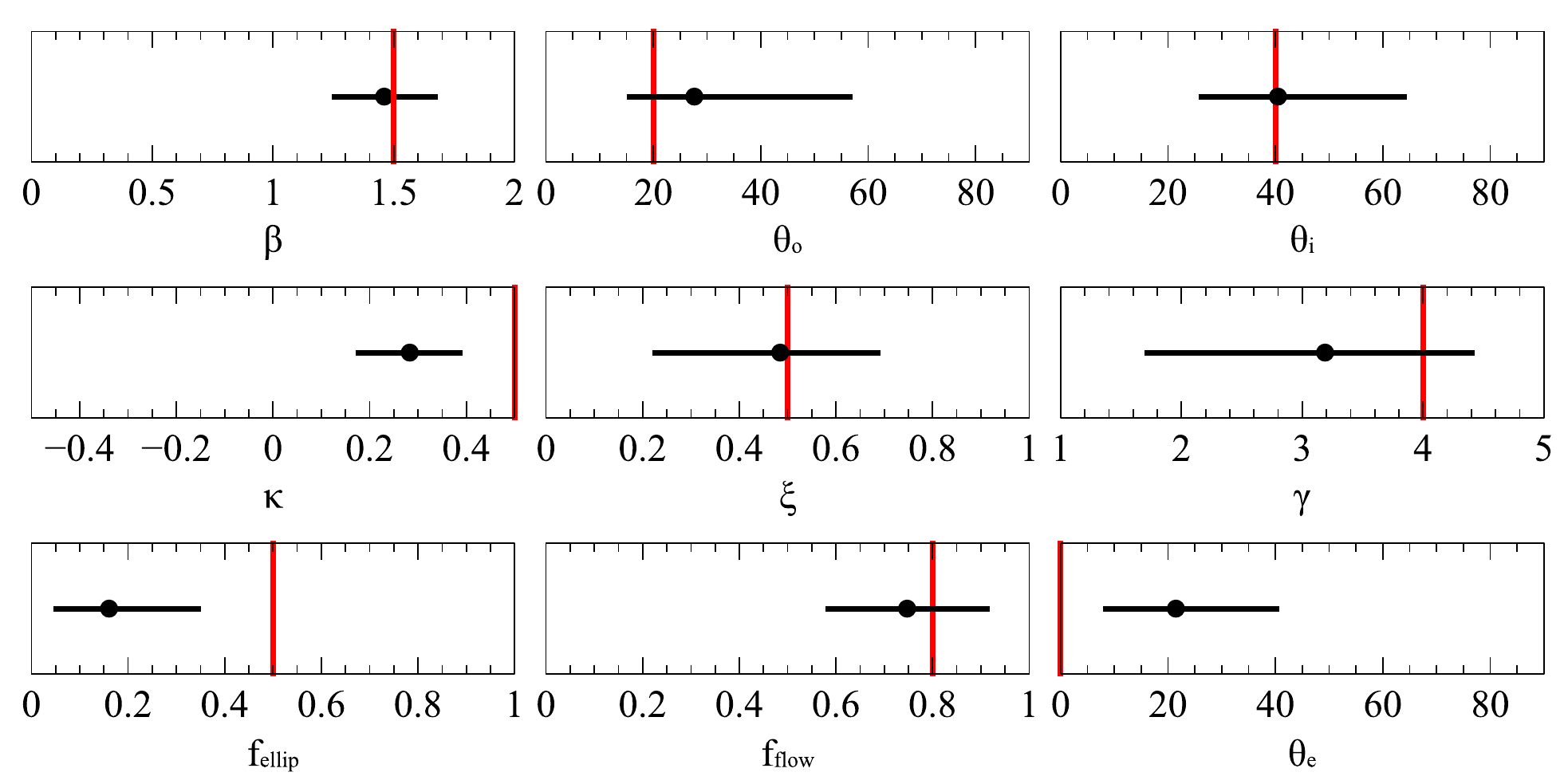}
\caption{Same as Fig.~\ref{inferred_sim1} but for the posterior probability distribution in Fig.~\ref{posterior_sim2}. The results shown are for Simulation 2.}
\label{inferred_sim2}
\end{figure*}

\begin{figure*}
\centering
\includegraphics[width=0.9\textwidth]{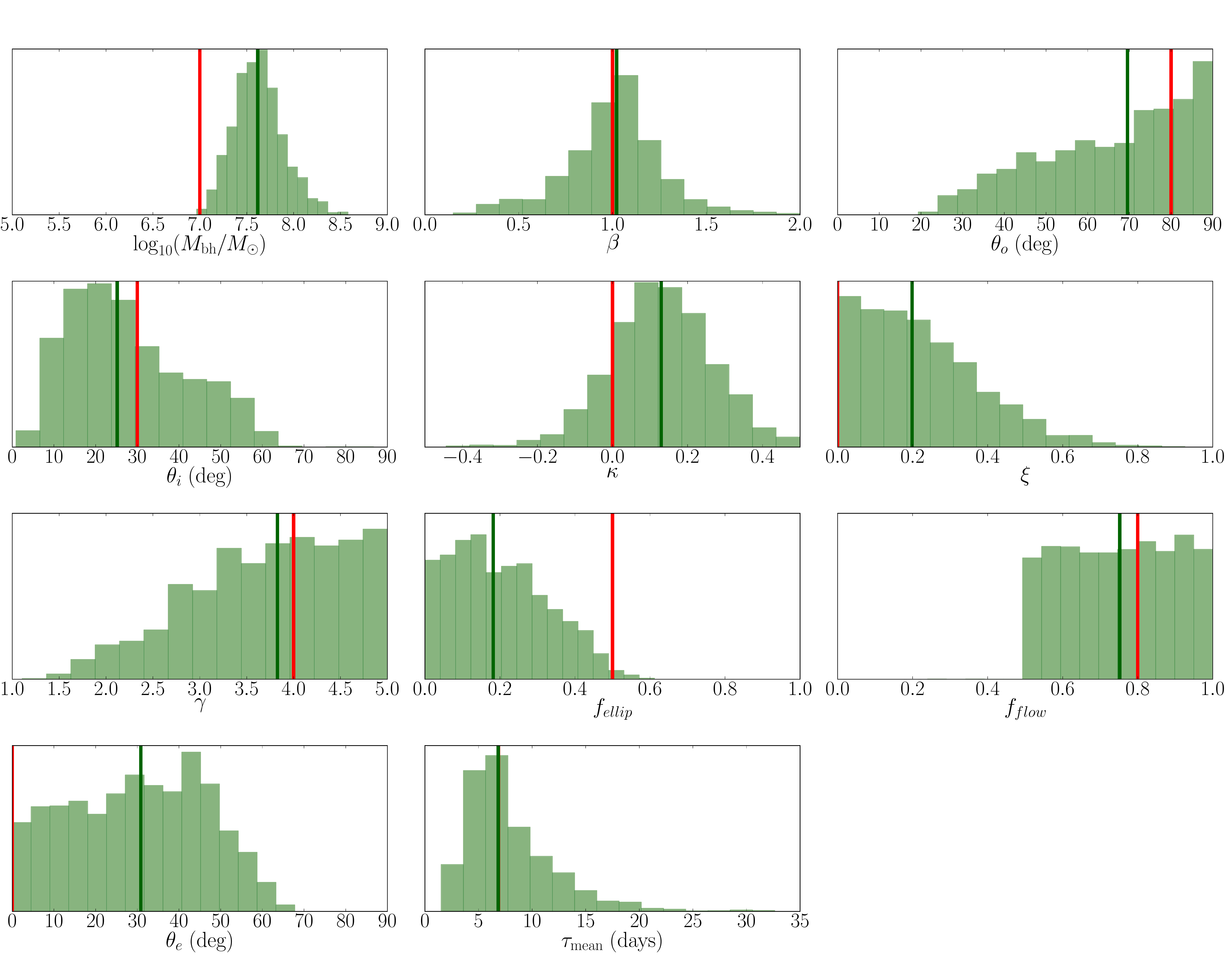}
\caption{Same as Fig.~\ref{posterior_sim1} but for different input parameters - see text. The results shown are for Simulation 3.}
\label{posterior_sim3}
\end{figure*}

\begin{figure*}
\centering
\includegraphics[width=0.8\textwidth]{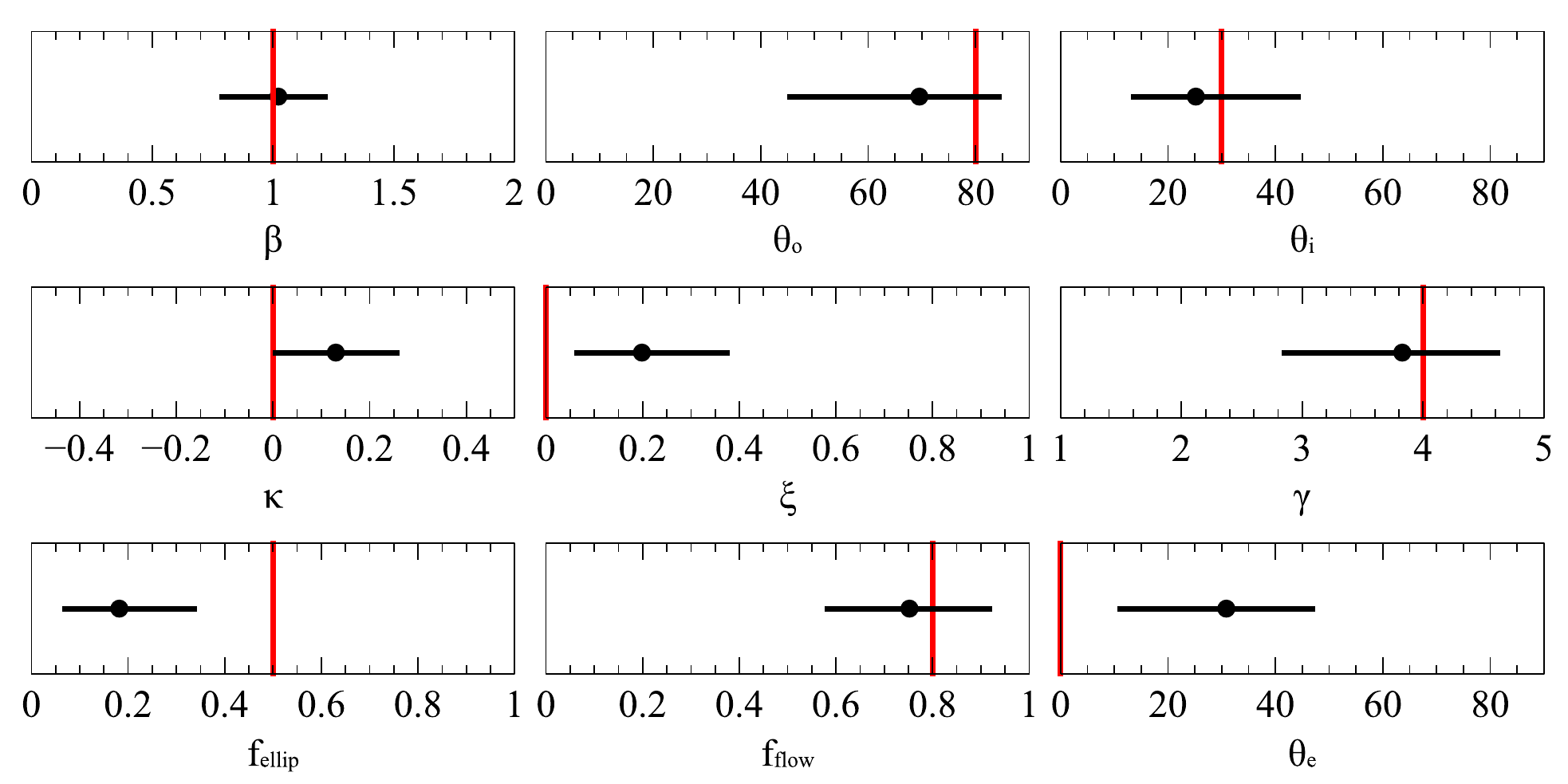}
\caption{Same as Fig.~\ref{inferred_sim1} but for the posterior probability distribution in Fig.~\ref{posterior_sim3}. The results shown are for Simulation 3.}
\label{inferred_sim3}
\end{figure*}

\begin{figure*}
\centering
\includegraphics[width=0.9\textwidth]{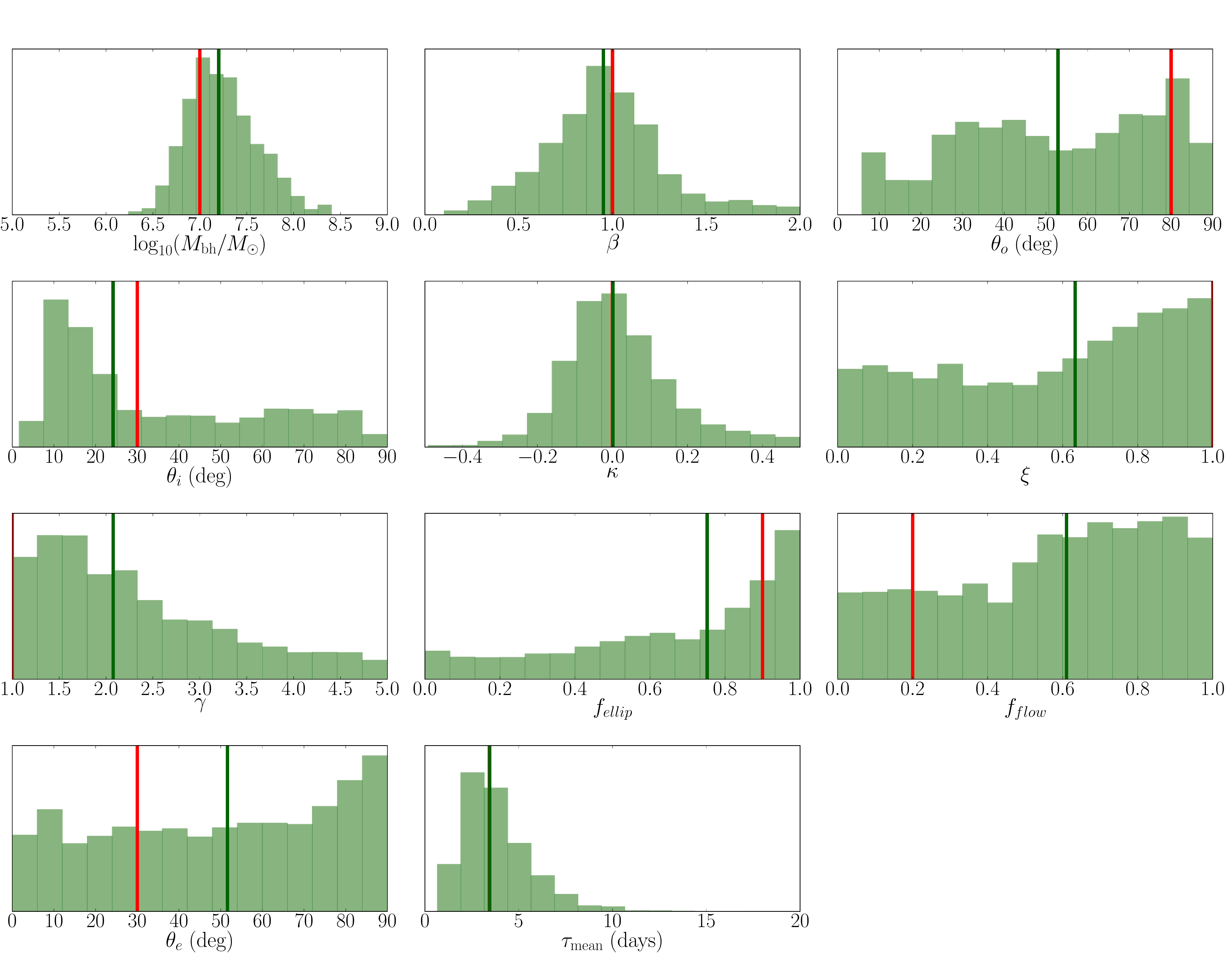}
\caption{Same as Fig.~\ref{posterior_sim1} but for different input parameters - see text. The results shown are for Simulation 4.}
\label{posterior_sim4}
\end{figure*}

\begin{figure*}
\centering
\includegraphics[width=0.8\textwidth]{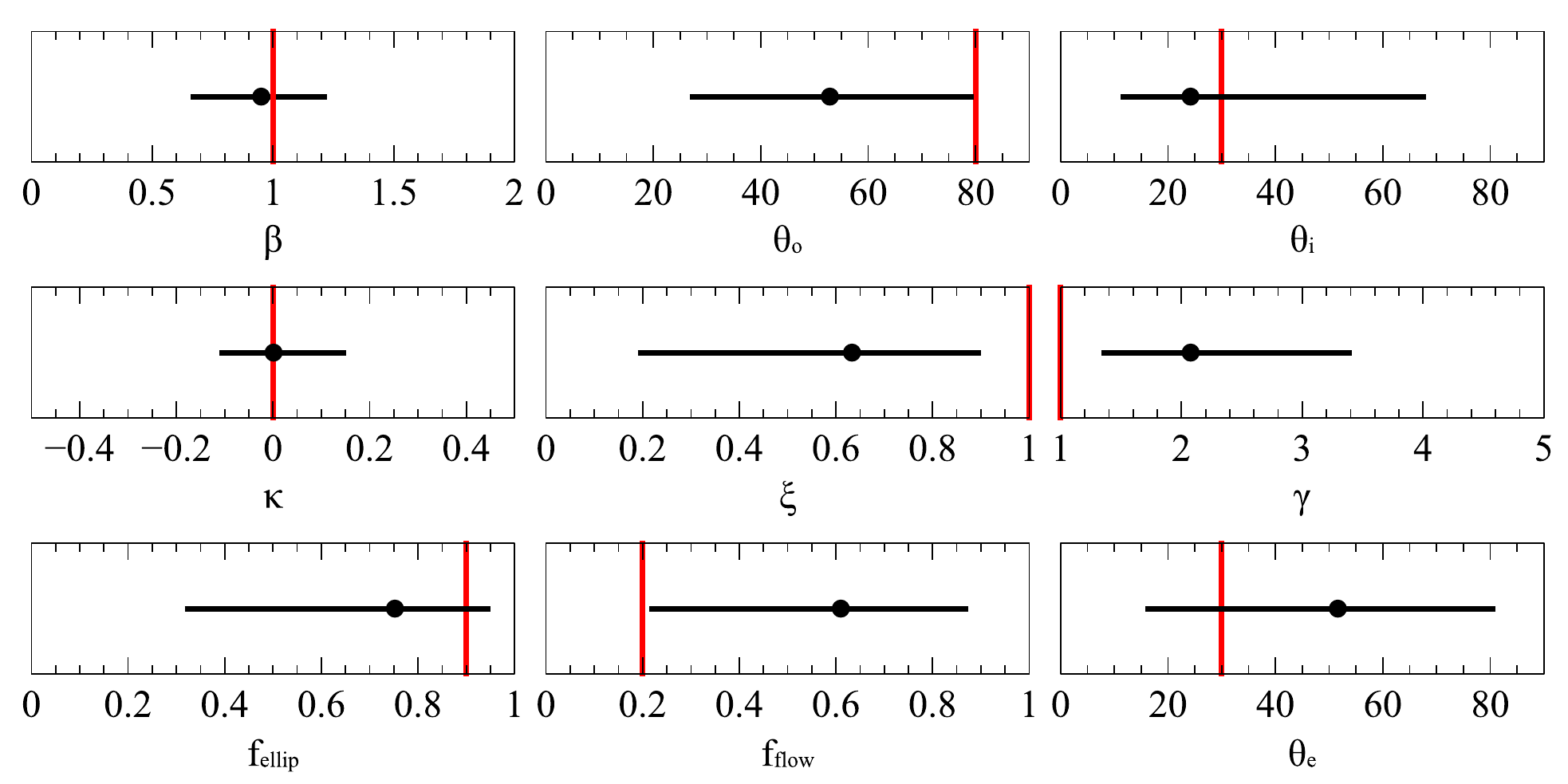}
\caption{Same as Fig.~\ref{inferred_sim1} but for the posterior probability distribution in Fig.~\ref{posterior_sim4}. The results shown are for Simulation 4.}
\label{inferred_sim4}
\end{figure*}

\begin{figure*}
\centering
\includegraphics[width=0.9\textwidth]{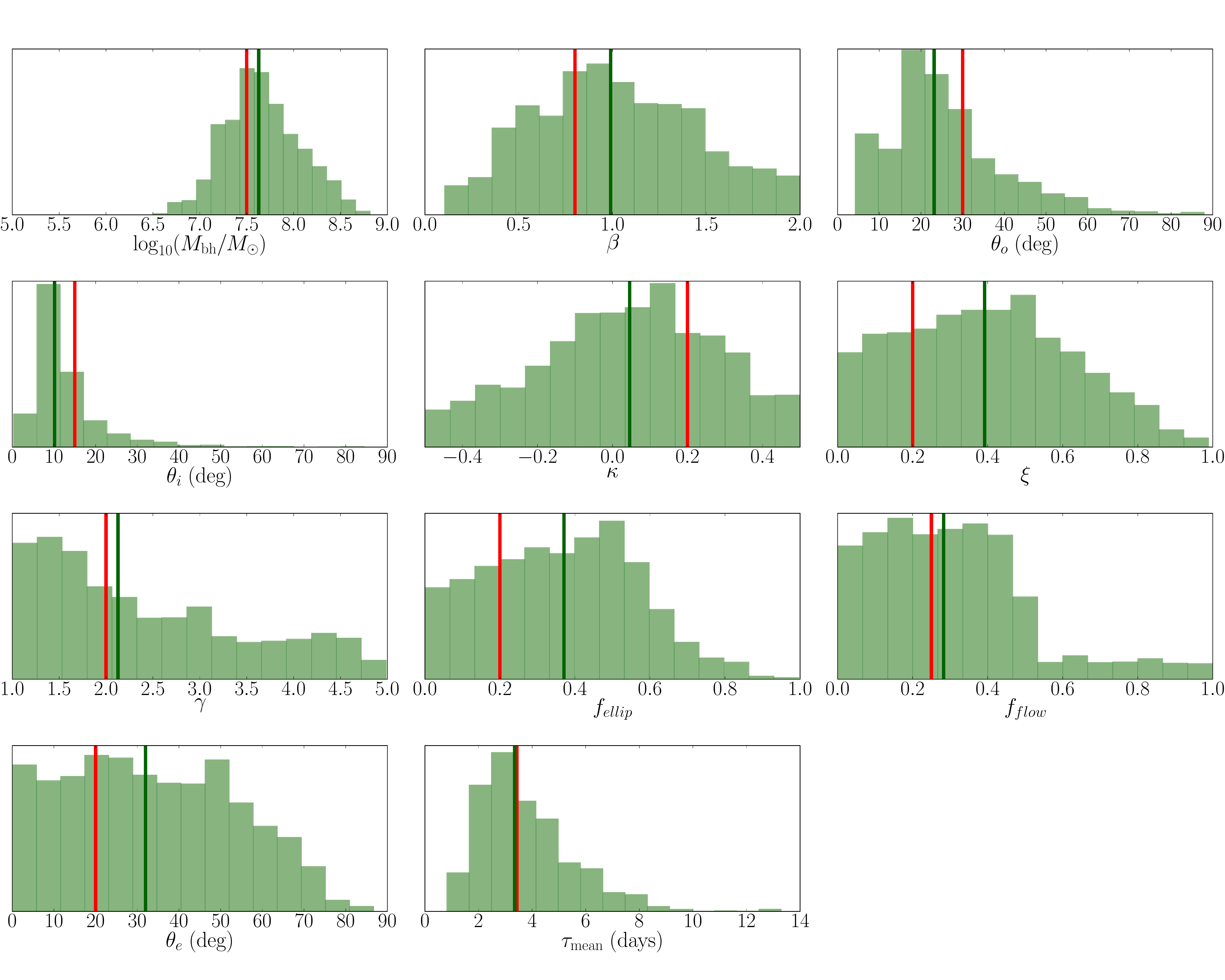}
\caption{Same as Fig.~\ref{posterior_sim1} but for different input parameters - see text. The results shown are for Simulation 5.}
\label{posterior_sim5}
\end{figure*}

\begin{figure*}
\centering
\includegraphics[width=0.8\textwidth]{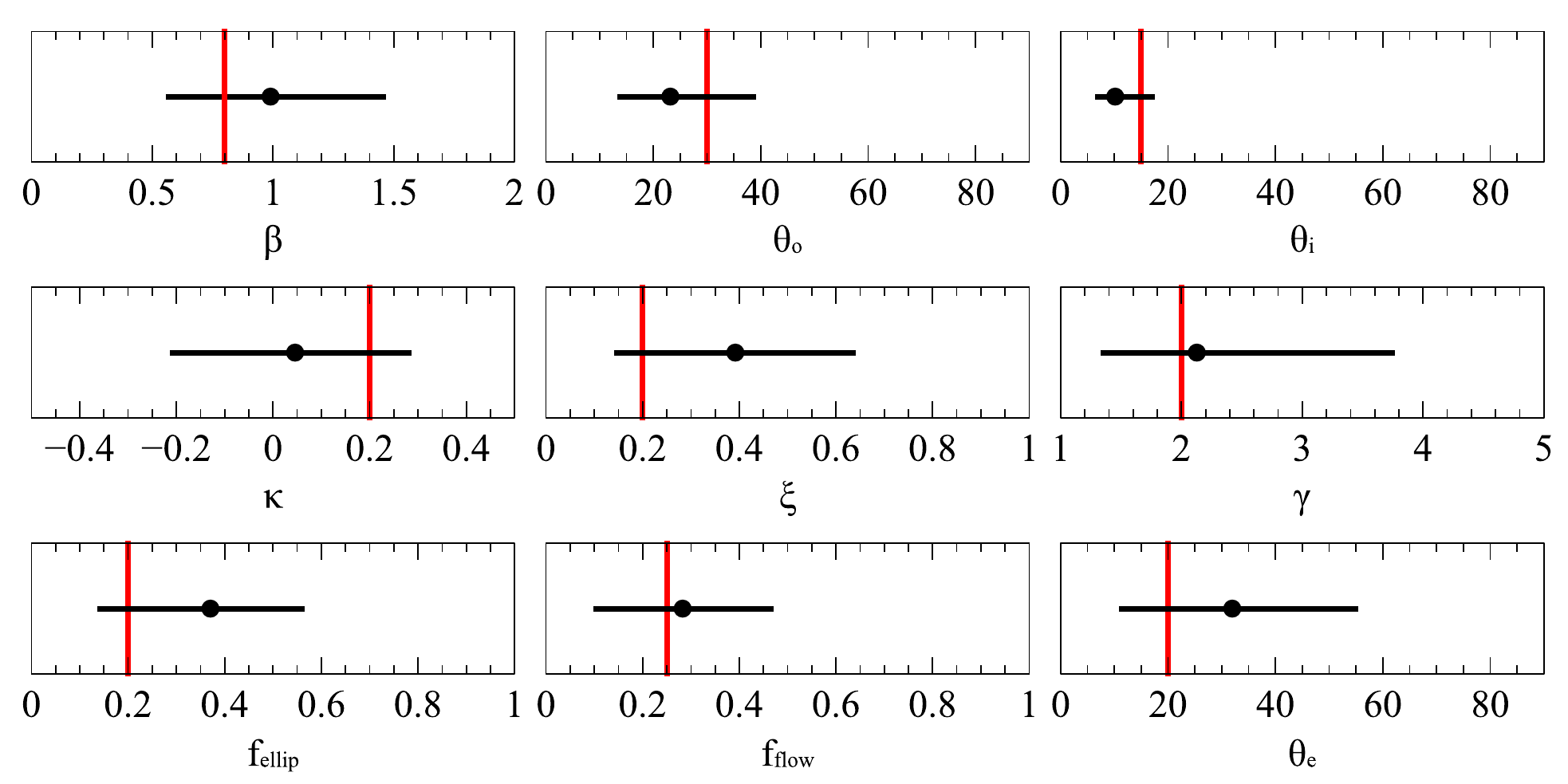}
\caption{Same as Fig.~\ref{inferred_sim1} but for the posterior probability distribution in Fig.~\ref{posterior_sim5}. The results shown are for Simulation 5.}
\label{inferred_sim5}
\end{figure*}

\clearpage
\onecolumn
\section{Additional figures}
\label{sec:appendix_extra_fig}
\begin{figure*}
\centering
\vspace{-20cm}
\includegraphics[width=0.9\textwidth]{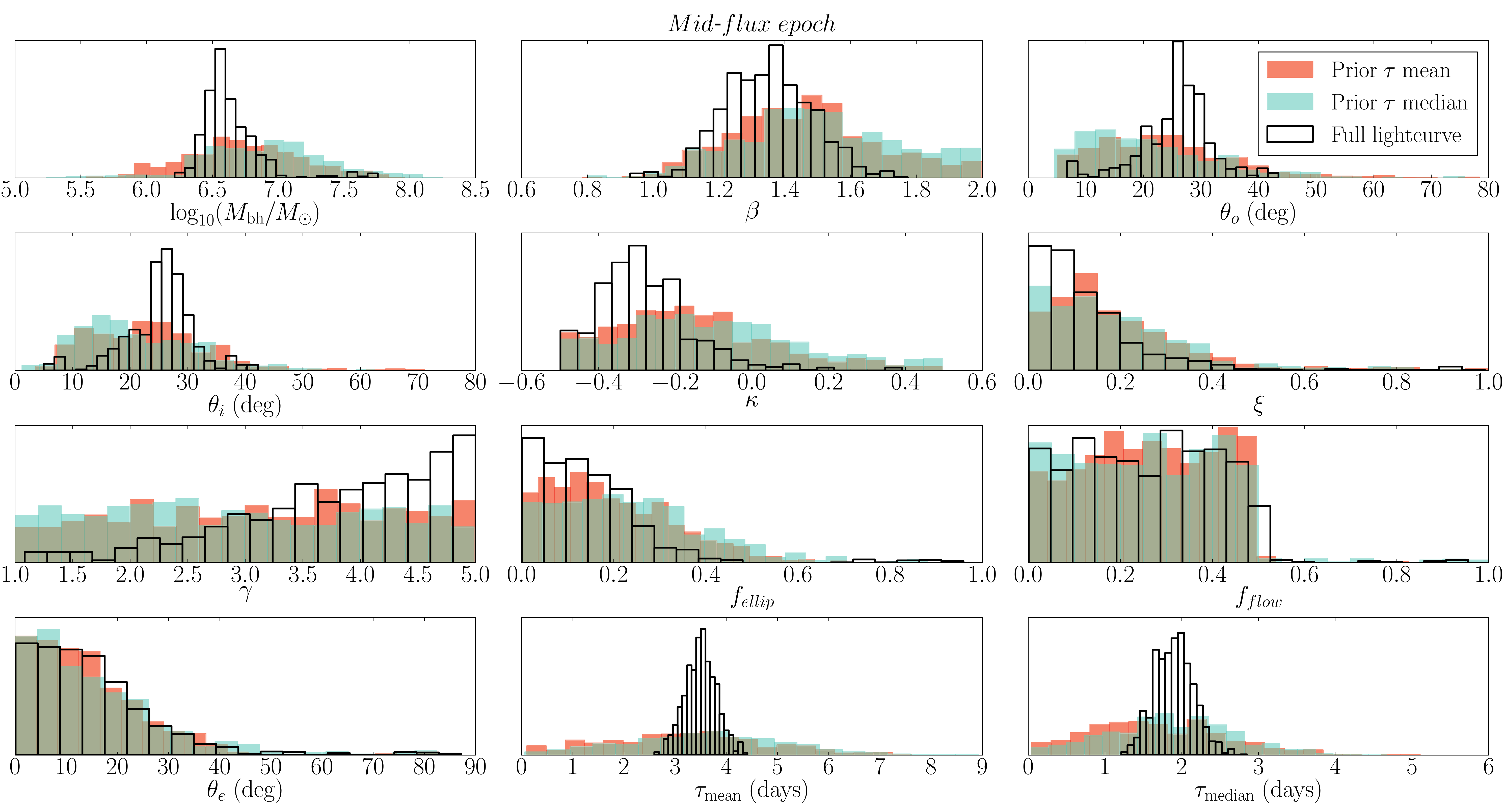}
\caption{Posterior probability distributions for the BLR geometry and dynamics parameters determined using the mid-flux epoch. Default modelling of the full light-curve without a Gaussian prior (\citealt{pancoast18}) is shown as the black solid line histogram. The result on the mid-flux epoch using a prior on $\tau_{\rm mean}$ is shown as a red filled histogram (same as red filled histogram of Fig.~\ref{3epochs_posterior}). Mid-flux epoch with a prior on $\tau_{\rm median}$ is shown as the filled cyan histogram.}
\label{tau_median_mean}
\end{figure*}

\begin{figure*}
\centering
\includegraphics[width=0.9\textwidth]{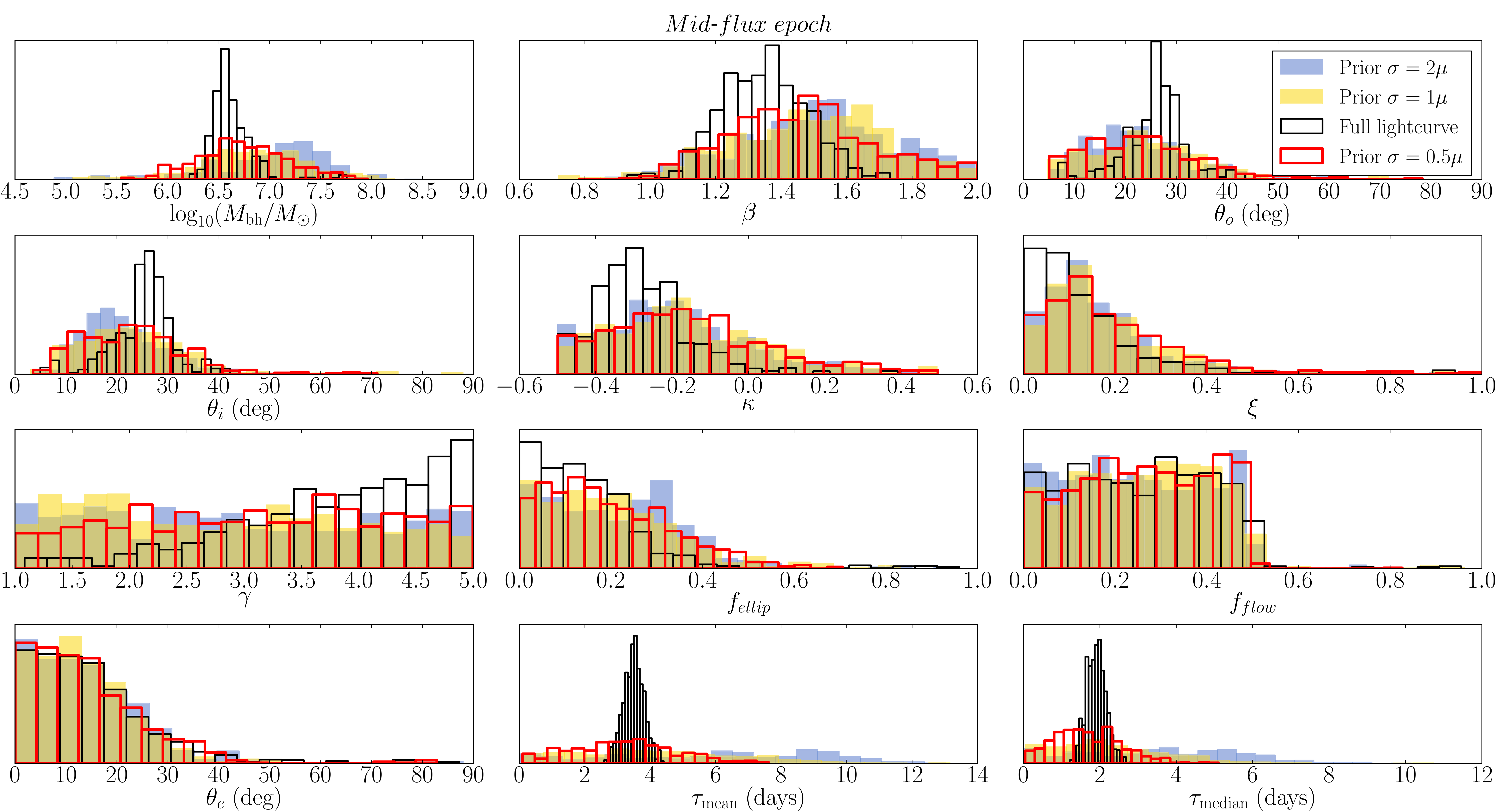}
\caption{Posterior probability distributions for the BLR geometry and dynamics parameters determined using the mid-flux epoch. Default modelling of the full light-curve without a Gaussian prior (\citealt{pancoast18}) is shown as the black solid line histogram. The result on the mid-flux epoch using a prior on $\tau_{\rm mean}$ and $\sigma_{\tau} = 0.5 \mu_{\tau}$ is shown as a red solid line histogram (same as red filled histogram of Fig.~\ref{3epochs_posterior}). The tests using $\sigma_{\tau} = 1 \mu_{\tau}$ and $\sigma_{\tau} = 2 \mu_{\tau}$ are shown as the filled yellow and blue histograms respectively.}
\label{tau_diff_sigma_posterior}
\end{figure*}

\begin{figure*}
\centering
\includegraphics[width=0.9\textwidth]{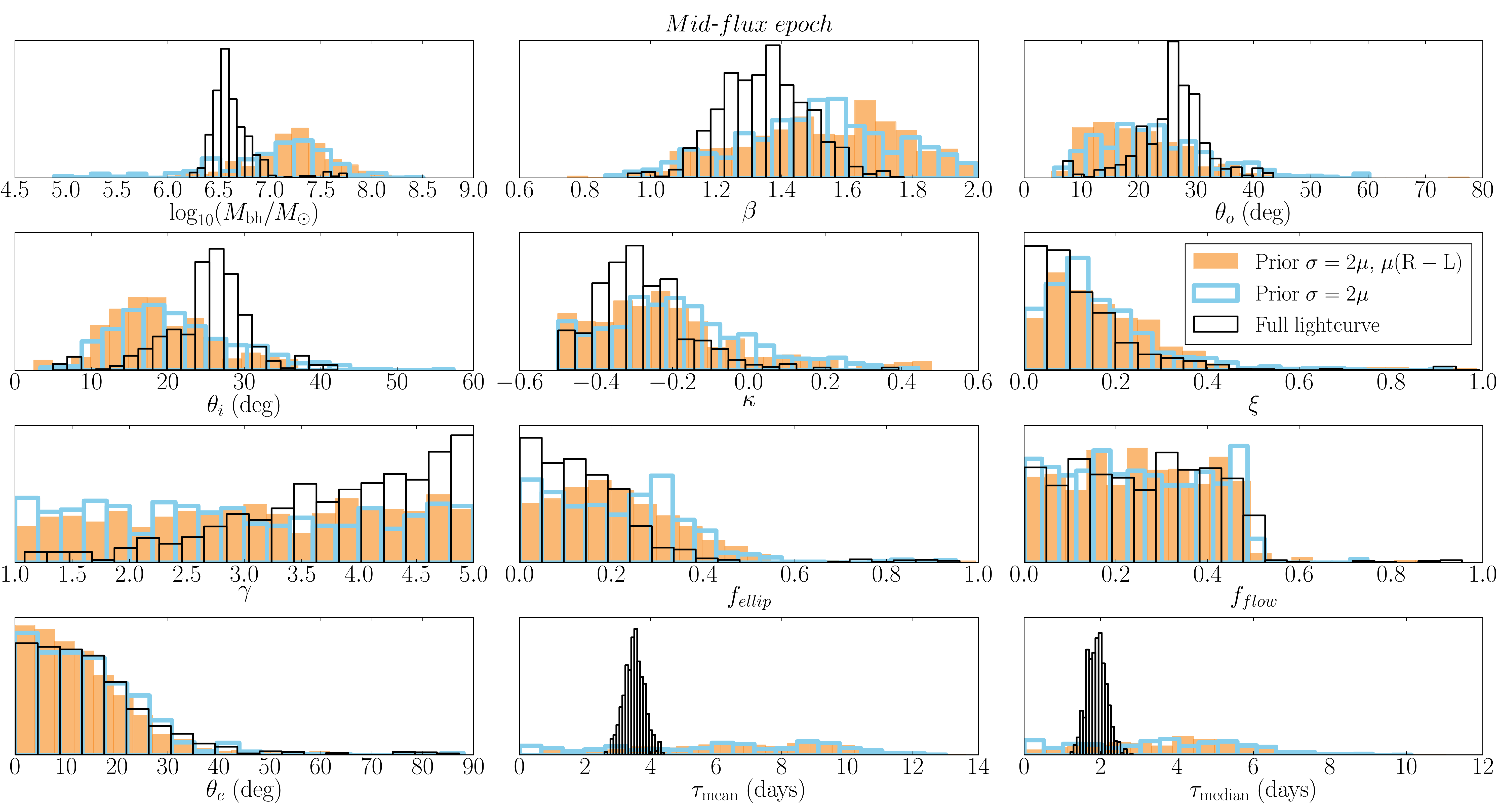}
\caption{Posterior probability distributions for the BLR geometry and dynamics parameters determined using the mid-flux epoch. Default modelling of the full light-curve without a Gaussian prior (\citealt{pancoast18}) is shown as the black solid line histogram. The result on the mid-flux epoch using a prior on $\tau_{\rm mean}$ with $\mu_{\tau} =  3.07$ days and $\sigma_{\tau} = 2 \mu_{\tau}$ is shown as a blue solid line histogram (same as blue filled histogram of Fig.~\ref{tau_diff_sigma_posterior}). The test using $\mu_{\tau} =  5.21$ days and $\sigma_{\tau} = 2 \mu_{\tau}$ is shown as the filled orange histogram.}
\label{tau_diff_sigma_mu_posterior}
\end{figure*}

\begin{landscape}
\centering
\begin{figure}
\includegraphics[width=1.4\textwidth]{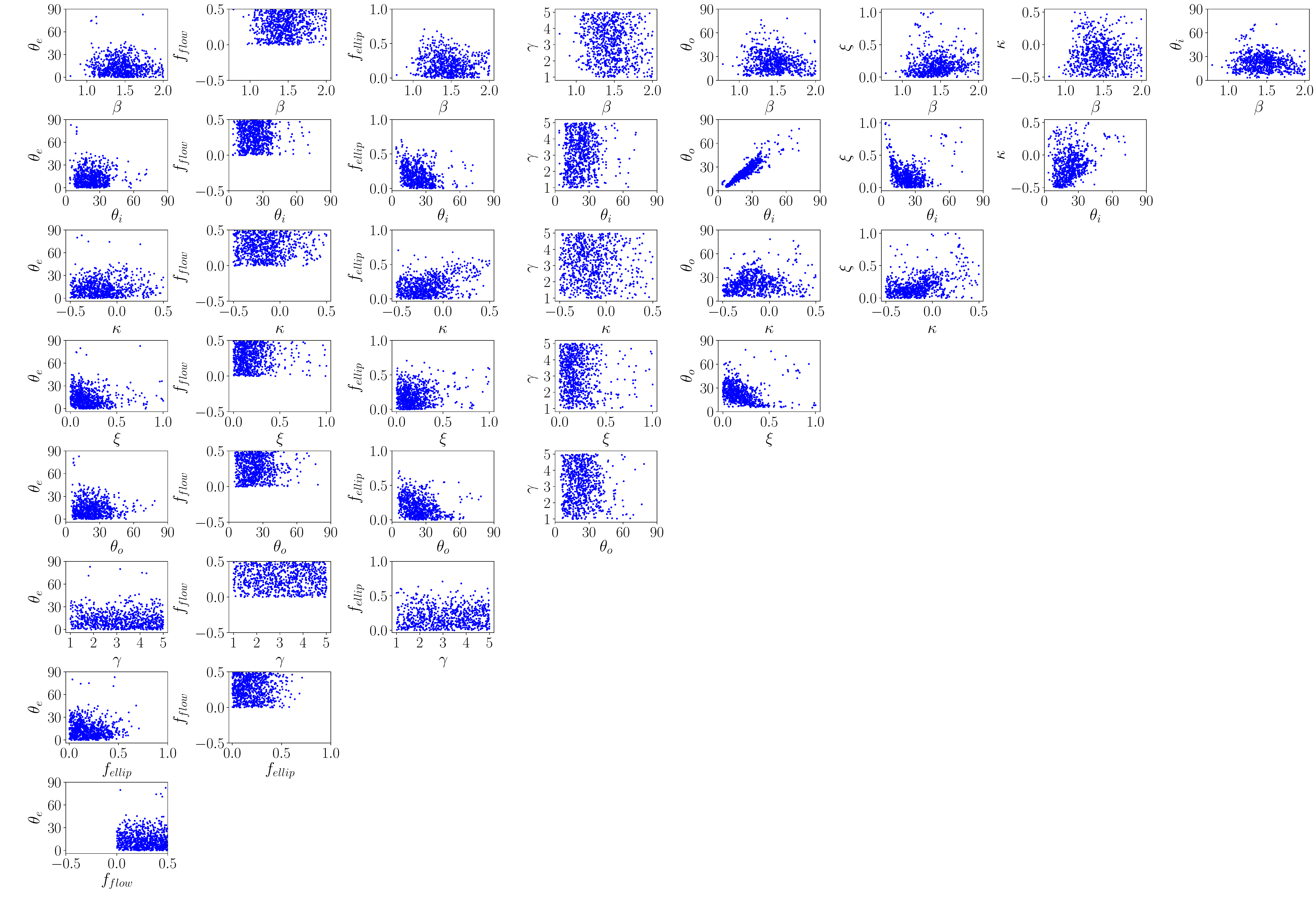}
\caption{Two dimensional posterior probability distributions for the BLR parameters.}
\label{corner_plot}
\end{figure}
\end{landscape}

\clearpage
\twocolumn
\section{Choosing the temperature for the posterior distribution}
\label{sec:appendix}

In this section we describe how the temperature parameter $T$ is chosen for each of our tests.

In BLR dynamical modelling a prior probability distribution is defined for each parameter. The knowledge on the parameters is updated by taking into account the observed data. In the case of our modified model, the information is contained primarily in the spectrum. The posterior probability distribution of the parameters is determined by applying Bayes rule to the prior distribution and the data. 
To determine the posterior distribution and the Bayesian evidence (the posterior normalisation), the model requires probing a complex multi-dimensional probability distribution. The tool used for this purpose is Diffusive Nested Sampling implemented with the algorithm DNest3 (\citealt{brewer09}, \citealt{brewer_dnest}). DNest3 starts with a `sample', which is a set of points in the parameter space (particles) drawn from the prior distribution. A sample is defined, and the particle with lowest likelihood is excluded and replaced with a new particle with a higher likelihood. The increasing likelihood levels are recorded, as smaller and smaller domains in terms of likelihood constraints are generated. To draw samples from the prior probability distribution, DNest3 uses a Markov Chain Monte Carlo method. The iteration stops when a maximum number of samples or the target likelihood level are reached. 
After DNest3 finishes its iteration of drawing and rejecting particles from the prior at increasing levels of likelihood, it is possible to do some of the analyses in post-processing. One has access to the sample distribution at each likelihood level and the possibility of defining a temperature $T$ to soften the likelihood function, i.e. to divide the logarithm of the likelihood function by a factor $T$. Setting $T$ is equivalent to increasing the uncertainties in the data so that the model can more easily reproduce the input spectra. This is useful for the case when the data have a high degree of complexity that is beyond the scope of what the model can achieve.

Here we give an example to illustrate how we make choices for the results in this paper. We use the mid-flux epoch with $\mu_{\tau}$ = 3$\times$ R-L, $\sigma_{\tau}$ = 0.2 dex for the prior on $\tau_{\rm median}$ (test [10] of Table~\ref{table_results}) as an example and analyse how the sample is distributed. Each particle is a point in the multi-dimensional parameter space. 

In Fig.~\ref{convergence_parameters_low_flux} we show the convergence distributions of a set of three parameters (black hole mass $M_{\rm BH}$, inclination angle $\theta_{i}$ and the mean time delay $\tau_{\rm mean}$) as an example. Convergence distributions show how the samples are distributed as a function of increasing likelihood level for each parameter. The $y - $axis is the parameter value and the $x - $axis shows the increasing likelihood levels from left to right. The levels in the x-axis translate into a logarithm of likelihood increasing with the number of the level so that 0 is the prior and particles at higher likelihood levels are to the right in the figure. At the top of each figure we show a Log$_{e}$(X) horizontal axis which is the natural logarithm of the compression of each level, with respect to the prior. For this particular test we can see that above a level $\sim$ 60 horizontal streaks start to appear. This may occur because the code is not finding solutions at the higher levels and is stranded in local solutions. It may be finding a solution and not being able to move from that solution or it may be slow to converge. In our modelling we limit the levels analysed to those below the start of the streaks, to ensure that the parameter space is being properly sampled. We do this by increasing the temperature. 
Increasing the temperature $T$ in post-processing means that the samples used to determine the posterior distribution will be from lower likelihood levels. 
For the Gaussian likelihood function used in our model, choosing a value of $T$ is equivalent to multiplying the spectral flux uncertainty by $\sqrt(T)$.

The vertical lines in Fig.~\ref{convergence_parameters_low_flux} indicate the level at which the posterior weight distribution peaks. The posterior weight distribution as a function of Log(X) is shown in the top panel of Fig.~\ref{convergence_parameters_low_flux}. Due to the shape of the posterior weight distribution, the posterior will be determined from samples that are from the immediate lower and higher levels around each vertical line. The three lines correspond to values of, $T = 1$ in green, $T = 3$ in black and $T = 10$ in red. In this test we choose $T = 3$ which is a compromise to ensure that the posterior distribution will be determined from levels below the start of the streaks while still choosing samples at relatively high levels. 
For the single epoch tests in this work, we use either $T = 1$ or $T = 3$. For the full light-curve, for example in our test where the Gaussian prior is added with $\sigma_{\tau} = 0.5 \mu_{\tau}$ and $\sigma_{\tau} = 3 \mu_{\tau}$, we use $T = 65$ and $T = 80$ respectively, which is similar to what \cite{pancoast18} used ($T = 65$) in their latest modelling of the full light-curve of Arp 151. 
In all cases we choose the minimum temperature that ensures the samples within the posterior weight distribution are converged.

\begin{figure*}
\includegraphics[width=1.0\textwidth]{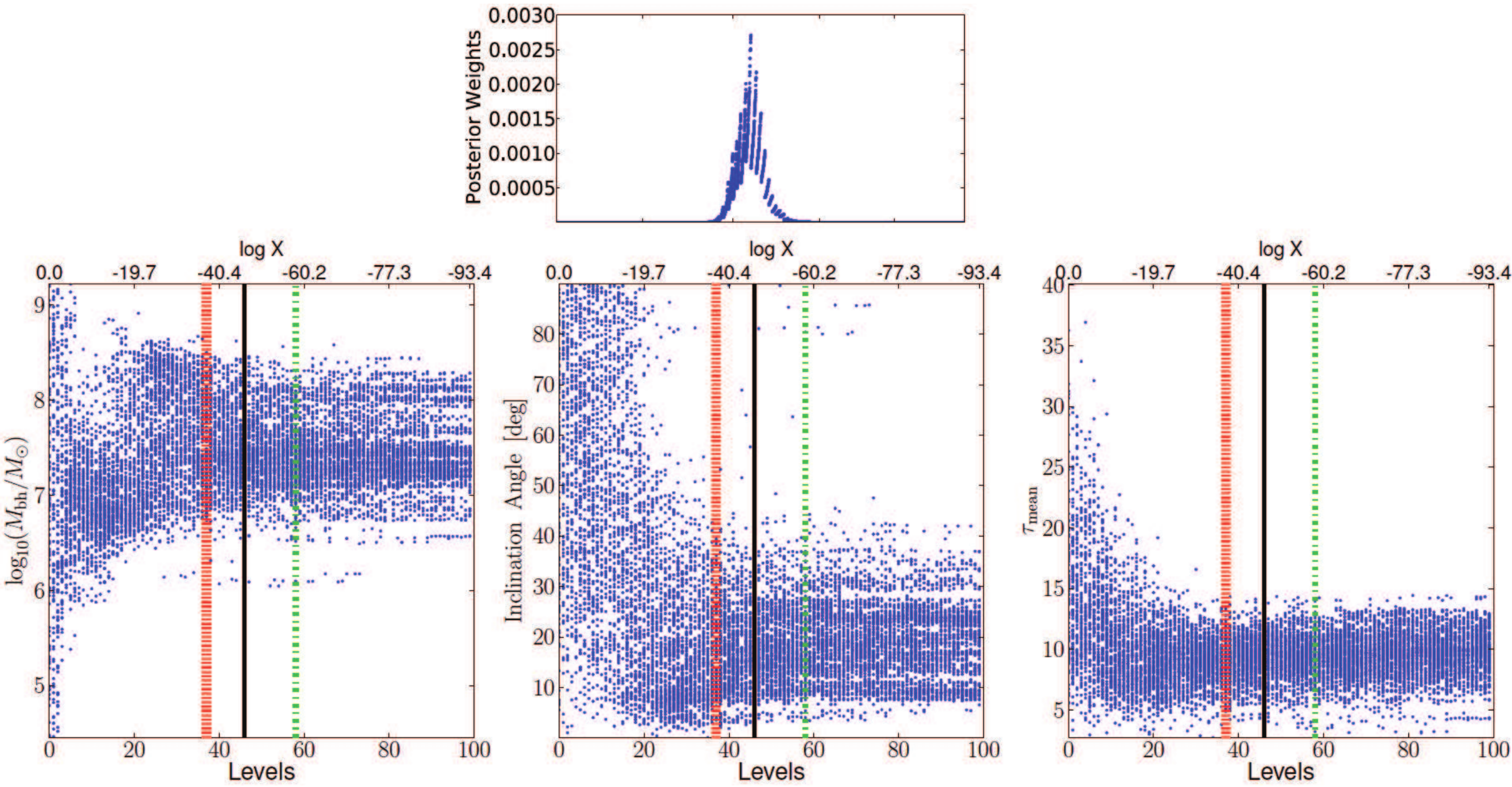}
\caption{Bottom panels show an example of the convergence distributions for three parameters in the modelling. Vertical lines indicate the level at which the posterior weight distribution peaks, for three temperature values: $T = 1$ (green dot-dashed line), $T = 3$ (black solid line) and $T=10$ (red dotted line). The top panel shows the posterior distribution for $T = 3$ with a peak matching the vertical black line of the bottom middle panel.}
\label{convergence_parameters_low_flux}
\end{figure*}



\bsp	
\label{lastpage}

\end{document}